
\input harvmac
\input diagrams.tex
\newcount\figno
\figno=0
\def\fig#1#2#3{
\par\begingroup\parindent=0pt\leftskip=1cm\rightskip=1cm\parindent=0pt
\global\advance\figno by 1
\midinsert
\epsfxsize=#3
\centerline{\epsfbox{#2}}
\vskip 12pt
{\bf Fig. \the\figno:} #1\par
\endinsert\endgroup\par
}
\def\figlabel#1{\xdef#1{\the\figno}}
\def\encadremath#1{\vbox{\hrule\hbox{\vrule\kern8pt\vbox{\kern8pt
\hbox{$\displaystyle #1$}\kern8pt}
\kern8pt\vrule}\hrule}}
\def\underarrow#1{\vbox{\ialign{##\crcr$\hfil\displaystyle
 {#1}\hfil$\crcr\noalign{\kern1pt\nointerlineskip}$\longrightarrow$\crcr}}}
%
\overfullrule=0pt

%
\def\tilde{\widetilde}
\def\bar{\overline}
\def\CN{{\cal N}} 
\def\CE{{\cal E}}
\def\Z{{\bf Z}}
\def\T{{\bf T}}
\def\S{{\bf S}}
\def\R{{\bf R}}
\def\Q{{\bf Q}}

\font\zfont = cmss10 
\font\litfont = cmr6

\def\bigone{\hbox{1\kern -.23em {\rm l}}}
\def\ZZ{\hbox{\zfont Z\kern-.4emZ}}
\def\half{{\litfont {1 \over 2}}}

\def\Re{{\rm Re ~}}

\def\p{\partial}

\def\T{{\bf T}}
\def\R{{\bf R}}
\def\C{{\bf C}}
\def\one{{\cal O}}
\def\mod{{\rm mod}}

\lref\ads{ M.F. Atiyah, H. Donnelly, and I.M. Singer, 
``Eta invariants, signature defects of cusps, and values 
of L-functions,'' Ann. Math. {\bf 118} (1983) 131.}
\lref\abs{M.F. Atiyah, R. Bott, and A. Shapiro, 
``Clifford Modules,'' Topology {\bf 3} (1964) 3.}
\lref\ahss{M.F. Atiyah and F. Hirzebruch, ``Vector Bundles 
and Homogeneous Spaces,'' Proc. Symp. Pure Math. {\bf 3} (1961) 53.}
\lref\asv{M. F. Atiyah, I. M. Singer, ``The index of 
elliptic operators: V'' Ann. Math. {\bf 93} (1971) 139.}
\lref\aps{M. F. Atiyah, V. Patodi, and I. M. Singer,
``Spectral asymmetry and 
Riemannian geometry.'' Math. Proc. Camb. Phil. Soc. 
{\bf 77} (1975) 43; Math. Proc. Camb. Phil. Soc. 
{\bf 78} (1975) 405; Math. Proc. Camb. Phil. Soc. 
{\bf 79}(1976) 71.}
\lref\bredon{G.E. Bredon, {\it Topology and Geometry}, Springer GTM 139.}
\lref\dfive{
V.M. Buhstaber, ``Modules of Differentials in The Atiyah-Hirzebruch
Spectral Sequence. I'', Math. USSR Sb. {\bf 7} (1969) 299; 
``Modules of Differentials in The Atiyah-Hirzebruch
Spectral Sequence. II'', Math. USSR Sb. {\bf 12} (1970) 59.}
\lref\cheegersimons{J. Cheeger and J. Simons, ``Differential Characters 
and Geometric Invariants,'' in {\it Geometry and Topology}, J. Alexander 
and J. Harer eds., LNM 1167.}
\lref\dougfiol{M. R. Douglas and B. Fiol, ``$D$-Branes And Discrete Torsion,
2,'' hep-th/9903031.}
\lref\fw{D.S. Freed and E. Witten, ``Anomalies in String Theory 
with D-branes,'' hep-th/9907189.}
\lref\fh{D.S. Freed and M. J. Hopkins, 
``On Ramond-Ramond Fields and $K$-Theory,''
hep-th/0002027.}
\lref\harris{B. Harris, ``Differential Characters and the Abel-Jacobi 
Map'', in {\it Algebraic $K$-theory: Connections with Geometry and 
Topology},
J.F. Jardine and V.P. Snaith eds., Nato ASI Series C: Mathematical 
and Physical Sciences  -- Vol. 279, Kluwer Academic Publishers, 
1989.} 

\lref\horava{P. Horava,`` $M$-Theory as a Holographic Field Theory,''
hep-th/9712130, Phys.Rev. {\bf D59} (1999) 046004;
P. Horava and M. Fabinger, ``Casimir Effect Between World-Branes in 
Heterotic $M$-Theory,'' hep-th/0002073.}
\lref\hull{C.M. Hull, ``Massive String Theories From 
$M$-Theory and F-Theory,''
JHEP {\bf 9811} (1998) 027, hep-th/9811021.}
\lref\milstash{J.W. Milnor and J.D.
Stasheff, {\it Characteristic Classes}, Princeton Univ. Press, 1974.}
\lref\topten{E. Witten, ``Topological Tools In Ten-Dimensional Physics'',
in {\it Unified String Theories}, 1985 Santa Barbara 
Proceedings, M. Green and D. Gross, eds. World Scientific 1986.}
\lref\vwoneloop{C. Vafa and E. Witten, ``A One Loop Test of
String Duality'',  Nucl.Phys. {\bf B447} (1995) 261, hep-th/9505053.}
\lref\fourflux{E. Witten, ``On Flux Quantization in $M$-Theory
and the Effective Action,'' hep-th/9609122; Journal of
Geometry and Physics, {\bf 22} (1997) 1.}
\lref\imwitten{E. Witten, ''Five-Brane Effective Action In $M$-Theory'',
J. Geom. Phys. {\bf 22} (1997) 103, hep-th/9610234.}
\lref\adscft{E. Witten, ``AdS/CFT
Correspondence and Topological Field Theory'',\hfill\break
JHEP {\bf 9812} (1998) 012, hep-th/98120112.}
\lref\baryons{E. Witten, ``Baryons and Branes in Anti de Sitter 
Space, JHEP {\bf 9807} (1998) 006, hep-th/9805112.}
\lref\duality{E. Witten, ``Duality Relations among
Topological Effects in String Theory,'' hep-th/9912086.}
\lref\kapustin{A. Kapustin, ``D-branes in a Topologically
Nontrivial B-field,'' hep-th/9909089.}
\lref\landweber{P. Landweber and R. Stong, ``A Bilinear Form for Spin
Manifolds'',
Trans. Amer. Math. Soc. {\bf 300} (1987) 625.}

\lref\massey{W.S. Massey, ``On the Stiefel-Whitney Classes of a
Manifold II'', Proc. AMS {\bf 13} (1962) 938.}
\lref\spingeometry{H.B. Lawson and M.-L. Michelsohn,
{\it Spin Geometry}, Princeton 1989.}
\lref\selfduality{G. Moore and E. Witten, ``Self-duality, RR fields
and $K$-Theory,'' hep-th/9912279.}
\lref\stong{R. Stong, ``Calculation of $\Omega_{11}^{spin}(K(Z,4))$'' 
in 
{\it Unified String Theories}, 1985 Santa Barbara 
Proceedings, M. Green and D. Gross, eds. World Scientific 1986}
\lref\thom{R. Thom, ``Quelques Propri\'et\'es Globales Des
Vari\'et\'es Diff\'erentiables'', Comment. Math. Helv. {\bf 28}
(1954) 17.}
\lref\ght{M. B. Green, C. M. Hull, and P. K. Townsend, ``$D$-Brane
Wess-Zumino Actions, $T$-Duality, and The Cosmological Constant,'' Phys.
Lett. {\bf B382} (1996) 65, hep-th/9604119.}
\lref\bouwknegt{P. Bouwknegt and V. Mathai,  ``$D$-Branes, $B$ Fields
and Twisted $K$-Theory,'' JHEP {\bf 0003} (2000) 007, hep-th/0002023.
}
\lref\as{M. F. Atiyah and G. Segal, to appear.}

\Title{\vbox{\baselineskip12pt\hbox{hep-th/0005090}
\hbox{IASSNS-HEP-00/39}}}
{\vbox{\centerline{$E_8$ Gauge Theory, }
\medskip
\centerline{and a Derivation of $K$-Theory from
$M$-Theory   }}}
\centerline{Duiliu-Emanuel Diaconescu}
\smallskip
\centerline{\it School of Natural Sciences, Institute for Advanced Study}
\centerline{\it Olden Lane, Princeton, NJ 08540, USA}
\smallskip
\centerline{Gregory Moore}
\smallskip
\centerline{\it Department of Physics, Rutgers University}
\centerline{\it Piscataway, New Jersey, 08855-0849}
\smallskip
\centerline{Edward Witten$^*$}
\smallskip
\centerline{\it Dept. of Physics, Caltech, Pasadena, CA 91125}
\smallskip\centerline{\it and CIT-USC Center for Theoretical
Physics, USC, Los Angeles CA}
\medskip


\noindent
The partition function of Ramond-Ramond $p$-form fields in Type IIA
supergravity on a ten-manifold $X$
contains subtle phase factors that are associated with
$T$-duality, self-duality, and the relation of the RR fields
to $K$-theory.  The analogous partition function of $M$-theory on
$X\times \S^1$ contains subtle phases that are similarly
associated with $E_8$ gauge theory.  We analyze the detailed phase
factors on the two sides and show that they agree, thereby
testing $M$-theory/Type IIA duality as well as the $K$-theory formalism
in an interesting way.  We also show that certain $D$-brane states
wrapped on nontrivial homology cycles are actually unstable, 
that $(-1)^{F_L}$ symmetry in Type IIA superstring theory depends in general
on a cancellation between a fermion anomaly and an anomaly of
RR fields, and that Type IIA superstring theory with no wrapped branes
is well-defined only on a spacetime with $W_7=0$.

\smallskip
\noindent
$^*$ On leave from Institute for Advanced Study, Princeton, NJ 08540.
\Date{May 4, 2000}
\newsec{Introduction }

The Ramond-Ramond (RR) $p$-form fields of Type II superstring theory
are rather subtle objects, even in a limit in which they are treated
as free fields.  The subtlety arises in part from the fact that
they are self-dual, and hence difficult to  understand fully from a
Lagrangian point of view, and in part from the fact that the
$D$-branes that they couple to have a $K$-theory interpretation,
and hence the RR fields themselves must be interpreted in $K$-theory.
A framework for incorporating self-duality in the $K$-theory framework
has been proposed \duality\ and used to
resolve puzzles associated with global brane anomalies for Type IIA
\selfduality.   A unified approach to brane anomalies
for both Type IIA and Type IIB in the $K$-theory framework has also been
proposed \fh.

In the present paper, we will focus primarily on the Type IIA
theory.  The RR partition function for Type IIA on a ten-manifold $X$
can be written
as a sum over the fluxes or periods of the RR fields, which are
forms $G_0, G_2,$ and $G_4$ of even order.  (The $G_{2p}$ of $p>4$
are the electric-magnetic duals of these.)  In the sum, subtle
phase factors appear which are associated with the $K$-theory
interpretation of RR fields, and thereby with $U(N)$ gauge theory.

One can also attempt to compare Type IIA on $X$ to $M$-theory on
$X\times \S^1$,  or more generally on a circle bundle over $X$ if
$G_2\not= 0$.  In the $M$-theory framework, $G_4$ is identified with
a component of the
 $M$-theory four-form, and $G_2$ with the first Chern class of the circle
bundle.  $G_0$ has no known $M$-theory origin.  In the sum over
periods of $G_2 $ and $G_4$ in $M$-theory, there appear subtle phase
factors.  Locally, these arise from the familiar Chern-Simons interaction
of eleven-dimensional supergravity, but to define the phase factors
globally is a subtle story that involves use of $E_8$ gauge theory
\fourflux.

The purpose of the present paper is to compare the partition function of the
RR fields as computed in Type IIA using self-duality and $K$-theory
to the corresponding partition function computed in $M$-theory.
We will find a nice match, which is a satisfying test of the
$M$-theory/Type IIA duality and of the $K$-theory and $E_8$ gauge theory
formalisms.

Throughout this paper, we keep the metric on $X$
fixed.  Moreover, for comparing   to $M$-theory we take this
metric large in string units.
We perform quantum mechanics only for the RR fields and only in the free
field approximation.  Everything will come down to comparing the
phases that appear on the two sides.

When the metric of $X$ is scaled up  by $g\to t g$, the action
$\int d^{10}x \sqrt g |G_{2p}|^2$   for the Type IIA RR
$2p$-form  scales
as $t^{5-2p}$.  This implies that there is a sensible approximation
of keeping only $G_4$, whose periods have the smallest action,
a sensible and better approximation of including
$G_4$ and $G_2$, and finally one can do a complete calculation with
$G_0$, $G_2$, and $G_4$ all included.
To keep things relatively simple, while also (as it turns out)
exhibiting most of the significant issues, we consider first
the case of the contribution of $G_4$.  We review the phase
factor in the sum over $G_4$ flux in $M$-theory in section 2
and specialize to compactification on a circle in section 3. Then
we  introduce
some necessary mathematical notions in section 4, and 
explain their role in Type II superstring theory
in section 5.  In section 6, we 
analyze the $M$-theory
partition function with $G_4$ only, and in section 7 we demonstrate
its agreement with the Type IIA partition function computed using
self-duality and $K$-theory.
In sections 8 and 9, we extend the analysis to include $G_2$,
again showing complete agreement between the two sides.
In section 10, we extend the Type IIA analysis to include $G_0$,
but as $G_0$ has no known interpretation in $M$-theory,
we are not able to compare the complete partition function to an $M$-theory
result.
In section 11, we explain a puzzle concerning the incorporation
of the Neveu-Schwarz $H$-field and $S$-duality, and in section
12, we speculate about possible physical interpretation of
some elements of the discussion.

Our notation is summarized in Appendix A.
Appendix B contains details needed to complete one of the computations.
Appendix C contains an outline of the Atiyah-Hirzebruch spectral
sequence (AHSS) that compares $K$-theory to cohomology and underlies
some of our considerations. Appendix D contains a technical 
argument for the existence of manifolds with nontrivial 
$W_7$.

\newsec{Review of the Phase of the $M$-Theory Effective Action}

We begin by recalling the origin in $M$-theory of the subtle phases
that will be the focus of the present paper.

As predicted by Nahm's theorem \ref\nahm{W. Nahm,
``Supersymmetries And Their Representations,'' Nucl. Phys.
{\bf B135} (1978) 149.}, eleven-dimensional
supergravity has a three-form field $C$, whose field strength we will
denote as $G=dC$.  Ever since the original construction of the classical
theory
\ref\julia{E. Cremmer, B.  Julia, and J. Scherk, ``Supergravity Theory
In 11 Dimensions,'' Phys. Lett. {\bf 76B} (1978) 409.},
it has been known that the Lagrangian contains a Chern-Simons
interaction, roughly
\eqn\polygo{\int_Y C \wedge G \wedge G,}
where $Y$ is the $M$-theory spacetime.

$M$-theory has fivebranes,
and  $G$ can have nonvanishing periods.  Hence $C$ is not globally
defined as a three-form, and one must ask whether the Chern-Simons
coupling is globally defined.
One might expect that, after determining the correct value of the
quantum of $G$ flux, the Chern-Simons interaction would be globally
well-defined mod $2\pi $.  For this, it should be an integral
multiple of
\eqn\nolygo{{1\over (2\pi)^2}\int_Y C \wedge G \wedge G,}
with $C$ and $G$ normalized so that the periods of $G$
are integer multiples of $2\pi$.

It turns out, however, that the Chern-Simons coupling of $M$-theory
is smaller than this by a factor of 6 \fourflux.  So how can the
theory be consistent?  First of all, there are a variety of gravitational
corrections to the classical Chern-Simons interaction.  There
is a coupling $\int C\wedge I_8(R)$, where $I_8(R)$ is an eight-form
constructed as a quartic polynomial in the Riemann tensor $R$.  
In addition,
there is a gravitational correction to the quantization law of $G$, 
by virtue
of which the periods of $G/2\pi$ are not in general integers.  Rather,
the general condition is that for any four-cycle $U$ in spacetime,
\eqn\olbo{\int_U{G\over 2\pi}={1\over 2}\int_U\lambda ~{\rm mod}~\Z.}
Here $\lambda$ is defined as follows.  The first Pontryagin class
of a spin manifold is always divisible by 2, and there is a characteristic
class $\lambda$ (represented in de Rham cohomology by
$\tr\,R\wedge R/16\pi^2$) such that $2\lambda = p_1(Y)$.  Thus, the periods
of $G/2\pi$ are integral or half-integral for cycles on which $\lambda$
is even or odd.

\def\Pf{{\rm Pf}}
There is, accordingly,
 an integral characteristic class, which we will call $a$,
\foot{For an explanation of our notation see appendix A. }
   that can be represented
in de Rham cohomology as 
\eqn\geea{
G/2\pi = a  -\lambda/2
.}
  The full Chern-Simons coupling
of $M$-theory is associated with
the twelve-form
\eqn\ilogo{B_{12}={2\pi \over 6}\left(a-{\lambda\over 2}\right)
\left(\left(a-{\lambda\over 2}\right)^2-{1\over 8}\left(p_2(Y)-\lambda^2\right)
\right).}  Here we have taken into account that $G/2\pi$ corresponds
to $a-\lambda/2$. Thus, \ilogo\ corresponds to $(2\pi/6)(G/2\pi)^3$ plus
corrections related to $CI_8(R)$.

In the usual fashion, one can try to define the Chern-Simons coupling
in eleven dimensions
by integrating $B_{12}$ over a twelve-manifold.  Indeed, by a result of
Stong \stong, there exists a twelve-dimensional spin
manifold $Z$ with boundary $Y$ over which $a$ extends.
We interpret the Chern-Simons integral as a factor
\eqn\ploog{\exp\left(i\int_Z B_{12}\right) }
in the path integral.  We must ask a key question:
does this phase depend on the choice of $Z$?  Such a dependence would
not be physically acceptable, since the effective action should depend
only on the spacetime $Y$, and not on the auxiliary choice of a twelve-manifold
$Z$ that is introduced for convenience in computation.

Before considering the $Z$-dependence, let us note that
\ploog\ is  not quite the only problematic factor in the $M$-theory
path integral at long distance.  One must also worry about the Pfaffian
(or square root of the determinant) of the Rarita-Schwinger operator $D_{RS}$.
Though this operator is actually a more delicate
construction, for our purposes (which mostly involve index theory of
various sorts), $D_{RS}$ is the Dirac operator 
$\Dsl$ coupled to $TY - 3 \CO$. Here $TY$ is the tangent bundle
to $Y$, and $\CO$ is a trivial line bundle.  The subtraction of $3\CO$
accounts for 
the ghost fields required to fix the gauge invariances of the Rarita-Schwinger 
operator.\foot{
The tangent bundle $TZ$ to $Z$ splits near $Y$ as $TZ=TY\oplus \CO$,
so by the Rarita-Schwinger operator in twelve dimensions, we will mean
the Dirac operator coupled to $TZ-4\CO$.  On the other hand,
if $Y=\S^1\times X$ or more generally if $Y$ is a circle bundle
over $X$, then $TY$ can be replaced by $TX\oplus \CO$, and hence
in Type IIA, the Rarita-Schwinger operator (including the dilatino as well
as the ghosts)
is equivalent to a Dirac operator coupled to $TX-2\CO$.}   
This Pfaffian, which we will write as $\Pf(D_{RS})$, is real but
not positive definite; there is in general no natural definition
of its sign.  (Mathematically, $\Pf(D_{RS})$ is interpreted as a vector
in a Pfaffian line rather than a real number.)  The problematical
factors in the $M$-theory effective action are thus the product
\eqn\inno{\Pf(D_{RS}) \exp\left(i\int_ZB_{12}\right).}

Although our interest in the present paper will mainly be on the second
factor in \inno, we pause to explain how to deal with the first factor.
(This will enable us to fill in a gap in the discussion in \fourflux,
where the effective action was proved to be anomaly-free, but no absolute
definition of its phase was given.)  In the abstract there
is no natural definition of the sign of $\Pf(D_{RS})$, but once a
twelve-dimensional
spin manifold $Z$ of boundary $Y$ is chosen, there is a natural way
to define the sign.\foot{The following observation is in the spirit of
\ref\oldfreed{D. Freed, ``Determinants, Torsion and Strings'', 
Commun. Math. Phys. {\bf 107} (1986) 473, with an appendix by D. Freed 
and J. Morgan, p. 510.}, and the idea was explained to us by 
D. Freed in
commenting on \ref\otherwitten{E. Witten, ``World-Sheet Corrections
Via $D$-Instantons'', hep-th/9907041.}, where a similar treatment 
was given for the
heterotic string using the Dai-Freed theorem.}  Let $I_{RS}$ be the
index of the Rarita-Schwinger operator on $Z$, computed with
Atiyah-Patodi-Singer (APS) global boundary conditions \aps. 
This index
is always even in 12 dimensions (or more generally in $8k+4$ dimensions
for any $k$).  We define the phase of $\Pf(D_{RS})$ to be
$(-1)^{I_{RS}/2}$, or equivalently we define $\Pf(D_{RS})$ itself as
\eqn\tygo{\Pf(D_{RS})=(-1)^{I_{RS}/2}|\Pf(D_{RS})|,}
where the absolute value $|\Pf(D_{RS})|$ can be defined by zeta function
regularization and has no anomaly.  The rationale for this definition
is as follows.   As the metric on $Y$ is varied, a pair of
eigenvalues of $D_{RS}$ may pass through zero, in which case the sign of
$\Pf(D_{RS})$ should jump.  Precisely when this occurs, $I_{RS}$ jumps
by $\pm 2$, so the right hand side of \tygo\ changes sign precisely
when the left hand side should.  Hence, \tygo\ gives a continuously
varying (but $Z$-dependent) definition of  $\Pf(D_{RS})$.
If we use this definition, then the phase factor that must be considered
in the $M$-theory path integral
is
\eqn\rygo{\Phi(Z)=(-1)^{I_{RS}/2}\exp\left(i\int_Z B_{12}\right).}

Each factor here has been defined in a way that depends on $Z$.  How
can we prove that the product does not depend on $Z$?
The key \fourflux\ is to use $E_8$ gauge theory.
The role of $E_8$  does not fall completely out of the sky.
If $Y$ has a boundary, there is an $E_8$ vector supermultiplet propagating
on the boundary.  As shown in \ref\hw{P. Horava and E. Witten, 
``Heterotic and Type I String Dynamics from Eleven Dimensions'',
Nucl.Phys. {\bf B460} (1996) 506, hep-th/9510209; 
``Eleven-Dimensional Supergravity on a Manifold with Boundary'',
Nucl.Phys. {\bf B475} (1996) 94, hep-th/9603142.}, anomaly
cancellation along the boundary depends (among
other things) on an anomaly inflow from the bulk, along the general
lines considered in \ref\calhar{C. G. Callan, Jr. and J. A. Harvey,
``Anomalies and Fermion Zero Modes on Strings and Domain Walls'',
Nucl. Phys. {\bf B250} (1985) 427.}.
The anomaly inflow depends upon the fact that the Chern-Simons coupling
is not gauge-invariant on an eleven-manifold with boundary.  To cancel
the anomalies, the key fact is that $B_{12}$ has a relation to $E_8$ and
Rarita-Schwinger index theory that we will now recall.

The homotopy groups $\pi_i(E_8)$ vanish for $1\leq i\leq 14$ except
for $i=3$, where we have $\pi_3(E_8)=\Z$.  Consequently, an
$E_8$ bundle $V$ on a twelve-manifold (or any manifold of dimension less
than sixteen) is completely classified topologically
by a four-dimensional characteristic
class, which is represented in de Rham cohomology by $\tr \,F\wedge F/16\pi^2$.
(Here $F$ is the curvature of the bundle, and
$\tr $ is $1/30$ times the trace in the adjoint representation.)
In particular, we can pick $V$ so that its characteristic
class is the class $a$ associated with the $C$-field of $M$-theory.
It is convenient if we can pick the connection on $V$ so that
the differential form $\tr \, F\wedge F/16\pi^2$ is precisely
equal to the differential form $a=G/2\pi +\lambda/2$ that appears in
the definition of $B_{12}$.
But it is not really essential to be able to do this.
For any choice of connection on $V$, these two four-forms  will be equal
 for some
$C$-field in its given topological class.  If
the phase of the $M$-theory effective action
is well-defined (independent of the
choice of $Z$) for some $C$-field
in a given topological class, then it follows that this effective action
is well-defined for every $C$-field in that class.  For the change in the
Chern-Simons
coupling when $C$ is continuously
varied is given by a well-defined local integral over $Y$
with no global subtleties.

Let $I_{E_8}$ be the index of the Dirac operator $D_{E_8}$
on $Z$, coupled to the
$E_8$ bundle $V$, with APS global boundary conditions.
Like $I_{RS}$, it is always even.  On a twelve-manifold $Z$ without
boundary, $I_{E_8}$ and $I_{RS}$ are given by the index theorem
in terms of the integrals over $Z$ of  certain twelve-forms $i_{E_8}$ 
and
$i_{RS}$.
The crucial
property of $B_{12}$ is that it can be expressed
in terms of these forms \refs{\hw,\fourflux}:
\eqn\uju{{B_{12}\over 2\pi}= {i_{E_8}\over 2} +{i_{RS}\over 4}.}
Hence, on a twelve-manifold $Z$ without boundary, one has
\eqn\puju{\int_Z{B_{12}\over 2\pi}={I_{E_8}\over 2}+{I_{RS}\over 4}.}
Inserting this in \rygo, and using the fact that $I_{E_8}$ and $I_{RS}$
are even, one finds that $\Phi(Z)$ equals $+1$ for any closed twelve-manifold
$Z$, independent of $Z$.  Thus $\Phi(Z)$ is independent of $Z$ if $Z$ has no
boundary.

It now follows, with a little more work, that
 $\Phi(Z)$ is independent of $Z$ also if $Z$ has a given boundary
$Y$.  The idea
here is that if $Z$ and $Z'$ are twelve-dimensional spin-manifolds
with the same boundary $Y$, and $\bar Z= Z\cup (-Z')$ is the closed
twelve-manifold
obtained by gluing $Z$ and $Z'$ along their boundary with a reversed
orientation for $Z'$, then
\foot{The multiplicativity that leads to the following formula
is clear for the second factor in the definition of $\Phi(Z)$.  It  also
holds for the first factor, since $I_{RS}(\bar Z)
=I_{RS}(Z)+I_{RS}(Z')=I_{RS}(Z)-I_{RS}(-Z')$.
(Reversing the orientation of $Z'$ changes the sign of the index:
$I_{RS}(-Z')=-I_{RS}(Z')$.)
This can be proved by picking on $\bar Z$ a
metric that near the common boundary $Y$ of $Z$ and $Z'$
has a long collar looking like $Y\times J$, where $J$ is
an interval in $\R$ very long compared to the radius of $Y$.  We also
assume on $Y$ a generic metric such that $D_{RS}$ has no zero  modes on $Y$.
Then a Rarita-Schwinger zero mode on $\bar Z$ converges (as $J$ becomes
very long) to a sum of zero modes on the two sides.  This leads to the
additivity of $I_{RS}$.}
 \eqn\plopo{\Phi(\bar Z)=\Phi(Z)\Phi(Z')^{-1}.}
   So, as $\Phi(\bar Z)=1$,
it follows that $\Phi(Z)=\Phi(Z')$, as promised.

This shows the well-definedness of the phase of the $M$-theory effective
action.  For our computations, however, it will generally be
inconvenient  actually to pick a $Z$ with boundary $Y$.  Instead,
we will use an alternative expression that follows from the
APS theorem for the index of the Dirac operator on a manifold with boundary.

Consider a Dirac operator $D$ (such as $D_{E_8}$ or $D_{RS}$) on
the eleven-dimensional spin manifold $Y$.  Let $\lambda_i$ be its
eigenvalues (which are real).  Atiyah, Patodi, and Singer define the function
\eqn\olpo{\eta(s) =\sum_{\lambda_i\not= 0}({\rm
sign}\,\lambda_i)|\lambda_i|^{-s}.}
The sum
 converges for sufficiently large $s$ and
has an analytic continuation to $s=0$. The value $\eta(0)$ is commonly
called simply $\eta$.  Let $h$ be the number of
zero modes of $D$.  Now, suppose that $Y$ is the boundary of a spin
manifold $Z$ (over which any data, such as an $E_8$ bundle, used in
defining $D$
are extended), and let $I(D)$ be the index of the extended operator
$D$ on $Z$, defined with APS global boundary conditions.  Then the APS theorem
(Theorem 4.2 in \aps) asserts that
\eqn\yolpo{I(D)=\int_Z i_D-{h+\eta\over 2},}
where $i_D$ is the twelve-form whose integral on a closed
twelve-manifold would equal $I(D)$.

In our case, we want a formula for $\int_Z B_{12}/2\pi$, which is
related to index densities by \uju.  So from \yolpo\ we get:
\eqn\pooko{\int_Z{B_{12}\over 2\pi}={1\over 2}I_{E_8}+{1\over 4}I_{RS}
   +{h_{E_8}+\eta_{E_8}\over 4}+{h_{RS}+\eta_{RS}\over 8 }.}
If we insert this in the formula \rygo\ for the phase $\Phi$, then
the $I_{E_8}$ and $I_{RS}$ terms can be dropped (using the fact that
these indices are both even).  We get that the phase of the effective
action is
\eqn\olooko{\Phi=\exp\left(2\pi i\left((h_{E_8}+\eta_{E_8})/4
+(h_{RS}+\eta_{RS})/8\right)\right).}

This  formula for the phase  manifestly depends only
on the fields on $Y$ -- not on an extension to $Z$.  In general,
one might expect that it would be hard to use, since the $\eta$ invariant
is a rather subtle thing.  But it turns out that in the situations
that we will encounter in the present paper (circle fibrations with
the Neveu-Schwarz $B$-field set to zero, i.e. with $G$ pulled
back from 10 dimensions), the expression
\olooko\ for the phase is very useful.

Let us verify using this formula that the  integrand in the path integral
is a smooth function of the metric on $Y$.  In $8k+3$ dimensions,
the eigenvalues of the Dirac operator coupled to a real vector bundle
(such as the $E_8$ bundle $V$ or the tangent bundle
$TY$) have a two-fold degeneracy
that comes from complex conjugation.
As the metric on $Y$ is varied, a pair of zero modes of $D_{E_8}$
can pass through zero.  Whenever an eigenvalue passes through zero,
$\eta$ jumps by $\pm 2$.  So the jumps in $h_{E_8}+\eta_{E_8}$ are
by multiples of 4.  Clearly this causes no discontinuity in $\Phi$.
On the other hand, $h_{RS}+\eta_{RS}$ likewise jumps by $\pm 4$ when
a pair of eigenvalues of $D_{RS}$ pass through zero.  When this occurs,
$\Phi$ changes sign.   We must remember, however, that the problematic
factors in the path integral measure are not just $\Phi$ but the
product $\Phi\, |\Pf(D_{RS})|$ of $\Phi$ with the absolute value of the
Rarita-Schwinger Pfaffian.  When a pair of eigenvalues of $D_{RS}$ passes
through zero, the Pfaffian $\Pf(D_{RS})$ changes sign, and its
absolute value $|\Pf(D_{RS})|$ is continuous but not smooth.  The sign
change in $\Phi$ whenever $\Pf(D_{RS})$ has a zero of odd order is precisely
right so that the product $\Phi\,|\Pf(D_{RS})|$ varies smoothly.

\newsec{Reduction to Ten Dimensions and the Mod Two Index}

As explained in the introduction, our initial goal will be to consider
the case that $Y=X\times \S^1$, where $X$ is a ten-dimensional spin manifold,
and we use the supersymmetric (nonbounding) spin structure in the $\S^1$
direction.  Moreover, we assume that the $M$-theory $C$-field is pulled
back from a $C$-field on $X$.
For this situation, we hope to compare $M$-theory on $Y$ to Type IIA
on $X$.

Note that $X\times \S^1$ with $C$ a pullback from $X$
has an orientation-reversing symmetry.
Under this symmetry, the Chern-Simons coupling reverses sign, and
the phase $\Phi$ is complex-conjugated.  So in this situation, $\Phi$
must equal $\pm 1$.  As $\Phi$ takes values in a discrete set, it is
a topological invariant in this situation.   This is true for any value
of the  characteristic class $a\in H^4(X;\Z)$  of the $C$-field.

Let us see how this works out in terms of
\olooko.  The Dirac operator changes sign under
reflection of one coordinate, so the nonzero eigenvalues are in pairs
$\lambda,-\lambda$.  Hence the $\eta$-invariants are zero.
So in this situation, the phase is just
\eqn\oblooko{\Phi=\exp\left(2\pi i(h_{E_8}/4+h_{RS}/8)\right).}

Let us analyze $h_{E_8}$ and $h_{RS}$.  A zero mode of the Dirac operator
on $X\times \S^1$ must be constant in the $\S^1$ direction, so it is
equivalent to a zero mode of the Dirac operator on $X$.  Such zero
modes may have either positive or negative chirality, and we
have $h_{E_8}=h^+_{E_8}+h^-_{E_8}$, $h_{RS}=h^+_{RS}+h^-_{RS}$, where
$h^\pm$ are the numbers of positive and negative chirality zero modes.
Because complex conjugation reverses the chirality while mapping
the real bundles $V$ and $TX$ to themselves, we have $h^+_{E_8}
=h^-_{E_8}$, $h^+_{RS}=h^-_{RS}$.  So we can write the phase
as
\eqn\noblooko{\Phi=(-1)^{h^+_{E_8}}i^{h^+_{RS}}.}

In $8k+2$ dimensions, the number of positive chirality zero modes
of the Dirac operator with values in a real vector bundle is
a topological invariant mod 2 \asv.
Hence $h^+_{E_8}$ in particular is a topological invariant mod 2.  We will
denote this mod 2 invariant as $f(a)$.

Likewise, $h^+_{RS}$ is a topological invariant mod 2.
This is not enough to prevent jumping of the sign of $\Phi$, but we must
remember that whenever $h^+_{RS}$ is nonzero, the other factor
$|\Pf(D_{RS})|$, which multiplies $\Phi$, is zero.  If the Rarita-Schwinger
mod 2 index is zero, then $h^+_{RS}$ is generically zero and
the phase reduces to
\eqn\poblooko{\Phi_a=(-1)^{f(a)}.}
Even when the Rarita-Schwinger mod 2 index is nonzero, the $a$-dependence
of the phase of the effective action, which is what we will primarily
study in the present paper, is given by \poblooko.
(We will consider the effect of Rarita-Schwinger zero modes in 
section 3.3, where we compute the anomaly in $(-1)^{F_L}$.)
Note that $f(a)=0$ for $a=0$, for in this case the $E_8$ bundle
$V$ is 248 copies of a trivial bundle, and has a vanishing mod 2 index.

Our next goal will be to discover useful properties of the $E_8$ mod
2 index $f(a)$.

\subsec{Analysis Of The $E_8$ Mod 2 Index}

As will gradually become clear, it is hopeless to find an elementary
formula for $f(a)$.  However, it is possible to find a relatively
simple algebraic identity obeyed by $f(a)$, and to deduce what we need
from this identity.

An analogy with Type IIA and $K$-theory may be useful.
The Type IIA partition function can be expressed
as a sum over RR fluxes, which are classified
by $K(X)$.   An important role in writing the partition function
is played by a mod 2 invariant $j(x)$.
For $x\in K(X)$, $j(x)$ is defined as the mod 2 index with values
in the $KO$ (or real $K$-theory) class $x\otimes \bar x$.
There is no elementary formula for $j(x)$, but there is a useful
algebraic identity:
\eqn\lollypop{j(x+y)=j(x)+j(y)+\omega(x,y),}
where here $\omega(x,y)$ is defined as the ordinary index of the Dirac
operator with values in $x\otimes \bar y$.  There is an elementary
formula for this index (from the Atiyah-Singer index theorem)
so \lollypop\ says that the difference
$j(x+y)-j(x)-j(y)$ is a much more elementary
invariant than $j$ itself.
In proving \lollypop, one uses \duality\
the following property of the mod 2 index.
Suppose that $V$ is a real vector bundle whose complexification splits
as $W\oplus \bar W$, where $W$ is a complex vector bundle.  Let $q(V)$
denote the mod 2 index with values in $V$ and $I(W)$ the ordinary index
with values in $W$.  Then
\eqn\ollypop{q(V)=I(W)~{\rm mod}~2.}
(It is also true that $q(V)=I(\bar W)$ mod 2; indeed, in 
$8k+2$ dimensions,
$I(\bar W)=-I(W)$.)

\def\ch{{\rm ch}}
We need an analog of \lollypop\ for the $E_8$ mod 2 index.
First we introduce an important concept.
We will say that an element $a\in H^4(X;\Z)$ can be ``lifted to $K$-theory''
if there exists, for some $N$,
 a rank $N$ complex vector bundle $E$ with $c_1(E)=0$,
$c_2(E)=-a$.  The rationale behind this definition is that, in Type IIA
superstring theory, RR fields are classified topologically by an element
$x$ of $K(X)$.  The relation is
\eqn\umu{{G\over 2\pi}=\sqrt{\hat A}\,\,\ch\,x.}
For $G_0=G_2=0$, which is the case related to $M$-theory on $X\times
\S^1$, this implies in particular
\eqn\jumu{\eqalign{{G_4\over 2\pi}& = -c_2(x)\cr
                    {G_6\over 2\pi}& = {c_3(x)\over 2}.\cr}}
Thus, when $G_0=G_2=0$, $G_4/2\pi$ is in Type IIA always minus the
second Chern class of a $K$-theory element.
Consequently, a $G$-field
in $M$-theory with characteristic class $-a$ has a straightforward
interpretation in Type IIA only if $a$ can be lifted to $K$-theory.
We will eventually see that this restriction on $a$ arises in $M$-theory
in a more roundabout way.

If $a$ is minus the second Chern class of an $SU(N)$ bundle for some $N$,
we want a bound on how big $N$ must be.  To reduce the structure
group of an $SU(N+1)$ bundle $E$ to $SU(N)$, one needs to pick a section $s$
of $E$ that is everywhere nonzero.  One can scale $s$ so $|s|^2=1$
everywhere, in which case $s$ is a section of the bundle $U$ of unit
vectors in $E$.  The fibers of $U$ are copies of $\S^{2N+1}$.
Over a base space of dimension $k$, the obstruction to finding a section
of $U$ is controlled by $\pi_i({\S^{2N+1}})$ for $i<k$; these groups
vanish if $2N+1\geq k$.\foot{The reader may want to consult 
\refs{\topten} for an introduction to obstruction theory
for physicists.} So in particular, for $k=10$,
we can always reduce the structure group of an $SU(N)$ bundle to 
$SU(5)$.

Let $a,a'$ be elements of $H^4(X;\Z)$ that lift to $K$-theory.  From
what has just been said, we can assume that there are $SU(5)$ bundles
$E,E'$ with $c_2(E)=-a$, $c_2(E')=-a'$.  We want to compute
$f(a+a')-f(a)-f(a')$.
{}From the two $SU(5)$ bundles $E$ and $E'$, we can, using the existence
of an embedding  $(SU(5)\times SU(5))/\Z_5 \subset E_8$, construct
in a natural way an $E_8$ bundle whose characteristic class is $a+a'$.
We simply embed $E$ in the first $SU(5)$ and $E'$ in the second.
In the same way, replacing $E$ or $E'$ by a rank five trivial bundle,
we get an $E_8$ bundle with characteristic class $a$ or $a'$.

The decomposition of the $E_8$ Lie algebra under $SU(5)\times SU(5)$
is
\eqn\jurry{{\bf 248}=({\bf 24},{\bf 1})\oplus ({\bf 1},{\bf 24})
\oplus ({\bf 5},{\bf 10})\oplus ({\bf \bar 5},{\bf\bar {10}})
\oplus ({\bf 10},{\bf \bar 5})\oplus ({\bf \bar {10}},{\bf 5}).}
Here ${\bf 5}$ is the fundamental representation of $SU(5)$, ${\bf 10}$
is its second antisymmetric power, and ${\bf 24}$ is the adjoint
representation.
When we compute $f(a+a')-f(a)-f(a')$, the contributions of the
$({\bf 24},{\bf 1})$ and $({\bf 1},{\bf 24})$ cancel out.
The mod 2 index with values in $({\bf 5},{\bf 10})\oplus ({\bf\bar 5},
{\bf \bar {10}})$ is, from  \ollypop, the mod 2 reduction of
the ordinary index of the $({\bf 5},{\bf 10})$, and likewise
the mod 2 index with values in $({\bf 10},{\bf\bar 5})\oplus
({\bf\bar {10}},{\bf 5})$ is the mod 2 reduction of the ordinary
index with values in $({\bf 10},{\bf\bar 5})$.

Let $\wedge^2 E$, $\wedge^2E'$ denote the bundles associated to $E$ and
$E'$, respectively, in the ${\bf 10}$ of
$SU(5)$.  We need to compute the ordinary index with values in
$E\otimes \wedge^2E'\oplus
\wedge^2E\otimes \bar E'$.
To compute $f(a+a')-f(a)-f(a')$, we only want terms in the index
formula that involve Chern classes of both $E$ and $E'$.  In general,
in ten dimensions,
the terms in the index formula for $A\otimes B$ that involve Chern
classes of both $A$ and $B$ are (if $c_1(A)=c_1(B)=0$)
\eqn\olox{-\int_X{c_2(A)c_3(B)+c_3(A)c_2(B)\over 2}.}
We have
\eqn\nolox{c_2(\wedge^2 E)=3c_2(E),\,c_3(\wedge^2E)=c_3(E),}
and similarly for $E'$.
Using \olox\ and \nolox, the contribution to the index of
$E\otimes \wedge^2E'\oplus \wedge^2E\otimes \bar E'$ that survives
in $f(a+a')-f(a)-f(a')$ is
\eqn\pluxo{\int_Xc_2(E)c_3(E')~{\rm mod}~2.}

At first sight, there is something perplexing about this result.
We have partly characterized $E$ and $E'$
by requiring that $c_1(E)=c_1(E')=0$, $
c_2(E)=-a$, $c_2(E')=-a'$, but we have said
nothing about $c_3(E)$ and $c_3(E')$.  They are not uniquely determined
by the values of $c_1 $ and $c_2$.  However, for \pluxo\ to make sense,
it must be that $c_3(E')$ is uniquely determined mod 2 by the conditions
on $c_1(E')$ and $c_2(E')$.  Moreover, it must be that $c_3(E)c_2(E')
= c_2(E)c_3(E')$ mod 2.

To explain these points, we begin with the following fact.  Let $F$
be a complex vector bundle on $\S^6$.  Then from the index theorem,
the index $I(F)$ of the Dirac operator on $\S^6$ with values in $F$ is
\eqn\normo{I(F)=\int_{\S^6}{c_3(F)\over 2}.}
In particular, it follows that for a bundle $F$ on $\S^6$, $c_3(F)$
is always divisible by 2 -- congruent to 0 mod 2.

In general, for a rank $N$ complex vector
bundle $F$ on an arbitrary manifold $X$, $c_3(F)$ is not
necessarily congruent to 0 mod 2, but it is determined mod 2 in terms
of $c_1(F)$ and $c_2(F)$.  This can be  deduced from \normo\ if one
starts with a triangulation of $X$ and inductively constructs the
bundle $F$ on the $p$-skeleton for $p=0,1,2,\dots$.\foot{Here again
the reader may consult \topten\ for background on obstruction
theory.}  Suppose that
$F$ has been defined on the $(p-1)$-skeleton of $X$ and that one wishes
to define it on the $p$-skeleton.  Consider a particular $p$-simplex
over which one wishes to extend $F$. Topologically, it is a
$p$-dimensional ball ${\bf B}^p$ whose boundary is a sphere $\S^{p-1}$.
$F$ must be trivial on $\S^{p-1}$ or  no extension over ${\bf B}^p$ exists.
If $F$ is trivial on the boundary, $F$ can be extended over ${\bf B}^p$
but the extension may not be unique: given any one extension, the others
can be obtained from it by ``twisting'' by an arbitrary element of
$\pi_{p-1}(U(N))$, which (for large enough $N$) is nonzero precisely
if $p$ is even.  (The twist in the bundle is made by cutting out a small
$p$-ball from ${\bf B}^p$ and gluing it back in while making
a gauge transformation on the boundary by an element of $\pi_{p-1}(U(N))$.)
The twist shifts $c_{p/2}(F)$ on the simplex in question
by an amount equal to $c_{p/2}(F)$
for some $U(N)$ bundle on $\S^{p}$ 
(namely, the bundle made by cutting and
gluing using
the same element of $\pi_{p-1}(U(N))$).  This is the full indeterminacy
in $c_{p/2}(F)$ once $F$ is given on the $(p-1)$-skeleton.  So in particular,
setting
$p=6$ and using the fact that on 
$\S^6$, $c_3(F)$ is always even, it follows that
$c_3(F)$ is determined mod
2 in terms of $c_1(F)$ and $c_2(F)$, which completely determine $F$
on the five-skeleton.

When $c_1(F)=0$, the relation is $c_3(F)=Sq^2(c_2(F))$ mod 2, where
$Sq^2$ is a certain cohomology operation known as a Steenrod square.
We will give an introduction of sorts to Steenrod squares in section 4.1
and explain this formula.
 For the moment the reader can simply think of $Sq^2$ as 
a mysterious linear map from degree 4 to degree 6 cohomology. 
Setting $c_3(E')=Sq^2(E')$ mod 2, we can restate the result in
\pluxo\ as follows:
\eqn\jugoxo{f(a+a')=f(a)+f(a') +\int_X a \cup Sq^2 a'.}
A standard property of the Steenrod squares, explained in section
4.1 below, is that on a spin manifold,
\eqn\ugoxo{\int_X a \cup Sq^2 a'=\int_X Sq^2a \cup a'.}
So the right hand side of \jugoxo\ is symmetric in $a$ and $a'$, as it must be.

\subsec{The role of cobordism theory}

Our next goal is to show that \jugoxo\ actually holds for
arbitrary $a,a'\in H^4(X;\Z)$, whether or not they can be lifted
to $K$-theory. 
It follows from general considerations that $f(a)$ must be quadratic,
simply because we are working in a dimension -- 10 -- which is less
than three times four (where four is the degree of the class $a$).
In other words, the most elementary cubic function of $a$
would be $\int a \cup a\cup a$, which can be nonzero on a manifold
of dimension $12= 3 \times 4$.  All less elementary cubic functions
appear in dimensions still higher than 12, roughly because all
cohomology operations raise the degree of a cohomology class.  (As we
explain in section 4, the Steenrod squares are examples of cohomology
operations
raising the degree of a class.)  Functions of higher order than cubic
require a yet higher dimension.
Nevertheless, this argument leaves open the possibility  
that there is a relation of
the general form of \jugoxo\ with some more general
bilinear function on the right hand side. We will show 
this is not the case.  For this, we will use a rather abstract
argument based on cobordism theory. The reader may wish to
accept \jugoxo\ for all $a,a'$ and
skip this subsection. On the other hand, we have found
the techniques described below to be a powerful tool in
analyzing this and several related problems.

The mod 2 index of the Dirac operator coupled to a bundle $V$ on a
spin manifold  vanishes if $X$
is the boundary of a spin manifold over which $V$ extends.
So in particular, our $E_8$ mod 2 index $f(a)$ vanishes if
$X$ is the boundary of an eleven-dimensional spin manifold $Y$ over
which $a$ extends.
Similarly, consider the quantity
\eqn\bilin{
Q(a_1,a_2) = f(a_1+a_2) - f(a_1) - f(a_2).
}
It  is a function
of two degree four classes $a_1,a_2$, and vanishes
if both $a_i$ can be extended to an eleven-dimensional spin manifold
bounding $X$.  Invariants such as $f(a)$ and
$Q(a_1,a_2)$ are related to
a generalized (co)homology theory known as (co)bordism theory.
We now briefly explain these terms. 

Recall that an $n$-manifold $M_n$ is cobordant to zero if there
exists an $(n+1)$-manifold $B_{n+1}$ such that $\p B_{n+1}=M_{n}$.
We can ask that $B_{n+1}$ carries structures carried by $M$, and thus
we can
define, for example, the spin bordism groups $\Omega_n^{spin}$,
where $M_n$ is spin and $B_{n+1}$ is required to be spin.  An element
of $\Omega_n^{spin}$ is a spin manifold, which is considered to be zero
if it is the boundary of a spin manifold.
To define the group structure of $\Omega_n^{spin}$,
manifolds (representing elements of $\Omega_n^{spin}$) are added
by taking their disjoint union.

Now let $X$ be a topological space. We can define the 
``bordism groups of $X$'' by taking sets of pairs $(M_n, \mu)$ 
where   $M_n$ is equipped with a continuous
map $\mu:M_n \rightarrow X$. The equivalence relation on pairs 
is defined  by declaring that $(M_n,\mu)$ is cobordant to
zero if there is a bounding manifold $B_{n+1}$ {\it together with}
an extension of $\mu$, i.e. $\tilde \mu: B_{n+1} \rightarrow X$.
The resulting bordism group is denoted by $\Omega_n(X)$. If $M_n$ 
carries a structure we can then require that $B_{n+1}$ also carry 
this structure. For example if  $M_n, B_{n+1}$ are required to be spin
we define $\Omega_n^{ spin}(X)$.
The map $X \rightarrow \Omega_n(X)$ defines a generalized homology
theory. The spin bordism groups $\Omega_n^{spin}$
with no $X$ specified
can be regarded as $\Omega_n^{spin}(X)$ for $X$ a point.
$\tilde \Omega_n(X)$ is defined as the kernel of the natural
map from $\Omega_n(X)$ to $\Omega_n$ in which one
``forgets'' $X$ (or maps $X$ to a point).  A class $(M_n,\mu)$ in
$\Omega_n(X)$ represents  an element of
 $\tilde \Omega_n(X)$ if (forgetting $\mu$) $M_n$ is a boundary.

Let us now interpret $f(a)$ in cobordism theory.
It is a $\Z_2$-valued
function $f(a)$ of a single cohomology  class
$a$, which vanishes when $a$ extends to
a spin manifold $B$ bounding $X$.  $f(a)$ 
is (therefore) additive on disjoint unions.
To give a degree four
class $a$ on $X$ is to give a map
$\mu: X \rightarrow K(\Z,4)$ to the universal space $K(\Z,4)$ that
classifies four-dimensional cohomology.
The class $a\in H^4(X;\Z)$
can be extended to a bounding manifold $B$ -- ensuring that $f(a)=0$ --
if and only if
the map $\mu$ can be extended to a map from $B$ to $K(\Z,4)$.
In this case, the pair $X,a$ is zero as an element of $\Omega_{10}^{spin}
(K(\Z,4))$.
Thus, $f(a)$ can be regarded as a $\Z_2$-valued invariant
of $\Omega_{10}^{spin}(K(\Z,4))$
or more precisely as an
element of ${\rm Hom}(\Omega_{10}^{spin}(K(\Z,4)),\Z_2)$. Moreover,
since $f(0)=0$, $f$ can be viewed as an element of
${\rm Hom}(\widetilde{\Omega}_{10}^{spin}(K(\Z,4),\Z_2)$.
Replacing $\Omega$ by $\tilde \Omega$ leads to important
simplifications in the computation described below.

The quotient
group $\tilde \Omega^{spin}_{10}(K(\Z,4))$ has been computed
 by Stong \stong\ and shown to be $\Z_2\times \Z_2$. Thus, there are two
independent $\Z_2$-valued invariants of the pair $(X,a)$.
One such invariant is elementary:
\eqn\lopo{v(a)=\int a \cup w_6=\int a \cup Sq^2\lambda}
(here $w_i$ are the Stiefel-Whitney class of $X$; the two formulas
are equivalent because on a spin manifold $w_6 = Sq^2\lambda$).
In particular, $v(a)$ is a linear function of $a$:
$v(a+a')=v(a)+v(a')$.  Our invariant $f(a)$ is not linear,
as we see from \jugoxo, so it is the ``second'' invariant of
$\tilde \Omega^{10}(K(\Z,4))$.

Now that we have put $f(a)$ in the context of cobordism theory, let us 
turn to the bilinear identity for $f$. The object
 $Q(a_1,a_2)=f(a_1+a_2)-f(a_1)-f(a_2)$ is a homomorphism 
from $ \Omega_{10}(K(\Z,4) \times K(\Z,4))$ to $\Z_2$.  
Moreover,  $Q$ vanishes if either $a_1$ or $a_2$
is zero.  This means that we can replace $K(\Z,4)\times K(\Z,4)$
by $K(\Z,4) \wedge K(\Z,4)$, where $\wedge$ denotes the
smashed product of two spaces
\eqn\smashed{
X \wedge Y = X \times Y / (X \vee Y).}
($X\wedge Y$ is obtained from $X\times Y$ by picking
points $p\in X$ and $q\in Y$, and collapsing $p\times Y\cup Z\times q$
to a point.  An arbitrarily selected point is often denoted $\{*\}$.)
It is often inconvenient to work directly with
the smash product $\Omega(X\wedge Y)$
and technically more convenient to work with
relative bordism groups $\Omega(X \times Y, X \vee Y)$.
In general, relative bordism groups $\Omega_n(X,A)$
 are defined as above by allowing $M_n$ to have a
nonempty boundary and considering maps of pairs
$f(M_n, \p M_n) \rightarrow (X,A)$; in other words,
$f$ maps the boundary of $M_n$ to $A$.  There is a natural
notion of when such a pair should be considered cobordant to zero.

Now, putting the above remarks together and
taking into account that we also want
spin manifolds,  our quantity $Q(a_1,a_2)$ in
\bilin\ is a homomorphism to $\Z_2$ of  the relative bordism group
\eqn\relbord{
{\Omega}_{10}^{spin}(K(\Z,4)\times K(\Z,4),
K(\Z,4)\times\{*\}\cup
\{*\}\times K(\Z,4)).}
We would therefore like to compute the group \relbord.

The computation of groups in ``generalized (co)homology theories''
is greatly facilitated by the ``Atiyah-Hirzebruch spectral sequence.''
We will explain in greater depth how this works for $K$-theory in 
appendix C. 
Here we simply note that we can regard $K(\Z,4) \times K(\Z,4) \rightarrow
K(\Z,4)$ (the projection map to the second factor)
as a (rather trivial) fibration, and apply
\ref\rmsw{R.M. Switzer, 
{\it Algebraic Topology -- Homotopy and Homology},
Springer-Verlag 1975.}
remark 2, pg 351. See also
\ref\dyer{E. Dyer, {\it Cohomology Theories}, 
W.A. Benjamin, 1969.}, ch. 1
sec. B.
This allows us to construct the groups from a spectral sequence
whose $E^2$ term is
\eqn\bordss{
E^2_{p,q} =
{\widetilde H}_p(K(\Z,4);
{\widetilde \Omega}^{spin}_q(K(\Z,4)) )}
Now the differentials act as $d^r: E^r_{p,q} \rightarrow E^r_{p-r,q+r-1}$
and thus change total degree by one. Thus we are interested in
the above groups for $p+q =9,10$ (to compute kernels) and $p+q=10,11$
to compute images. At this point a very lucky fact occurs:
the groups ${\widetilde H}_p(K(\Z,4); G)$ for $G=\Z,\Z_2$
and ${\widetilde \Omega}^{spin}_q(K(\Z,4)) )$   are very
sparse in low dimensions! Indeed, from Eilenberg-MacLane 
\ref\eilmac{S. Eilenberg
and S. MacLane, ``On The Groups $H(\Pi, n)$, II'', Ann. of Math.
{\bf 58} (1953) 55-106.}
we get $\widetilde{H}_i(K(\Z,4);\Z)=0 $ for $i=0,1,2,3,5,7$
while $\widetilde{H}_4(K(Z,4);\Z)=\Z$, and 
$\widetilde{H}_6(K(\Z,4);\Z)=\Z_2$.
Similarly the groups ${\widetilde \Omega}^{spin}_q(K(\Z,4)) )$
have been computed by Stong to be $0$ for $0 \leq q < 4$, and $\Z$ for $q=4$.
This is all we need for the present computation, which is moreover 
facilitated by considering the diagram in 
\diagram[PostScript=dvips,tight,height=2.0em,width=2.0em]
   & \vLine &   &    &    &    &    &  &  & & \\
 \hbox{\hskip 15pt 9} &       &  0 &
&    &    &    &  &  &  & \\
   &       &   & 
\luTo(3,2)^{d^6} 
&  &    &  &  & & & \\
 \hbox{\hskip 15pt  4} &       &   &  
& {\bf Z}   & 
{\bf Z}/2 & & & & & \\
   &       &   &    &    &     & \luTo(3,2)^{d^5}&  & & \\
\hbox{\hskip 15pt 0} &       &   &    &    &     &    & 
& 0 & & \\
   &\HmeetV&  \hLine &    &    &     &    &    &  & \\
   &       & \raise 10pt \hbox{ 0} &    & 
\raise 10pt \hbox{ 4}  
& \raise 10pt \hbox{ 6}  &    & & 
\raise 10pt \hbox{ 11} & & \\
\enddiagram
\centerline{\vbox{\baselineskip12pt
\advance\hsize by -1truein\noindent
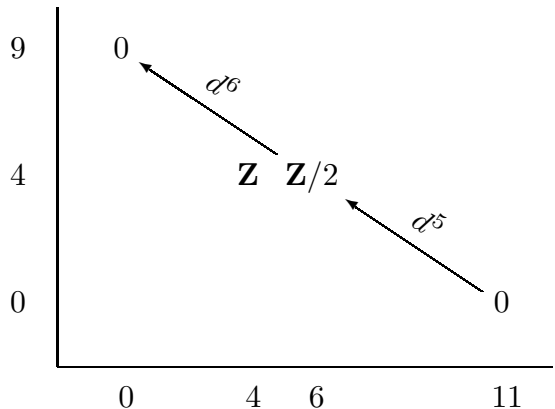
\footnotefont{\bf Caption:}
The $E^2$ term in the spectral sequence 
computation of cobordism invariants relevant to the bilinear 
identity. Differentials act from the diagonal $p+q$ to the 
diagonal $p+q-1$. Since the cobordism groups are simple in 
low dimensions, the differentials are trivial and we can 
read off the answer. }}
The only nonzero group on the diagonal $p+q=10$ is $E^2_{6,4}=0$.
All groups on the diagonals $p+q=9,11$ vanish. Thus we conclude
\eqn\compgroup{
{\Omega}_{10}^{spin}(K(\Z,4)\times K(\Z,4),
K(\Z,4)\times\{*\}\cup
\{*\}\times K(\Z,4))= \Z_2
}
so there is only one nontrivial cobordism invariant!

One example of such an invariant is $\int_X a_1 \cup Sq^2 a_2$.
In section 4.1 we give an example -- $X= \S^2\times \S^2\times {\bf CP}^3$ --
for which it is nonzero.  So it is the unique invariant.  $Q(a_1,a_2)$
is not identically zero, for we have seen that it coincides with
$\int_X a_1\cup Sq^2 a_2$ when $a_1$ and $a_2$ can be lifted to $K$-theory
(which is the case for all classes on $X$).  So $Q(a_1,a_2)
=\int a_1\cup Sq^2a_2$ in general.

\subsec{Parity Symmetry}

We conclude this section by analyzing a symmetry of $M$-theory that
has been obscured in our formalism.   (One might return to this discussion
after reading section 4.1.)
Parity symmetry of $M$-theory acts by orientation reversal, together
with $G\to -G$, which in terms of
$a=G/2\pi +\lambda/2$ is $a\to -a+\lambda$.
For $Y=\S^1\times X$, there is hence a parity symmetry
that acts by reflection of $\S^1$ together with $a\to -a+\lambda$.
In Type IIA superstring theory, this symmetry is interpreted as $(-1)^{F_L}$.
It plays an important role in the structure of the theory, and one does
not expect it to be anomalous.

We will now demonstrate that there is an apparent anomaly, due to 
gravitational instantons, in the 
$(-1)^{F_L}$ symmetry in IIA theory. 
We will then show, as an application of the 
bilinear identity \jugoxo,  that this anomaly is 
cancelled by a nontrivial effect in the RR sector.

 Consider Type IIA superstring theory on a ten-dimensional spin manifold
$X$.  Denote by $q(V)$ the mod two index of the Dirac operator coupled 
to a real vector bundle $V$. (For greater precision, we will 
sometimes denote the mod 2 Dirac index on $X$
with values in a real bundle $V$ 
as $q(V;X)$.) In particular, the relevant mod two index for the 
Rarita-Schwinger operator is 
$q(TX)$.  For a given set of background fields on $X$,
let $n_L$ and $n_R$ denote the number of zero modes of
the gravitino fields $\psi_L$ and $\psi_R$ coming from the left-
or right-moving worldsheet Ramond sector; 
$n_L$ and $n_R$ are both congruent mod 2 to $q(TX)$.
  There is      an effective action proportional to 
$ \psi_L^{n_L}\psi_R^{n_R}$.                               
As $(-1)^{F_L}$ acts
as $\psi_L\to -\psi_L$, $\psi_R\to\psi_R$, the fermion measure $\mu$
transforms under $(-1)^{F_L}$ as
\eqn\ucup{\mu\to (-1)^{q(TX)}\mu.}
In the path integral the ghosts plus dilatino
make no net contribution to the transformation of the measure 
\ucup\  as they constitute an even number of chiral spin $1/2$ fields.
There is no problem in finding an $X$ with $q(TX)\not= 0$; $X={\bf HP}^2
\times T^2$ will serve as an example, as we discuss more fully below.
By deleting a point from $X$ and projecting it to infinity, we get a
candidate gravitational instanton in an asymptotically flat spacetime
with a fermion measure that is odd under $(-1)^{F_L}$.

It appears that we must either
 search for a physical principle that might forbid this
instanton, or accept that $(-1)^{F_L}$ is anomalous in asymptotically
flat spacetime and hence in anything to which it can be compactified.
Instead, we will demonstrate that this anomaly is canceled by an 
anomaly in the action of $(-1)^{F_L}$ on the Ramond-Ramond fields.
We use the $M$-theory framework.  The description by $E_8$ gauge theory
does not make manifest how the partition function of the $C$-field
transforms under a reflection of the circle together with $a\to\lambda-a$,
but can be used to determine this transformation.
The bilinear relation gives
\eqn\poko{f(\lambda)=f(a)+f(\lambda-a)+\int a\cup Sq^2(\lambda - a).}
According to a result of Stong \stong,
\eqn\noko{\int a\cup Sq^2a =\int a\cup Sq^2\lambda.}
Hence, 
\eqn\jonoko{(-1)^{f(\lambda-a)}=(-1)^{f(a)}(-1)^{f(\lambda)}.}
So under parity or $(-1)^{F_L}$, the RR partition function is
multiplied by $(-1)^{f(\lambda)}$.
 This factor will turn out to cancel the
anomaly of the fermions.
A quick way to demonstrate this cancellation 
is to use a relation to 
the heterotic string described at the end of this 
section. We will take a more leisurely route which 
brings out some useful information. 

$f(\lambda)$ is defined as the mod 2 index of an $E_8$ bundle
whose characteristic class is $\lambda$.  Such a bundle can be described
very simply: take the tangent bundle $TX$ of $X$, whose structure group
is $SO(10)$, and build from it an $E_8$ bundle using the chain of
embeddings $SO(10)\subset SO(10)\times SU(4)\subset E_8$.
The adjoint representation of $E_8$ decomposes under $SO(10)\times SU(4)$
as $({\bf 45},{\bf 1})\oplus ({\bf 1},{\bf 15})\oplus ({\bf 10},{\bf 6})
\oplus ({\bf 16},{\bf 4})$.  Since six copies of the ${\bf 10}$ or four
copies of the ${\bf 16}$ will not contribute to the mod 2 index,
and 15 copies of the ${\bf 1}$ are equivalent to a single copy,
we can replace the adjoint representation of $E_8$, for purposes
of computing the mod 2 index, with the
${\bf 45}\oplus {\bf 1}$ of $SO(10)$.
The bundle on $X$ that corresponds to this representation
is $\wedge^2TX\oplus {\cal O}$, where
${\cal O}$ is a trivial line bundle and $\wedge^2TX$ is the antisymmetric
product of $TX$ with itself.  We thus have $f(\lambda)=q(\wedge^2TX)
+q({\cal O})$, and hence, including also the fermion anomaly from \ucup,
the total sign change of the effective action under $(-1)^{F_L}$ is
\eqn\hokko{J=
(-1)^{q(\wedge^2 TX) + q(TX)+q({\cal O})}.}

We want to demonstrate that $J=1$ for all ten-dimensional spin manifolds $X$.
To do so, we will use the fact that $J$ is a spin cobordism invariant;
indeed, each factor in $J$ is separately equal to $1$ for any $X$ that is
the boundary of an eleven-dimensional spin manifold.  The group
$\Omega^{spin}_{10}$ is equal to $\Z_2\times \Z_2\times \Z_2$.  
One choice of three independent invariants is $q(\CO)$, $q(TX)$, 
and 
\eqn\oplk{K=\int_X w_4\cup w_6=\int_X \lambda \cup Sq^2
\lambda.} 
 (The examples we give will
show that these three invariants are independent.)

\def\TT{{\bf T}}
One can pick two generators of the spin bordism group to be of the form
$X  =Y
\times \TT^2$, for suitable $Y$ (with a supersymmetric
                   or unbounding
spin structure on $\TT^2$).  
On such a manifold,
$q({\cal O};X  )$ 
is equal to the mod 2 reduction of $I(\one ; Y  )$. 
\foot{We use $\CO$ for a trivial real or complex  line bundle. 
Which is meant should be clear from the context.  In general,  $I(x;X)$ denotes 
the ordinary index of the Dirac 
operator on $X$ coupled to the $K$-theory class $x$.}
This follows from 
the fact that a fermion zero mode on $X  $ is a constant on $\TT^2$
times a zero mode on $Y  $.  Likewise, on such a manifold, $q(TX;X  )$
equals the mod 2 reduction of $I(TY;Y  )$, the ordinary index of the Dirac
operator on $Y  $
with values in the tangent bundle.
   To show this, one uses the decomposition
of the tangent bundle $TX$ of $X$ as $TX
=TY\oplus {\cal O}\oplus {\cal O}$,
where ${\cal O}\oplus {\cal O}$ is the tangent bundle to $\TT^2$.
The two copies of ${\cal O}$ do not contribute to the mod 2 index,
so we need the mod two index of the Dirac operator on $Y\times \TT^2$
with values in $TY$.  As a zero mode must be a constant on $\TT^2$ times
a zero mode on $Y$, we get $q(TX;X)=I(TY; Y)~\mod ~2$.  Any manifold $Y\times \TT^2$
has $K=0$, since $w_4$ and $w_6$ are pullbacks from $Y$.

Now, let
 $Y_1$ be a spin manifold with Dirac index 1 and Rarita-Schwinger index 0,
and let $Y_2$ be a spin manifold with Dirac index 0 and Rarita-Schwinger
index 1.  Such manifolds exist. The spin cobordism group in 
eight dimensions is known to be $\Z \oplus \Z $, generated by 
a manifold $Y_1$ such that $4 Y_1$ is spin cobordant to 
$K3 \times K3$ and by   $Y_2= {\bf HP}^2$.  Then $X_i=Y_i\times \TT^2$,
$i=1,2$, with unbounding (or RR) spin structure on
$\TT^2$, will serve as two generators of $\Omega^{spin}_{10}$.  For
these generators, we have in view of the remarks in the last paragraph
$q({\cal O};X_1)=1$, $q(TX_1;X_1)=0$, and
$q({\cal O};X_2)=0$, $q(TX_2;X_2)=1$.

For examples of this kind, we can readily use index theory to show that
the total anomaly factor $J$ is trivial.  Indeed, by reasoning as above,
we find that the mod 2 index                          on $X=
Y\times \TT^2$
with values in $\wedge^2TX$
  is the same as the ordinary index of the
Dirac operator on $Y$ with values in $\wedge^2TY\oplus {\cal O}$. 
(We use the decomposition $\wedge^2TX=\wedge^2TY\oplus 2 \,TY\oplus
{\cal O}$, and note that two copies of $TY$ do not contribute to the mod
2 index.)   For such manifolds, the total anomaly factor becomes
\eqn\jurry{J=(-1)^{I(\wedge^2TY)+I(TY)}.}
The index theorem shows that on any eight-dimensional
spin manifold $Y$, 
\eqn\eightspin{
\eqalign{I({\cal O};Y) & = \int_Y\hat A_8  \cr
I(TY ;Y) & = \int_Y\left( 248 \hat A_8  - \lambda^2\right)\cr
I(\wedge^2TY; Y)& = \int_Y\left(
 28 \hat A_8 + \lambda^2 \right)\cr}
}
In particular, $I(TY;Y)$ is congruent to $I(\wedge^2 TY ;Y)$ mod 2.  
 So
for this class of manifold, there is no anomaly in $(-1)^{F_L}$. 

For the third generator of $\Omega_{10}^{spin}$, we can take a manifold
$V_{1,1}$ defined as a hypersurface of degree $(1,1)$ in ${\bf CP}^2\times
{\bf CP}^4$.  This manifold has $K=1$.  It also has, as we demonstrate
in appendix B, $q({\cal O})=q(TX)=q(\wedge^2 TX)=0$.  So in particular,
the total anomaly vanishes for this manifold.  This completes the
demonstration that the combined $(-1)^{F_L}$ anomaly of the fermions plus
the RR fields always vanishes in Type IIA superstring theory at $G_2=0$.
We extend the analysis to backgrounds with nonvanishing  $G_2$ in section eight. 

Let us now make more explicit what is going on in the above discussion
for $X_2={\bf HP}^2\times \TT^2$, where  an anomaly of the RR fields
cancels a fermion anomaly.  The first key point is that $H_4(X_2;\Z)=\Z$, generated
by  $U={\bf HP}^1\subset {\bf HP}^2$.  Moreover, $\int_U\lambda=1$.
Hence, it is impossible for the four-form flux $\int_U G/2\pi$ to vanish.
The quantization condition
(as derived in \fourflux\ from world-volume anomalies) is
\eqn\kilp{\int_U{G\over 2\pi}=\half\int_U\lambda~{\rm mod}~\Z.}
So the least action $G$-field on $X_2$ has $\int_UG/2\pi =\pm 1/2$.  The two possibilities
are exchanged by parity or $(-1)^{F_L}$,
and as they contribute to the partition function
with a relative sign, there is an anomaly.
If instead the $\lambda$ class is divisible by two, then the $G$-field
can vanish.  A configuration with $G=0$, if it exists,
 is parity-invariant, and makes a nonvanishing
contribution to the sum over $G$-field fluxes, so if such a configuration
 exists, there is no parity anomaly in the partition function
of the $M$-theory $C$-field, or of the RR fields in the Type IIA description.

\bigskip\noindent{\it Application To The Heterotic String}

The result we have just found has an interesting application
to the $E_8\times E_8$ heterotic string.  

Consider the $E_8\times E_8$
heterotic string on a ten-dimensional spin manifold $X$.
Let $a$ and $b$ be the characteristic classes of the 
two $E_8$ bundles.  Anomaly cancellation of the heterotic
string requires that $a+b=\lambda$.

We want to show that the total number of fermion zero modes of the heterotic
string is even.  Otherwise, the heterotic string on $X$ would be anomalous.
Since the gluinos are chiral fermions in the adjoint representation
of $E_8$, the number of gluino zero modes is $f(a)+f(b)$ mod 2.  As we have
proved above using Stong's formula, $f(a)+f(b)=f(a)+f(\lambda-a)=f(\lambda)$.
The number of Rarita-Schwinger zero modes (including ghosts and
dilatinos) is $q(TX)$ mod 2.  So cancellation of the anomaly
requires that $f(\lambda)=q(TX)$ mod 2, as we have shown above.

In fact, this result is a special case of a general statement
about the heterotic string. In general,
given any diffeomorphism $\varphi:X\to X$ with a lift of $\varphi$ to the spin bundle
of $X$ and to the $E_8$ bundles, 
the effective action of the heterotic string is invariant
under $\varphi$ \topten.  In case $\varphi=1$ and we take $\varphi$ to act on the spin
bundle as multiplication by $-1$ and trivially on the $E_8$ bundles, invariance of the effective
action is equivalent to the statement that the total number of fermion zero modes is even.

\newsec{Topological Background}

  The present section is
devoted to explaining topological background that will be important
in the rest of the paper.  Essentially everything in this section
will be used later, though it is possible to proceed to section 5 without
digesting everything presented here.

\subsec{A Crash Course on Steenrod Squares}

We have encountered a new topological operation -- $Sq^2$ -- that has
not previously played much role in physics.
We will here, with no attempt at completeness, try to motivate
the properties of Steenrod squares that are needed in the present paper.
Some references with further useful material and details are 
\refs{\milstash,\bredon}.

Suppose that $X$ is a manifold and $Q$ a codimension $k$
submanifold whose normal bundle $N$ is oriented.  Let $b=[Q]$
be the Poincar\'e dual cohomology class to $Q$.  $b$ is a $k$-dimensional
cohomology class and is represented in
de Rham cohomology by a $k$-form delta function supported on $Q$
and integrating to 1 over the normal directions to $Q$.

Let $w_i(N)\in H^i(Q;\Z_2)$ be the
Stieffel-Whitney classes of $N$.  We want to consider the objects
that we can loosely call $ w_i(N) \cup b \in H^{k+i}(X;\Z_2)$.
Informally, this cup product makes sense because, while $w_i(N)$ is only
defined in a small neighborhood of $Q$,\foot{It is defined to begin
with on $Q$ and can then be pulled back to a small tubular neighborhood
of $Q$.}  $b$ anyway has its support in that
neighborhood.  The precise formal way to define $w_i(N)\cup b$ is
as $i_*(w_i(N))$, where $i:Q\to X$ is the inclusion and $i_*$ the
push-forward.  We will settle for informal expressions like $w_i(N)
\cup b$.

Suppose we are given $b\in H^k(X;\Z)$.  We represent it as the
Poincar\'e dual to a submanifold $Q$.\foot{In general, there are 
obstructions
to doing this, but some odd multiple of $b$ can always be so 
represented,
and that is good enough for our purposes, since
the Steenrod squares annihilate classes that are divisible by 2.}
  Then we define
\eqn\olpo{Sq^i(b) = w_i(N)\cup b\in H^{k+i}(X;\Z_2).}
It can be shown that, as an element of $H^{k+i}(X;\Z_2)$, $Sq^i(b)$
is independent of the choice of a submanifold $Q$ dual to $b$.
Not proving this will be a major gap in our presentation.

Thus, $Sq^i$, as we have defined it so far, is a linear  map
\eqn\lopo{Sq^i:H^k(X;\Z)\to H^{k+i}(X;\Z_2).}
$Sq^i$ is read ``square $i$.''

Actually, the $Sq^i$ can be extended to linear  maps  $H^{k}(X;\Z_2)
\to H^{k+i}(X;\Z_2)$.  Given $\bar b\in H^k(X;\Z_2)$, one represents
$\bar b$ as the Poincar\'e dual of
a submanifold $Q$ (whose normal bundle is not necessarily
orientable, so that its Poincar\'e dual is only defined mod 2)
and defines again
\eqn\nolpo{Sq^i(\bar b)=w_i(N)\cup \bar b.}
If $\bar b$ is the mod 2 reduction of an integral class $b$,
then  these definitions imply $Sq^i(b)=Sq^i(\bar b)$.

\def\spinc{{\rm Spin}^c}
The $Sq^i$ for odd $i$ can actually be defined as maps
to $H^{k+i}(X;\Z)$; $Sq^i(y)$ is always two-torsion, that is,
$2 Sq^i(y)=0$.
We will define the integer-valued $Sq^i$ only for $i=1$ (which we consider
later) and $i=3$.
  The Stieffel-Whitney
class $w_3$ has a canonical integral lift $W_3$, 
which is the obstruction
to 
${\rm Spin}^c$
 structure.  It arises by considering the coefficient
sequence
\eqn\lopok{0\to \Z\underarrow{2}\Z\underarrow{r} \Z_2\to 0,}
where the first map is multiplication by 2 and  $r$ is reduction mod 2.
This leads to a long exact sequence
\eqn\kolop{\dots H^2(Q;\Z)\underarrow{r}
H^2(Q;\Z_2)\underarrow{\beta} H^3(Q;\Z)\underarrow{2}
H^3(Q;\Z)\dots}
where $\beta$ is called the ``connecting homomorphism'' or the
Bockstein.
One defines
\eqn\yurry{W_3(N)=\beta(w_2(N)).}
  Exactness of \kolop\ implies
that $2W_3(N)=0$, and that $W_3(N)=0$ precisely if $w_2(N)$ can be
lifted to a class in $H^2(Q;\Z)$. (This can be used to show
 that $W_3(N)$ is the obstruction
to having a $\spinc$ structure on $N$.)   To
define
\eqn\unumu{Sq^3:H^k(X;\Z)\to H^{k+3}(X;\Z),}
we simply use $W_3$ instead of $w_3$:
\eqn\bumumu{Sq^3(y)=W_3 \cup y.}

It can be shown that $W_3$ reduces mod 2 to the Stieffel-Whitney
class $w_3$, so $Sq^3$ understood as a map to $H^{k+3}(X;\Z)$ reduces
 mod 2 to $Sq^3$ as defined previously.
Because of relations such as this one, the different $Sq^i$ maps
are compatible, and it usually causes  no confusion to use the same
name $Sq^i$ for slightly different maps.

The $Sq^i$ have many useful properties that we will need.
For example, suppose that $b$ is a $k$-dimensional cohomology class, dual to
a codimension $k$ manifold $Q$.  The normal bundle $N$ has rank $k$,
so $w_i(N)=0$ for $i>k$.  It follows that
\eqn\ucx{Sq^i(b)=0~{\rm for}~i>k.}

To derive the next property of $Sq^i$ recall 
(see, e.g. 
\ref\steenrod{N. Steenrod, {\it The topology of fibre bundles}, 
Princeton, 1951.}) 
that the Stieffel-Whitney classes of a bundle can be 
defined using obstruction theory. In this approach one 
finds that for $N$ of real rank $k$, $W_k(N)$ is the Euler class 
for $k$ odd, while $w_k(N)$ is the mod 2 reduction
of the Euler class for $k$ even.
The Euler class of the normal bundle of $Q$ is represented by the
zero set of a generic section $s$ of the normal bundle, or equivalently by
$Q\cap Q'$ where $Q'$ is obtained from $Q$ by displacing it by the
section $s$. $Q'$ is homologous to $Q$, and $Q\cap Q'$ is dual to
$b\cup b$.  So
\eqn\bucx{Sq^k(b)=b\cup b}
for a $k$-dimensional class $b$. The cup product 
$b\cup b$ is also written as $b^2$,
and this formula is actually the reason
that the $Sq^i$ are called ``squares.''

Next, let us work out a formula
for $Sq^i(b\cup b')$, where $b$ and $b'$ are classes of degree $k$
and $k'$.  We suppose that  $b$ is Poincar\'e dual to $Q$ and $b'$ is
Poincar\'e dual to $Q'$,
and that $Q$ and $Q'$ intersect transversely in a codimension $k+k'$ manifold
$Q''$.  Then $Q''$ is dual to $b\cup b'$, and the normal bundles on $Q''$ 
are related
by
\eqn\normbu{N(Q'')=N(Q)\vert_{Q''} \oplus N(Q')\vert_{Q''} .}
{}From \normbu, it follows that

\eqn\orbu{w_i(N(Q''))=\sum_{j=0}^iw_j(N(Q))\cup w_{i-j}(N(Q')).}
Hence, we deduce the Cartan formula
\eqn\bormbu{Sq^i(y\cup y')=\sum_{j=0}^iSq^j(y)\cup Sq^{i-j}(y').}

Next, suppose that  $X$ is an orientable manifold of dimension
$n$.  Given  ${\bar b}\in H^{n-1}(X;\Z_2)$, we wish to show that
\eqn\hobo{Sq^1(b)=0.}
In fact, ${\bar b}$ is dual to a compact one-manifold $Q$, which is 
inevitably
a union of circles and hence orientable.
The Stieffel-Whitney classes of $X$, restricted to $Q$, are related
to those of $Q$ and of the normal bundle $N$ by
\eqn\uzz{(1+w_1(X)+w_2(X)+\dots)=(1+w_1(Q)+w_2(Q)+\dots)(1+w_1(N)+w_2(N)+
\dots),}
so for instance
\eqn\buzz{w_1(X)=w_1(Q)+w_1(N),~~w_2(X)=w_2(N)+w_1(Q)w_1(N)+w_2(Q).}
Since $X$ and $Q$ are
orientable, we have $w_1(X)=w_1(Q)=0$; hence $w_1(N)=0$, and \hobo\ follows.

Now suppose that $X$ is a spin manifold of dimension $n$ and
${\bar b}\in H^{n-2}(X;\Z_2)$.  We want to prove that
\eqn\nobo{Sq^2(b)=0.}
$b$ is dual to a not necessarily orientable
two-manifold $Q$.  Every two-manifold has
$w_1(Q)^2+w_2(Q)=0$.  Since $X$ is spin ($w_1(X)=w_2(X)=0$), the
relations \buzz\ imply $w_2(N)=0$, and \nobo\ follows.
Next consider $b\in H^{n-3}(X;\Z)$.
$b$ is dual to an oriented three-manifold $Q$; such a manifold
has $w_1=w_2=w_3=W_3=0$.
So we get
\eqn\ploog{Sq^1(b)=Sq^2(b)=Sq^3(b)=0.}
For a final result of this kind, consider $b\in H^{n-4}(X;\Z)$.
Then $b$ is dual to a four-manifold, and as four-manifolds are
$\spinc$, one has
\eqn\noloss{Sq^3(b)=0.}
 Some of the above formulas have an important generalization
known as  the Wu formula \milstash.
In general, on any $n$-manifold $X$, 
for  ${\bar b}\in H^{n-i}(X; \Z_2)$, one has  
$Sq^i({\bar b}) = V_i \cup {\bar b}$
where  $V_i$ is a polynomial in Stieffel-Whitney classes
 known as the Wu class.  From the above, we have $V_1=w_1$ and $V_2
=w_1^2+w_2$.

Suppose we are given $\bar b\in H^k(X;\Z_2)$, dual to a manifold
$Q$.  Then $w_1(N)$ vanishes if and only if $N$ is orientable;
but orientability of $N$ is the condition that $\bar b$ is the reduction
of an element of $H^k(X;\Z)$.  Indeed, $Q$ is dual to an element $b$ of
$H^k(X;\Z)$ (which can be reduced mod 2
to give an element $\bar b$ of $H^k(X;\Z_2)$) precisely if its normal bundle
is orientable.
So
$Sq^1(\bar b)=0$ if $\bar b$ is the reduction mod 2 of an integral class.
More generally, it can be shown that
\eqn\sonx{Sq^1(\bar b)=\beta(\bar b),}
where $\beta$ is the Bockstein map $\beta:H^k(X;\Z_2)\to H^{k+1}(X;\Z)$
derived from the coefficient sequence \lopok.
It fits in the exact sequence
\eqn\jumbly{\dots H^k(X;\Z)\underarrow{r}H^k(X;\Z_2)\underarrow{\beta}
H^{k+1}(X;\Z)\dots,}
with $r$ the mod 2 reduction.
The relation $Sq^1=\beta$ defines
 $Sq^1$ as a map
\eqn\unvu{Sq^1:H^{k}(X;\Z_2)\to H^{k+1}(X;\Z).}
 Exactness of the sequence \jumbly\ implies
that for $\bar b\in H^k(X;\Z_2)$,
\eqn\cux{Sq^1(\bar b)=0~{\rm if~and~only~if}~\bar b = r(b)}
for some $b\in H^k(X;\Z)$.

Let us consider the object $Sq^1Sq^2(b)$ for $b\in H^k(X;\Z)$.
Using the definition  $Sq^2(b) = w_2(N)\cup b$, this is
$Sq^1Sq^2(b)=Sq^1(w_2(N)\cup b)$.  Using the Cartan formula and \cux,
we have $Sq^1(w_2(N)\cup b)=Sq^1(w_2(N))\cup b=W_3(N)\cup b$
(where in the last step we use \yurry).  But $Sq^3(b)=W_3(N)\cup b$
and so we have for $b\in H^k(X;\Z)$
\eqn\tomop{Sq^3(b)=Sq^1Sq^2(b).}
Hence, $Sq^3(b)= 0$ if and only if there exists $c\in H^{k+2}(X;\Z)$
with $c=Sq^2(b)$ mod 2.  For (as we learned in the previous
paragraph) this is the condition for $Sq^1 $ to
annihilate $Sq^2(b)$.

The relation $Sq^3=Sq^1Sq^2$
is a special case of a system of relations among products
of Steenrod squares called the Adem relations.  Another special case
of the Adem relations that we will need is
\eqn\lomop{Sq^3Sq^3=0.}
In fact, $Sq^3(Sq^3y)$ is represented by $Sq^3(W_3 \cup y)$.  Using the
Cartan formula and the fact that $Sq^1$ annihilates an integral class,
this is $Sq^3W_3 \cup y + W_3\cup  Sq^3 y$, and vanishes since
$Sq^3W_3=W_3\cup W_3$, $Sq^3y=W_3 \cup y$, and
$2 Sq^3 W_3=0$.

We now have the tools to verify some assertions made in section 3.1.
For $X$ a spin manifold
of dimension $n$, given $a\in H^k(X;\Z)$, $a'\in H^{n-k-2}(X;\Z)$,
we want to show that
\eqn\bunvu{\int Sq^2 a \cup a' =\int a\cup Sq^2a'.}  For $k=4$, $n=10$,
this was asserted in section 3.1.
In fact, from \nobo\ we have $Sq^2(a\cup a')=0$.
In view of the assertion in \cux, $Sq^1a=Sq^1a'=0$, so the Cartan formula
\bormbu\ gives $Sq^2(a\cup a')=Sq^2a\cup a'+a\cup Sq^2a'$.  Putting
these facts together, we have $Sq^2a\cup a'+a\cup Sq^2a'=0$ and
(as $Sq^2$ is only defined mod 2) this implies \bunvu.

Now let $F$ be a complex vector bundle.  We wish to show that
\eqn\redform{c_3(F)=c_1(F)c_2(F) + Sq^2c_2(F)~ {\rm mod}~2.}
For $c_1(F)=0$, this assertion was made in section 2.2.
In proving such a statement, it suffices, by the splitting principle,\foot{
See, for example \ref\bottu{R. Bott and L. Tu, {\it Differential Forms
In Algebraic Topology}, Springer-Verlag 1982.}.}
to consider the case that $F=\oplus_{i=1}^n{\cal L}_i$ is a direct sum of
line bundles.  (This is proved by finding a fiber bundle $Z$ over $X$,
such that $F$ pulls back on $Z$ to a sum of line bundles, and such that
if a cohomological statement like \redform\ holds on $Z$, it must also
hold on $X$.)  Let $b_i=c_1({\cal L}_i)$.  We have
\eqn\redid{\eqalign{c_1(F) & =\sum_ib_i \cr
                    c_2(F) & = \sum_{i<j}b_ib_j \cr
                    c_3(F) & = \sum_{i<j<k}b_ib_jb_k \cr
                    Sq^2c_2(F) & =\sum_{i\not= j}b_i^2b_j~{\rm mod}~2.\cr}}
The last of these formulas is proved using the Cartan formula together
with $Sq^2(b_i)=b_i^2$. \redform\ is a straightforward consequence
of these formulas.

\bigskip\noindent{\it Examples}

Since our discussion has been somewhat abstract, we will here
give a few examples.

First we want to give an example in which the bilinear form
$\int a \cup Sq^2 b$ that appears in the bilinear relation for
the $E_8$ mod 2 index is nontrivial.  For this, quite elementary
examples suffice.  We take, for example, $X=\S^2\times \S^2 \times
{\bf CP}^3$.  The second cohomology groups of the three factors are generated
by classes $d_1$, $d_2$, $d_3$.  The integral cohomology of $X$ is
generated by the $d_i$ with the relations $d_1^2=d_2^2=0$, $d_3^4=0$.
Let $a=d_1\cup d_3$, $b=d_2\cup d_3$.  We have $Sq^2(a)=d_1\cup d_3^2$,
$Sq^2(b)=d_2\cup d_3^2$. (The right hand sides of these formulas are reduced
mod 2 as $Sq^2$ is a map to the $\Z_2$-valued cohomology.)
 To prove these relations, one uses the Cartan
formula, the fact that $Sq^1$ annihilates an integral class, and the
fact that for $d$ a two-dimensional class, $Sq^2(d)=d\cup d=d^2$.
So we have
\eqn\urru{\int a \cup Sq^2 b = \int Sq^2 a\cup b =\int d_1\cup d_2\cup
d_3^3=1}
 Note
that $Sq^3(a)=Sq^3(b)=0$, as $Sq^2(a)$ and $Sq^2(b)$ have integral
lifts, namely $d_1\cup d_3^2$ and $d_2\cup d_3^2$.
In this example, all of the cohomology of $X$ can
be lifted to $K$-theory, since the cohomology ring
is generated by two-dimensional
classes, and any two-dimensional class can be lifted to $K$-theory by
finding a suitable line bundle.

It is a bit trickier to give an example in which $Sq^3$ is nontrivial.
Take $X={\bf RP}^7\times {\bf RP}^3$.  Then the $\Z_2$-valued cohomology
of the two factors are generated, respectively, by one-dimensional
classes $a$ and $b$ with $a^8=b^4=0$.  Let $h=Sq^1(a\cup b)$, where
we understand $Sq^1$ as a map to the integral cohomology of $X$, so
$h\in H^3(X;\Z)$.
We can evaluate the mod 2 reduction of $h$ by interpreting $Sq^1$
as a map to the $\Z_2$-valued cohomology.  With this interpretation
of $Sq^1$, we have $Sq^1(c)=c\cup c=c^2$ for any one-dimensional class
$c$.  Using this and the Cartan formula, we see that the mod 2 reduction
of $h$ is $a^2\cup b+a\cup b^2$, so in particular $h$ is nonzero.
We can also evaluate the mod 2 reduction of $h\cup h$; it is
$a^4\cup b^2\not=0$, so in particular $h\cup h$ is nonzero.
\foot{However, $h\cup h$ is two-torsion.  Indeed, given any two
integral classes $h$ and $h'$ of odd degree, one has $h\cup h'=-h'\cup h$
or $h\cup h'+h'\cup h=0$.  Setting $h'=h$, we get $2 h\cup h=0$,
so $h\cup h$ is always two-torsion.}  As $h\cup h=Sq^3h$, we see
that $Sq^3 h \not= 0$.

We want to give an example of $Sq^3$ acting nontrivially
on $H^4(X;\Z)$.  For this, we begin with a five-manifold
$Q$ constructed as a ${\bf CP}^2$ bundle over $\S^1$, with ${\bf CP}^2$
undergoing complex conjugation as one goes around the $\S^1$.  This
example was discussed in \fw,
and is not $\spinc$, that is $W_3(Q)\not= 0$.
(In fact, in this example, $W_3(Q)$ is Poincar\'e dual to $L={\bf CP}^1\times
p$, where $p$ is a point in $\S^1$ and ${\bf CP}^1$ a linearly embedded
subspace of ${\bf CP}^2$.)
One can construct
an $SO(3)$ bundle $N$ over $Q$ such that the total space $U$ of the bundle
is spin.  (Explicitly, one can take $N$ to be the direct sum of the
nontrivial real line bundle over $\S^1$ and the standard
complex line bundle ${\cal O}(1)$ over ${\bf CP}^2$ regarded as
a rank two real bundle.)   Now, 
embed $Q$ in $U$ as the zero section  and
let $h\in H^3(U;\Z)$
be Poincar\'e dual to $Q$.
The normal bundle to $Q$ is $N$, and as $W_3(N) = W_3(Q)\not= 0$,
one has $Sq^3(h)\not =0$; in fact, $Sq^3(h)$ is Poincar\'e dual
to $L$ (embedded in $U$ via $L\subset Q\subset U$; note that as a submanifold
of $U$, $L$ has codimension six and so is dual to a degree six cohomology class)
because in the cohomology of $Q$, $L$ is dual to $W_3(N)$.

Now set $X=U\times \TT^2$ and $a=h\cup b$, where $b\in H^1(\TT^2;\Z)$ is any
class not divisible by two.  So $a\in H^4(X;\Z)$.
As all $Sq^i$ annihilate $b$
($Sq^1$ annihilates $b$ as $b$ is an integral class, and the higher
$Sq^i$ do so for dimensional reasons), we have by the Cartan formula
$Sq^3a= h \cup h \cup b\not=0$.  In this example, $X$ 
is not compact.  If desired, one can compactify $X$ without modifying
the discussion by adding a point at infinity to each $\R^3$ fiber of
$U\to Q$, replacing the $\R^3$ bundle by an $\S^3$ bundle.

Finally, we mention that if one does not wish to restrict
to ten-manifolds, there is a set of ``universal'' examples,
namely the cohomology of the Eilenberg-MacLane spaces $K(\Z,n)$ 
themselves. They are ``universal'' because any cohomology 
class on $X$ is uniquely associated to the  homotopy class of a
map $f: X \rightarrow K(\Z,n)$. The cohomology of the spaces 
$K(\Z,n)$ is built from some basic generators and 
certain ``cohomology 
operations'' such as $Sq^i$.

\subsec{Torsion Pairings}

We will here describe another important bit of
topological background.

We work on an oriented manifold $X$ of dimension $n$.  For $a\in H^k(X;\Z)$,
$c\in H^{n-k}(X;U(1))$, there is a cup product $a\cup c\in H^{n}(X;U(1))=U(1)$.
This gives a pairing which we denote as
\eqn\olop{(a,c) = \int_X a \cup c.}
One version of Poincar\'e duality is the statement that this pairing
is a Pontryagin duality between $H^k(X;\Z)$ and $H^{n-k}(X;U(1))$.

Now consider the short exact sequence of coefficient groups
\eqn\poik{0\to \Z\underarrow{i}\R\underarrow{r}U(1)\to 0.}
Here $i$ is the embedding of $\Z$ in $\R$, and $r$ maps the
real number $t$ to $\exp(2\pi it)\in U(1)$.  The associated
cohomology sequence reads
\eqn\noik{\cdots \rightarrow~  H^s(X;\R)\underarrow{r}
H^s(X;U(1))\underarrow{\beta} H^{s+1}(X;\Z)\underarrow{i}
H^{s+1}(X;\R)~\rightarrow \cdots.}
A class $b\in H^{s+1}(X;\Z)$ is torsion if and only if $i(b)=0$.
Exactness of \noik\ says that this is the condition for the existence of
$c\in H^s(X;U(1))$ with $\beta(c)=b$.

Now suppose we are given $a\in H^k(X;\Z)$ and a torsion
class $b\in H^{n-k+1}(X;\Z)$.  Because $b$ is torsion, there
exists $c\in H^{n-k}(X;U(1))$ such that $\beta(c)=b$.  In general,
the pairing $\int_X a\cup c$ depends on the choice of $c$ and not only on $b$.
However, the indeterminacy in $c$ is (according to \noik) $c\to c+r(e)$
where $e\in H^{n-k}(X;\R)$.  Because the cup product of a torsion class
with a real class is zero, the pairing $(a,c)$ is unaffected by the
indeterminacy in $c$ if $a$ is  a torsion class.

So for torsion classes $a$ and $b$, there is a well-defined torsion
pairing
\eqn\ono{T:H^k_{tors}(X;\Z)\times H^{n-k+1}_{tors}(X;\Z)\to U(1),}
defined by $T(a,b)= \int a \cup c$ where $\beta(c)=b$.
Equivalently, $a$ being torsion, there is $c'$ with $\beta(c')=a$,
and we can define $T(a,b)=\int c'\cup b$.
Poincar\'e duality can be used to prove that $T$ is a Pontryagin
duality between $H^k_{tors}$ and $H^{n-k+1}_{tors}$.
In the text below we often switch between ``additive'' and 
``multiplicative'' notation for abelian groups. When we use 
additive notation we will consider $T$ to be valued in $\R/2\pi\Z$. 
Which convention is used will be clear from the context. 

Here is a typical example where we will use the torsion pairings.
For $X$ of dimension 10 and $a,b\in H^4(X;\Z)$, we have met in
section 3 the bilinear form
\eqn\iki{\phi(a,b)=\int_X a \cup Sq^2 b.}
If $a$ is a torsion class, then $\phi$ can be interpreted
as a torsion pairing, as follows.  We have $\beta(Sq^2b)=Sq^1Sq^2b
=Sq^3b$ by the Adem relations. Now, $Sq^3b$ is a torsion class, and
running through the definition of $T$, we see that for $a$ torsion, we have
$\phi(a,b)=T(a,Sq^3b)$.  We have already proved that $\phi(a,b)$ is
symmetric, so it follows that $T$ is symmetric;
for torsion classes $a,b$, we have
\eqn\inop{T(a,Sq^3b)=T(b,Sq^3a).}

We conclude with some technical observations that will be useful in sections 6 and 7.
For $a$ a torsion class in $H^4(X;\Z)$ and $b$ any
class in $H^4(X;\Z)$, consider   $<a,b>=T(a,Sq^3b)$.
We will show that $<a,b>$ establishes a duality between certain spaces.
Note that though $T$ is $U(1)$-valued in general, as $Sq^3b$ is
two-torsion, $T(a,Sq^3b)$ takes values in the subgroup $\{\pm 1\}$ of
$U(1)$, which is isomorphic to $\Z_2$; so we will consider $T(a,Sq^3b)$ to
be $\Z_2$-valued.

Let $A=Sq^3(H^4(X;\Z))$, that is, $A$ is the subgroup of $H^7(X;\Z)$
consisting of elements of the form $Sq^3 b$ for $b\in H^4(X;\Z)$.
Let $B=Sq^3(H^4_{tors}(X;\Z))$; that is, $B$ is the subgroup of $A$ consisting
of elements $Sq^3b$ where $b$ is torsion.

Let $\Upsilon_0$ be the kernel of $Sq^3:H^4_{tors}(X;\Z)\to H^7(X;\Z)$.
It consists of torsion classes $b$ such that $Sq^3b=0$, i.e., such 
that  $Sq^2b$ has
an integral lift.  $\Upsilon_0$ has a subspace that we will call $\Upsilon$,
which consists of torsion classes $b$ such that $Sq^2b$ has an integral
lift which moreover is {\it torsion}.  As $\Upsilon$ is a subspace of $\Upsilon_0$,
$V=H^4_{tors}/\Upsilon$ has $W=H^4_{tors}/\Upsilon_0$ as a quotient: $W=V/(\Upsilon_0/\Upsilon)$.

$A,B,V$, and $W$ are all vector spaces over the field $\Z_2$.
(For $A$ and $B$ this is obvious; for $V$ and $W$ it requires the
observation that for any torsion class $c$, $2c\in \Upsilon$, since
$Sq^2(2c)=0$.)  Moreover, since $Sq^3$ (by definition) maps
$W$ injectively to $H^7(X;\Z)$, $W$ is isomorphic to its image, which
is $B$.

The pairing $<a,b>$ is nondegenerate as a map $W\times W\to \Z_2$
or equivalently $W\times B\to \Z_2$.  For this, we just need to know
that for every torsion class $a$ with $Sq^3a\not=0$ (so that $a$ represents
a nonzero element of $W$),
there is a torsion class $b$ with $<a,b>\not= 0$.
Since $<a,b>=T(Sq^3a,b)$ for $a$ and $b$ torsion, this is true
by nondegeneracy of the torsion pairings.  Thus, there is a natural
duality (as well as a natural isomorphism) between $B$ and $W$.

We claim that in addition, $V$ is dual to  $A$ by the pairing
that to $a\in V$ and $Sq^3c\in A$ assigns the value $T(a,Sq^3c)$.
(This pairing is well-defined because if $a\in\Upsilon$, so $Sq^2a$ can
be lifted to a torsion integral class, then for any integral class $c$,
$0=\int Sq^2a\cup c=\int a\cup Sq^2 c = T(a, Sq^3c)$.)
First, we must show that for all $a\in V$, there is $Sq^3c\in A$ with
$T(a,Sq^3c)\not=0$.  If $Sq^3a\not= 0$, the nondegeneracy of the torsion
pairing gives us a torsion class $c$ with $0\not= T(Sq^3a,c)=T(a,Sq^3c)$.
If $Sq^3a=0$ and
$a$ is a nonzero element of $V$,
then $Sq^2a$ can be lifted to an integral class, but this class cannot
be chosen to be a torsion class (or to be congruent mod 2 to a torsion
class).  So by ordinary
integer-valued Poincar\'e
   duality, there is $c\in H^4(X;\Z)$ (not  a torsion class)
with
$\int Sq^2 a \cup c\not= 0$ mod 2, and hence $\int a \cup Sq^2 c\not= 0$.
This last expression equals $T(a,Sq^3c)$.
This shows that for any nonzero $a\in V$, there is $Sq^3c \in A$
with $T(a,Sq^3c)\not= 0$.
Conversely, given
any $c\in H^4(X;\Z)$, if $Sq^3c\not= 0$, then there exists
 $a\in H^4_{tors}(X;\Z)$ such that that $T(a,Sq^3c)\not=0$.

So to summarize, $V$ is dual to $A=Sq^3(H^4(X;\Z))$, and the
quotient space $W=V/(\Upsilon_0/\Upsilon)$
 of $V$ is dual to the subspace $B=Sq^3(H^4_{tors}(X;\Z))$
of $A$.
This induces a duality
\eqn\ununu{(\Upsilon_0/\Upsilon)^*\cong A/B.}
According to the definitions,
$\Upsilon_0/\Upsilon$ consists of torsion classes $c$ that obey $Sq^3c=0$ and
so can be lifted to $K$-theory, modulo those whose lift to $K$-theory
is a torsion class.  In other words, given $c\in \Upsilon_0$, we have $c\in \Upsilon$
if and only if $Sq^2c$ can be lifted to an integral torsion class
$d$ (which can be the third Chern class of a $K$-theory lift of $c$).

\subsec{ A Note on Poincar\'e Duality In $K$-Theory}

The result \ununu\ has a relation to Poincar\'e duality (not used in the
rest of the paper) that we will
briefly explain.

In cohomology theory, one has the lattices $S=H^4(X;\Z)/H^4_{tors}$
and $T=H^6(X;\Z)/H^6_{tors}$.  They are dual to each other
by Poincar\'e duality. In passing to $K$-theory, one loses
certain classes in $S$ that are not annihilated by $Sq^3$.  $S$ is replaced
by a sublattice $S'$ (defined presently) that is of some finite index $n$.
 Poincar\'e
duality holds in $K$-theory just as it does in cohomology.  To maintain
the duality between $S$ and $T$, if one loses classes in $S$, one must
gain classes in $T$; $T$ must be replaced by a lattice $T'$, containing
$T$, such that $T'/T$ has the same index as $S/S'$.  In fact, $T'/T$
must be dual to $S/S'$.

  How does this happen?  We will give a brief
explanation, without any attempt at completeness.
The analog of $S$ in $K$-theory is the  group $S'$ of
 classes in $K(X)$ that are
torsion if restricted to the three-skeleton of $X$ modulo classes that
are torsion if restricted to the four-skeleton.  Given $a\in S$,
$a$ corresponds to an element of $S'$ if and only if, after possibly
adding to $a$ a suitable torsion class,
one can achieve $Sq^3a=0$ so that
$a$ can be lifted to $K$-theory.  The class in $S'$ determined by
$a$ is invariant under adding a torsion class to $a$ (and vanishes
if $a$ is torsion).  Thus $S/S'=A/B$.

The analog of $T$ in $K$-theory is the group $T'$ of
 classes that are torsion
if restricted to the five-skeleton of $X$ mod classes that are torsion
if restricted to the six-skeleton.  If $a\in H^4(X;\Z)$ is torsion
and can be lifted to $K$-theory, but one cannot take its lift
to be torsion (thus, $Sq^2a $ can be lifted to an integral class,
but not a torsion integral class), then the lift of $a$ to $K$-theory
corresponds to an element of $T'$, though $a$ does not correspond
to an element of $T$.  $T$ is the sublattice of $T'$ consisting of elements
that are trivial (and not just torsion) if restricted to the five-skeleton.

As a result, one has $T'/T=\Upsilon_0/\Upsilon$.  Given the conventional Poincar\'e duality
between $S$ and $T$ and the $K$-theory duality between $S'$ and $T'$,
the duality between $\Upsilon_0/\Upsilon$ and $A/B$ is a consequence.

We have described a mechanism for ``losing'' classes in $S$ in going
to $K$-theory, and for ``gaining'' classes in $T$.  There is no
analogous gain of classes in $S$ in going to $K$-theory, because
every torsion element of $H^2(X;\Z)$ can be lifted to torsion in $K$-theory
(by finding a suitable line bundle).  There is no analogous loss of
classes in $T$ because of \noloss.  There is no further mechanism
(involving higher AHSS differentials) for ``losing'' classes in $S$
because, on dimensional grounds, there is no further mechanism for
``gaining'' classes in $T$.

\newsec{Type II Superstrings, Cohomology, and $K$-Theory}

We encountered Steenrod squares in analyzing $M$-theory phases
in section 3, but they also enter in Type II superstring theory.
Understanding this will help us understand what we should aim
for in analyzing the $M$-theory partition function in section 6.

\subsec{Role of $Sq^3$}

We first think in terms of $D$-branes.  Given $b\in H^k(X;\Z)$,
we want to ask if there exists a $D$-brane state
such that its RR $k$-form charge is $b$ and the $r$-form
charges vanish for $r<k$.  
For this, we pick a submanifold $Q$ of spacetime that is Poincar\'e dual
to $b$, and try to wrap a $D$-brane on $Q$.  Such a $D$-brane state
will automatically have the desired $k$-form charge, and, as $Q$ is of
codimension $k$, it will have vanishing $r$-form charges for $r<k$.
Depending on the  Chan-Paton gauge field on the brane,
there may be RR $s$-form charges for $s>k$.

There is, however,   an obstruction to wrapping a $D$-brane
on $Q$ \fw: such a $D$-brane exists if and only
if the normal bundle to $Q$ (or equivalently, as $X$ is spin,
$Q$ itself) is $\spinc$.   In other words, the condition is $W_3(Q)=0$.
But $W_3(Q)=0$ implies $Sq^3(b)=0$.  In other words,
a $D$-brane whose lowest nonvanishing RR charge is $b$ exists
only if $Sq^3(b)=0$.

Since $D$-brane charge is measured by $K$-theory, finding a $D$-brane
whose lowest nontrivial brane charge is $b\in H^k(X;\Z)$ means finding a
$K$-theory
class $x$ ($x$ is in $K(X)$ or $K^1(X)$ for even or odd $k$)
such that $x$ is trivial on the $(k-1)$-skeleton of $X$, and the
obstruction to trivializing it on the $k$-skeleton is measured by $b$.
We call such an $x$ a lift of $b$ to $K$-theory.  The Atiyah-Hirzebruch
spectral sequence (AHSS) gives a systematic framework for relating 
cohomology
to $K$-theory  and determining
what cohomology classes can be lifted to $K$-theory \ahss.  In the AHSS,
the first approximation to $K(X) $ is 
(for $X$ of dimension $n$)
\eqn\kipp{E_2=\oplus_{2s\leq n} H^{2s}(X;\Z),}
and the first approximation to $K^1(X)$ is
\eqn\jipp{E_2^1 =\oplus_{2s+1\leq n}
H^{2s+1}(X;\Z).}  Thus the starting approximation is the one
in which $D$-brane
charge is just measured by cohomology.
Then one considers $Sq^3:E_2\leftrightarrow E_2^1$.
Since $(Sq^3)^2=0$ (as we mentioned in \lomop), one can define the
cohomology groups of $Sq^3$ acting on $E_2$ and $E_2^1$, respectively.
We call these cohomology groups $E_3$ and $E_3^1$; they
give the second AHSS approximations to $K(X)$ and $K^1(X)$, 
respectively.
In the AHSS, there is a sequence of higher order corrections, converging
eventually to a ``graded version'' of $K(X)$ and $K^1(X)$.
(This is explained in detail in appendix C.)
They are constructed from a series of ``differentials'' 
$d_r:H^k(X;\Z)\to
H^{k+r}(X;\Z)$,
where $r=3,5,7,\dots$ and the first differential is
 $d_3=Sq^3$.  The image of $d_r$ consists
of torsion classes that  in general have $p$-primary pieces for all primes
that divide $(r+1)!$  For example, $d_3$ is annihilated by
multiplication by 2 (as $2Sq^3=0$),
and $d_5$ has 2-torsion and 3-torsion.

On dimensional grounds, the only higher AHSS differential that might
be nontrivial on a ten-manifold is $d_5$. However, 
considerations of Poincar\'e duality show that $d_5$ annihilates the even-dimensional cohomology
of a ten-manifold $X$ (see the last sentence in section 4.3),
so    $d_5$ is not very important for understanding
the Type IIA $K$-theory theta function studied in the present paper.
It is possible that $d_5$ will  
 have a nontrivial action on $H^3(X;\Z)$, in which case it
would play a role in understanding the $K^1$ theta function of Type IIB.
Classes of the form $d_5(x)$ in general have both 2-torsion and 
3-torsion. It can be shown that the 2-primary part $d_5'$ is of order 4 and that
$(2d_5')(x) = Sq^5(x)$ ($d_5'$ itself is defined in terms of 
``secondary operations'') while the 3-torsion part of $d_5(x)$ 
involves a mod-3 version of the Steenrod operations
\dfive.\

Differentials beyond $d_5$ vanish on a ten-manifold.
Indeed, any class $b$ in $H^1(X;\Z)$ or $H^2(X;\Z)$ can be lifted
to $K$-theory.  For instance, for $b\in H^2(X;\Z)$, we can find a complex
line bundle ${\cal L}$ with $c_1({\cal L})= b$, and then
${\cal L}-{\one}$ (with ${\one}$ a trivial line bundle) will do
as a $K$-theory lift of $b$.  So we only have to consider $b\in H^k(X;\Z)$
for $k\geq 3$.  Then $d_rb=0$ for $r\geq 7$, since there is no torsion
in $H^{10}(X;\Z)$ and the higher cohomology of $X$ is zero.

Since RR fields, like RR charges, are classified by $K$-theory,
there is an analog of all this for RR fields.  In fact, this analog
will play the major role in the present paper.
The following example will enable us to tie together some of the points
that we have explained.  Consider Type IIA superstring theory and
ask whether, for some given $b\in H^4(X;\Z)$, there exists an RR field
with $G_0=G_2=0$, and $G_4/2\pi = b$.
For this, we must find a $K$-theory lift of $b$.  Equivalently,
we must find
a  complex vector bundle $E$ with $c_1(E)=0$,
$c_2(E)=-b$; then the desired RR field is associated with the $K$-theory class
$x=(E,F)$ (or $E-F$), where $F$ is a trivial bundle with the same rank as $E$.
For in this case $\sqrt{\hat A}\,\,\ch(x)=b+\dots$ where the $\dots$
are classes of degree $\geq 6$ and we have mapped $b$ into 
$H^4(X;{\bf Q})$ (thus 
losing torsion information). 
The role of subtracting $F$ is to cancel $G_0$; $G_2$ is zero because
$c_1(E)=0$.
If $E$ exists, then $c_3(E)$ is an integral class with $c_3(E)=Sq^2c_2(E)
=Sq^2b$
mod 2; we call it an integral lift of $Sq^2c_2(E)$.
  The existence of such an integral lift of $Sq^2b$ means
that $Sq^1Sq^2b=0 $ or
\eqn\unu{Sq^3b=0.}
So we see again in this particular example that
$Sq^3b$ is an obstruction to lifting a cohomology
class to $K$-theory.

\subsec{ Instability Of Some $D$-Branes}

We will now give an application of this formalism to exhibit
a new physical effect involving $D$-branes.
We know \fw\ that certain $D$-branes that would be allowed if
$D$-brane charge were measured by cohomology are actually not
allowed because $D$-brane charge is really measured by $K$-theory.
We will now show a flip side to this: certain $D$-branes that do exist
and would be stable if $D$-brane charge were measured by cohomology
are actually unstable, in fact, they are in the topologically trivial
component of the field space.

Suppose we are
given $c_0\in H^{k-3}(X;\Z)$ with $Sq^3c_0\not= 0$, and
set $c=Sq^3c_0$.  To be definite in the terminology, we will assume
that $k=2n$ is even.  Now the relation
\eqn\imo{Sq^3c_0=c}
can be read in two ways.  It asserts that $c_0$, not being annihilated
by $Sq^3$, is not an element of the cohomology of $Sq^3$, and so cannot be
lifted to an element of $K^1(X)$.  But
it also says that $c$, while annihilated by $Sq^3$ (since $Sq^3Sq^3=0$),
is trivial as an element of the $Sq^3$ cohomology --  so $c$ can be lifted
to $K$-theory but the lift is zero.  The first statement
 means that $c_0$ is not the
lowest brane charge of any $D$-brane.  In trying to construct
such a $D$-brane, one would run into the anomaly studied in \hw, which
we will soon look at from a different point of view.  The second statement 
means that while there exists a $D$-brane whose lowest brane charge
is $c$, this $D$-brane is unstable.  This last statement is the novel
one that we now wish to explain.

First let us recall a standard construction of a $K$-theory class
on a sphere $\S^{2n}$.  It is equivalent to construct a $K$-theory
class on $\R^{2n}$ with a trivialization at infinity.  For this
we take a pair $(E,F)$ of trivial bundles of equal rank $N=2^{n-1}$, 
together
with the usual tachyon condensate (first considered in various
examples in 
\ref\sen{A. Sen, ``Stable Non-BPS States in String Theory'',
JHEP {\bf 9806} (1998) 007, hep-th/9803194; `` Stable Non-BPS Bound 
States of BPS D-branes'', JHEP {\bf 9808} (1998) 010, hep-th/9805019;
``Tachyon Condensation on the Brane Antibrane System'', JHEP {\bf 9808}
(1998) 012.}):
\eqn\onso{T={\vec\Gamma\cdot \vec x\over \sqrt{1+|\vec x|^2}}.}
Near infinity, $T$ is a unitary matrix that defines a generator
of $\pi_{2n-1}(U(N))$; as a result, the pair $(E,F)$ with this tachyon
condensate at infinity is a generator of the compactly supported $K$-theory
of $\R^{2n}$ or equivalently a generator of $K(\S^{2n})$.

Now let us recall how this works in a global situation.  (The following
construction is due to Atiyah, Bott, and Shapiro \abs; for an 
explanation for
physicists, see section four of 
\ref\uwitten{E. Witten, ``$D$-Branes and $K$-Theory'', JHEP 
{\bf 9812} (1998) 019, hep-th/9810188.}.)
We start with a class $c\in H^{2n}(X;\Z)$ and find a $\spinc$ manifold
$Q$ dual to $c$.   
We want to describe a $D$-brane wrapped on $Q$ in terms of $K$-theory.
For this, we  lift $c$ to $K$-theory by constructing
a suitable $K$-theory class supported near $Q$.  We set 
$E=S_+(N)$, $F=S_-(N)$ where $S_\pm(N)$ are positive and negative
chirality $\spinc$ bundles of the normal bundle $N$ of $Q$.  We pull
back $E$ and $F$ to a small neighborhood $W$ of $Q$ in $X$; topologically,
we can think
of $W$ as the total space of the normal bundle $N$.  Then in each
fiber of $W\to Q$, we use the formula \onso\ to define the tachyon field.
This describes the $D$-brane state near $Q$ with a trivialization
(tachyon condensation that brings us to the vacuum) away from the immediate
neighborhood of $Q$. 
Denote by $W'$ the neighborhood with $Q$ omitted, 
and similarly, denote by $X'$ the complement of $Q$ in $X$.  For a complete
description, one extends $E$ and $F$ over $X$ (perhaps after a replacement
$(E,F)\to (E\oplus G,F\oplus G)$ for some bundle $G$) in such a way that $T$ extends over
$X'$ as a unitary map $T:E\to F$.  The importance of extending $T$ is
that if one cannot extend $T$ over $X'$, one will end up with additional
$D$-branes somewhere else away from $Q$, where unitarity of $T$ breaks down.

Using an additional bit of physics, the discussion we are about to give
can be simplified somewhat.  The simplifying fact is that actually,
a $D$-brane system wrapped on $Q$ with bundles $(E,F)$ is only allowed
if the bundles $E$ and $F$ are isomorphic when restricted to $Q$.
Otherwise, one cannot solve the equations for the RR fields
\selfduality.  (This is a $K$-theory version of a statement at the level
of cohomology that the Euler class to the normal bundle of a brane
must vanish \imwitten; otherwise,
 the equation for the appropriate RR or NS $p$-form
field  that couples magnetically to the brane in question would have no
solution.)  This means that, after possibly replacing $(E,F)$
by $(E\oplus G,F\oplus G)$, we can assume that $E$ and $F$ are trivial in $W$.
As a result, we
can interpret the tachyon field topologically as a map $T:W'\to U(N)$
(for some large $N$).
%
%
Thus, the whole content of the $D$-brane state is captured by a $U(N)$-valued
function on $W'$, just as in the example on $\R^{2n}$, it was captured
by a $U(N)$-valued function on the complement of the origin in $\R^{2n}$.

Now let us ask under what conditions 
a $D$-brane wrapped on $Q$, constructed as above, can decay even though
the homology class of $Q$ may be nontrivial.
This can happen if the $K$-theory class of the $D$-brane is zero.
If so, the $D$-brane can decay, by a process that involves nucleation
of $9-\bar 9$-brane pairs in the intermediate state, to exploit the fact
that, modulo creation and annihilation of such pairs, $D$-brane states
are completely classified by their class in $K(X)$.

The $D$-brane wrapped on $Q$ is trivial in $K(X)$
if the map $T:W'\to U(N)$ can be extended to a map
$T:X'\to U(N)$.  For in this case, we can extend $E$ and $F$ as 
trivial bundles over $X'$ while also extending
the tachyon field as a unitary map between them.  We end up with
a trivial class $(E,F)\in K(X)$ since $E$ and $F$ are both trivial.  
By contrast, if $T$ did not extend over $X'$ as a map to $U(N)$,
then to extend $T$ we would need to extend $E$ and $F$ over $X'$ as suitable nontrivial
bundles (chosen so that $T$ can be extended), 
and we would end up with a nonzero $K$-theory class $(E,F)$.
That is what happens for stable $D$-brane states.

\bigskip\noindent{\it Extension Of $T$}

Under what conditions can we extend $T:W'\to U(N)$ to $T:X'\to U(N)$?
As we will see, this will happen if there is $c_0\in H^{2n-3}(X;\Z)$ with 
$Sq^3c_0=c$.  In fact, as $c_0$ is an odd degree cohomology class,
one can try to lift it to an element of $K^1(X)$.  The lift will fail,
as $Sq^3c_0\not = 0$, and the failure will give us, as we will see
momentarily, the desired extension of $T$ over $X'$.

An element of $K^1(X)$ can be described by a map $V:X\to U(N)$ (for
some large $N$).  Let us try to construct such a map associated with
$c_0$.  We will use obstruction theory (see \topten\ for a review
for physicists).  We begin by triangulating $X$.  The class $c_0
\in H^{2n-3}(X;\Z)$ defines
(up to a certain equivalence relation)
an integral-valued function on the set of $(2n-3)$-simplices; this function
adds
up to zero for any collection of $(2n-3)$-simplices that comprise  the
boundary of a $(2n-2)$-simplex.

We define 
$V$ inductively on the $p$-skeleton (the union of all the $p$-simplices)
for $p=0,1,2,\dots$.  At the $p$-th step, $V$ has been defined on the 
$(p-1)$-skeleton, and we wish to define it on the $p$-skeleton.  
Each $p$-simplex is topologically
a $p$-dimensional ball $B^p$ with boundary $\S^{p-1}$ made from
$(p-1)$-simplices; $V$ has already
been defined on the boundary.  If $V$, restricted to the boundary,
is non-trivial in $\pi_{p-1}(U(N))$, it has no extension over $B^p$.
If $V$ is trivial on the boundary, its extensions over $B^p$ are
classified by $\pi_p(U(N))$.

To begin the induction, we define $V$ to
be identically 1 on the $(2n-4)$-skeleton.  To extend $V$ on the 
$(2n-3)$-skeleton, we need an element of $\pi_{2n-3}(U(N))=\Z$ for
each $(2n-3)$ simplex $B$.  We simply assign to $B$ the integer
determined by the cohomology class $c_0$.  In extending $V$ over
the $(2n-2)$-skeleton, there is  potentially an obstruction since
$\pi_{2n-3}(U(N))\not=0$.  However, the obstruction vanishes because
$c_0$ assigns the value $0$ to a sum of $(2n-3)$-simplices that make
up the boundary of a $(2n-2)$-simplex.  At the next step, there is
no obstruction to extending $V$ over the $(2n-1)$-skeleton, since
$\pi_{2n-2}(U(N))=0$.  In extending $V$ over the $2n$-skeleton, however,
there is a potential obstruction, associated with $\pi_{2n-1}(U(N))=\Z$.
The obstruction  assigns an integer to each $2n$-simplex.

It can be shown that this collection of integers defines an element of
$H^{2n}(X;\Z)$.  Moreover, this element is just $Sq^3c_0$.
This assertion is equivalent to the statement that $Sq^3c_0$ is the first
obstruction to lifting $c_0$ to an element of $K^1(X)$.  We have denoted
$Sq^3c_0$ as $c$.
The dual to $c$ is our manifold $Q$, and having $c$ as the obstruction
to extending $V$ means that $V$ can be extended over the complement
$X'$ of $Q$.

Thus, $V$ is the desired extension of $T$ whose existence shows
that a $D$-brane wrapped on $Q$ can be unstable.  
We have not above defined the topological type of $T$ in a completely
unique way, because (using different $\spinc$ structures) there can
be different $D$-brane states wrapped on $Q$.  They differ in their
$(2n+2k)$-form charges for $k\geq 1$.  However, $V$ coincides with
{\it some} $U(N)$ valued function
$T$ whose behavior near $Q$ is suitable to describe a $D$-brane wrapped on $Q$.
Existence of $V$ means that this $D$-brane state is unstable.
 Other $D$-branes wrapped on $Q$, if they
carry $(2n+2k)$-form charges that cannot be written as 
$Sq^3(\dots)$, are not completely unstable but can decay to $D$-branes
wrapped on manifolds of dimension less than that of $Q$.

\bigskip\noindent{\it Examples}

We will conclude by giving a few concrete examples of $D$-branes 
wrapped on non-trivial homology cycles that are nonetheless unstable.
Pursuing one of the examples considered in section 4.1,
we take $X={\bf RP}^7\times {\bf RP}^3$, with generators $a$ and
$b$ for $H^1({\bf RP}^7;\Z_2)$ and $H^1({\bf RP}^3;\Z_2)$.
$Sq^1a$ and $Sq^1b$ are two-torsion
integral classes that we will somewhat loosely call
$a^2$ and $b^2$. (Strictly speaking, as $a$ and $b$ are mod 2 classes,
their squares $a^2=a\cup a$ and $b^2=b\cup b$ are mod 2 classes; these have
integral lifts that we will also call $a^2$ and $b^2$.) 
We set $c=a^4\cup b^2$.  $c$ is dual to 
$B={\bf RP}^3\times {\bf RP}^1$, with the two factors linearly embedded
in the two factors of $X$.  
As $B$ is nontrivial in homology, it appears that a $D$-brane wrapped
on $B$ would be stable.  But actually, we have $c=h\cup h = Sq^3h$,
where $h=Sq^1(a\cup b)$ reduces mod 2 to $a^2\cup b+a\cup b^2$.
So some $D3$-brane wrapped on $B$ is unstable.

Similarly, we could set $c'=a^6\cup b^2$, which is dual to $B'={\bf RP}^1\times
{\bf RP}^1$.  As $c'=Sq^3c_0$ with $c_0=Sq^1(a^3\cup b)$, a $D$-brane
wrapped on $B'$ can again be unstable.

In the last example, since ${\bf RP}^1$ is a circle,
$B'$ is isomorphic to $\TT^2$.
By turning on a magnetic flux on $\TT^2$, we can endow a $D$-brane on
$B'$ with $-1$-brane charge, which takes values in $H^{10}(X;\Z)=\Z$.
As there is no torsion here, a nonzero class in $H^{10}(X;\Z)$ cannot
be written as $Sq^3(\dots)$.  This example thus also makes clear that the
correct statement is that {\it some} $D$-brane wrapped on $B'$ is completely
unstable, and {\it any} $D$-brane wrapped on $B'$ can decay to a collection
of $-1$-branes.

These have been Euclidean examples, so the unstable objects
are really instantons rather than physical states of branes.
For a real time example, we begin with the eight-manifold $U$
considered at the end of section 4.1 (constructed as an $\R^3$ bundle
over $Q$, or an $\S^3$ bundle if one prefers to compactify the fibers).
We set $X=U\times \S^1\times \R$, where $\R$ is the ``time'' direction.
Then in view of the remarks in section 4.1, a threebrane wrapped on
$L\times \S^1$ is unstable, even though $L\times \S^1$ is nontrivial
in homology.

More generally, we can set $X=Y\times \R$ for any nine-dimensional
spin manifold $Y$, with $\R$ still understood as the time direction.
$Sq^3$ in general can act non-trivially
on $H^3(Y;\Z)$ or $H^4(Y;\Z)$, but annihilates the other cohomology
groups.  (For example, it annihilates $H^5(Y;\Z)$ because of \noloss.)
The image of $Sq^3$ thus lies in $H^6(Y;\Z)$ and $H^7(Y;\Z)$, which
equal $H_3(Y;\Z)$ and $H_2(Y;\Z)$.  So the real time
 $D$-branes (as opposed to Euclidean signature instantons) that are 
destabilized by the mechanism that we have described are always 
two-branes or three-branes.

\newsec{Partition Function In $M$-Theory}

We are finally ready to analyze the partition function of
the $C$-field on $Y=X\times \S^1$.  Actually, we will only evaluate
the contribution from $C$-fields that are pulled back from $X$ --
corresponding to the RR field $G_4$ in the Type IIA description.
(The other modes would correspond in the Type IIA description to the
Neveu-Schwarz $B$-field, which is generally omitted in the present 
paper.)

We treat the $C$-field on $Y$ as a free field.  Its modes that are 
pulled
back from $X$ are classified by the characteristic class 
$a\in H^4(X;\Z)$.
For each $a$ there is a harmonic four-form $G_a$ of the appropriate
topological class, as in \geea, and the
 kinetic energy $|G_a|^2=\int G_a\wedge *G_a$ 
vanishes if and only if $G_a$ is torsion. 
The partition sum we wish to evaluate is
\eqn\polu{\sum_{a\in H^4(X;\Z)} (-1)^{f(a)}\exp(-|G_a|^2).}

Here we are summing over all $a\in H^4(X;\Z)$, while in the
corresponding Type IIA expression, we would only be summing over
those $a$'s that have a lift to $K$-theory.  As a preliminary
step towards comparing \polu\ with Type IIA, we want to re-express it
as a sum only over a restricted set of $a$'s.  The basic strategy
for this will be to use the fact that the kinetic energy $|G_a|^2$ is
invariant under $a\to a+b$ for torsion $b$ while the $M$-theory 
phase is not.  Hence, if
$(-1)^{f(a)}$ vanishes upon averaging over $a\to a+b$ for a suitable
set of torsion $b$, the contribution of $a$ can be omitted from
the partition sum.   Physically, averaging over $a\to a+b$ for torsion
$b$ amounts to deriving the torsion part of a Gauss's law constraint.

\subsec{ An  Anomaly}

As a preliminary step, we first average over $a\to a+2b$, where
$b$ is torsion.  To see what this does, we first use the bilinear
relation \jugoxo\ to get
\eqn\ono{f(a+2b)=f(a)+f(2b),}
where the bilinear term can be dropped as $Sq^2(2b)=2 Sq^2b=0$.

Now, for $f(2b)$ we can give a simple formula, whether $b$ is torsion
or not.  Note, in the context of the cobordism discussion in section
3.2,   that $f(2b)$ is a cobordism invariant and so must be a linear
combination of $f(b)$ and $v(b)=\int b\cup Sq^2\lambda$.  In fact,
the bilinear relation gives at once
\eqn\bono{f(2b)=f(b)+f(b)+\int b\cup Sq^2 b = 
\int b\cup Sq^2\lambda=\int b\cup w_6= v(b),}
where we used the fact that $2f(b)=0$ together with \noko.
(The interested reader can use the same technique to show that
$f(3b)=f(b)+v(b)$ and $f((n+4)b)=f(nb)$.)

In view of \ono, the  partition function transforms under $a\to a+2b$
(where $b$ is torsion)
by multiplication by $(-1)^{f(2b)}$.  The partition function therefore
vanishes unless $f(2b)=0$ for all torsion $b$.  From \bono, this means
that $\int b \cup
Sq^2\lambda=0$ for all torsion $b$. This integral is the
torsion pairing $T(b,Sq^3\lambda)$ described in section 3.3.
Its vanishing for all torsion $b$ is equivalent, by nondegeneracy of
the torsion pairing, to
\eqn\lopo{Sq^3\lambda=0.}

If $Sq^3\lambda\not=0$, then the partition function vanishes.
This vanishing cannot be removed by inserting local operators constructed
from the $C$-field (as these are not sensitive to torsion classes).
We interpret it as an anomaly in the theory.  An analogous
anomaly was studied in \duality.  The meaning of this anomaly for Type IIA
will be explained in section 7.

The physical effect of a torsion
$C$-field is precisely to give a phase to the contribution to the path
integral of a wrapped brane.  So, if one wishes, one could (just as in the
examples studied in \duality) remove the anomaly by introducing a wrapped
brane.  We will  explain this more fully in section 6.4  below, but 
for the moment we focus attention to the standard partition function. 

It remains to show that the anomaly we have described is a nontrivial 
restriction on 10-manifolds. The argument for this is somewhat abstract 
and can be found in appendix D.

\subsec{ Restriction On $Sq^3(a)$}

Henceforth we work on spin manifolds with $W_7=0$.

We want to get a restriction on the $a$'s that contribute to the partition
sum, by considering the behavior under $a\to a+c$ for $c$ torsion.
We have already imposed invariance under $a\to a+2b$ for torsion $b$,
so we can consider $c$ to lie in $L=H^4_{tors}/2H^4_{tors}$,
which is a vector space over the field $\Z_2$.

One's first thought might be that the contribution of $a$ vanishes
unless $f(a+c)=f(a)$ for all torsion $c$.  That would be correct
if $f(a+c)$ were linear in $c$, but it is not.  By iterating the bilinear
relation for $f$, one finds that
\eqn\uvu{f(a+c+c')=f(a)+f(c)+f(c')+\int a\cup Sq^2(c+c')+\int c \cup Sq^2 c'.}
The last term is the obstruction to $f(a+c)$ being linear in $c$.

To derive a useful constraint, we will sum over a restricted set of
$c$'s chosen so that $f(a+c)$ is a linear function on this set.
For this, we simply define $L'$ to be the subspace of $L$ consisting
of classes $c$ such that $\int c\cup Sq^2 c'=0$ for all torsion $c'$.
(Since $\int c\cup Sq^2 c' = \int Sq^2 c\cup c' = T(Sq^3c,c')$, the
condition for this is that $Sq^3c=0$.)
For $c,c'\in L'$, $f(c+c')=
f(c)+f(c')$,
so $f(c)$ is a linear function when restricted to $L'$.
The bilinear relation
\eqn\huvu{f(a+c) = f(a)+f(c)+\int  c\cup Sq^2 a,}
shows that $f(a+c)$ is likewise linear in $c$  for $c\in L'$.

The linear function $f(c):L'\to \Z_2$ can be extended (nonuniquely)
to a linear function on $L$, and hence by the Pontryagin duality of
the torsion pairing, there is a (nonunique)
$P\in H^7(X;\Z)$ with $f(c)=T(c,P)$ for all $c$.  We can write
\eqn\buvu{f(a+c)=f(a)+T(c,Sq^3a +P).}
$P$ is unique mod the addition of an element $P'=Sq^3c'$ for torsion $c'$
(since these are the elements for which $T(c,P')=0$ for all $c\in L'$).

The condition for
\eqn\olo{\sum_{c\in L'}(-1)^{f(a+c)}}
to be nonzero is that $f(a+c)$ is independent of $c$ for $c\in L'$.
In other words,  $T(c,Sq^3a+P)=0$ for all such $c$.

Let $M$ be the two-torsion subgroup
of $H^7_{tors}$.   $L$ and $M$ are vector spaces over the field $\Z_2$,
and the torsion pairing $T:H^4_{tors}\times H^7_{tors}\to \Z_2\subset U(1)$
induces a nondegenerate pairing or duality $T:L\times M\to \Z_2$.
$L'$ was defined as the subspace of $L$ orthogonal to
the subspace $M'=Sq^3(H^4_{tors})$  of $M$.  We write this
as $L'=(M')^\perp$.  Just as in the more familiar case of linear
algebra over $\R$, given dual vector spaces $L$ and $M$ and subspaces
$L'$, $M'$ with $L'=(M')^\perp$, one has also    $M'=(L')^\perp$.
So the fact that $T(c,Sq^3a+P)=0$ for all $c\in L'$ means that
$Sq^3a+P\in M'$, that is, $Sq^3a+P=Sq^3c'$ for some torsion $c'$.

The restriction on $a$ can thus be written
\eqn\nojo{Sq^3a = P~{\rm mod}~Sq^3(H^4_{tors}).}
(The sign of $P$ does not matter as $P$ is two-torsion.)

If $P$ is identically zero, this means that classes $a$ that contribute
to the $M$-theory partition function have the property that,
after perhaps adding a torsion class to $a$, $Sq^3a=0$.  
We decompose the sum over $a\in H^4(X;\Z)$ into a sum over equivalence
classes, where $a\sim a'$ if $a-a'$ is torsion.  For $P=0$, every
equivalence class that contributes to the $M$-theory partition function
has a representative that lifts to $K$-theory.  
Thus, the $M$-theory partition function 
 can be written
as a sum over $K$-theory classes.  This is the expected answer from the
Type IIA side -- though of course we need to show that
the Type IIA partition function precisely reproduces the $M$-theory
partition function.  This will be our goal in section 7.

If $P$ is nonzero, it may be that \nojo\ has no solution.  Then the
partition function vanishes and $M$-theory on $X\times \S^1$
is anomalous. We will eventually show in section 7.8 below 
that this occurs only if $W_7\not=0$, 
so it is not really a new anomaly. 
As a preliminary 
to that discussion, let us find the criterion for the existence 
of a solution to \nojo.
If $f(c)$ is nonzero for some $c$ such that $Sq^2c $ can be lifted
to a torsion class $d\in H^6(X;\Z)$ -- in other words, if $c$ belongs
to the space $\Upsilon$ introduced at the end of section 3 --
 then \nojo\ has no solution. We prove this as follows.
Suppose $a$ obeys \nojo; using our
freedom to add $Sq^3c'$ to $P$ for $c'$ torsion, we can assume $Sq^3a=P$.
Then \nojo\ implies (using the definition of $P$, as well as \nojo\
and formulas from section 3)
that $f(c)=T(c,P)=T(c,Sq^3a)=\int c\cup Sq^2 a
=\int Sq^2 c\cup a =\int d\cup a$.  But $\int d\cup a=0$ if $d$ is torsion,
as the cup product in integral cohomology vanishes for torsion classes.
Conversely, if $f(c)=0$ for $c\in \Upsilon$,
then $f(c)$ can be regarded as a linear form on the vector space
$V=H^4_{tors}/\Upsilon$
  studied at the end of section 3.  We showed there that the
dual of $V$ is $Sq^3(H^4(X;\Z))$, so that $P$ is an element of
$Sq^3(H^4(X;\Z))$, that is $P=Sq^3a$ for some $a\in H^4(X;\Z)$.
In summary, 
\eqn\nosolp{
P\in Sq^3H^4(X,\Z) ~ \Leftrightarrow ~ f(c) =0 ~{\rm for~all}~ c\in \Upsilon.
}

So \nojo\
has no solution, rendering
the $M$-theory  anomalous, precisely if the function $f$ is nontrivial
if restricted to $\Upsilon$.  We interpret $\Upsilon$ to consist of the part of the
torsion subgroup of $K(X)$ with first Chern class $c_1=0$.             
Indeed, an element $c\in \Upsilon$
has  a $K$-theory lift because $Sq^3c=0$; this lift can be chosen
to be an element $x$ with Chern classes $c_1(x)=0$, $c_2(x)=-c$, $c_3(x)=d$,
and as $c$ and $d$ are torsion, this is compatible with $x$ being torsion.
\foot{To prove that $x$ can be taken to be torsion, we must show that
also $c_4$ and $c_5$ can be taken to be torsion.  For this, we consider
the index of the Dirac operator with values in $x$.  It is simply
$c_5(x)/4!$  So $c_5(x)$ is a multiple of $4!$  A bundle on $\S^{10}$ or
equivalently a bundle that is trivial outside a small neighborhood
of a point $P\in X$ can have $c_5$ an arbitrary multiple of $4!$
By adding such a bundle (and subtracting a trivial bundle of the same
rank) we can set $c_5(x)=0$ without changing $c_i(x)$ for $i<5$.
By considering the index of the Dirac operator with values in
${\cal L}\otimes x$, we next learn that $c_4(x)$ is a multiple of
$3!$   If $c_4(x)=-3! \,[\Sigma]$, where $[\Sigma]$ is the class of a Riemann
surface $\Sigma\subset X$, then by adding to $x$ the $K$-theory class
of a $D1$-brane wrapped on $\Sigma$,
we can make $c_4(x)$ vanish.  Since $\Sigma$ is spin, this can be done
with a flat Chan-Paton bundle and so without changing $c_5(x)$.}

Type IIA is anomalous if $j(x)$, defined as the mod 2 index with values in
$x\otimes \bar x$, is nonzero for a torsion class $x\in K(X)$.  In section 7,
we will compare $f(c)$ to $j(x)$, and a special case of our result is
that if $c\in \Upsilon$ can be lifted to a torsion class $x\in K(X)$,
then $f(c)=j(x)$ (see eq. (7.33) below).  So this $M$-theory anomaly arises
precisely if the Type IIA theory is anomalous.  This is part of the
matching between these two theories.

It remains to consider the possibility that $P$ is nonzero
but \nojo\ has a solution.  This happens if $f(c)$ is nonzero on $L$
but annihilates the subspace $\Upsilon$.  In analyzing this case, we set
$S=H^4(X;\Z)/H^4_{tors}$, as at the end of section 4.2.

In this case, let $a_0$ be any solution of \nojo.  Then
the general solution is $a=a_0+b$, where (after perhaps adding
a torsion class to $b$) $b$ must obey $Sq^3b=0$.
In other words, the $M$-theory partition function is not written
as a sum over the sublattice $S'={\rm ker}\,Sq^3$ of $S$, which is
the sublattice of $S$ consisting of classes that have a $K$-theory lift.
Rather, it is a sum over a coset of $S'$ in $S$, namely the coset containing
$a_0$.

To compare to Type IIA, we will have to take account of the following.
The comparison to Type IIA is made not quite in terms of $a$ but
in terms of the four-form $G/2\pi$, which as we recall from \geea\ in section
2 is not quite $a$ but $a-\lambda/2$.  We want to compare the $M$-theory
field $G/2\pi$ to a Type IIA RR field $G_4/2\pi$.  The RR
forms of Type IIA are defined just as differential forms, so in making
this comparison, we should work mod torsion. 

Let us   fix a definite solution of \nojo, 
\eqn\defao{
Sq^3 a_0 = P .
}
Then we have 
\eqn\juko{{G\over 2\pi}
=a-\lambda/2=b+a_0-\lambda/2,}
where   $b$ is an arbitrary element of $S'$.
This formula says that $G/2\pi $ takes values in a coset of $S'$
in $\half S$, namely the coset
that is generated by $a_0-\lambda/2$.

Actually, we will need to 
be more precise than this.  Note that $\theta_M=2a_0-\lambda$ is
an element of $S'$, since $Sq^3(2a_0)=0$ and (to cancel an anomaly)
we have had to assume that $Sq^3\lambda=0$.  So the allowed fluxes $G/2\pi$
in $M$-theory take values in a coset of $S'$ in $\half S'$,
namely the coset generated by $\theta_M/2$.  (This is an improved
statement because, $\half S'$ being a sublattice of $\half S$,
there are fewer cosets of $S'$ in $\half S'$ than in $\half S$.)

Note that $\theta_M$ is not well-defined as an element of $S'$, since
the solution $a_0$ of 
\nojo\ is not uniquely determined.  But, as the ambiguity in $a_0$
consists of the possibility of adding to $a_0$ an element of $S'$,
$\theta_M$ is well-defined as an element of $S'/2S'$.

In comparing to Type IIA, we will among other things have to explain why the 
RR four-form flux takes values precisely in the coset of $S'$ in
$\half S'$ that has just been described.  
 
\subsec{ Contribution Of An Equivalence Class}

We want to describe the $M$-theory partition function as a sum over
equivalence classes of solutions of \nojo.  We consider two solutions
$a$ and $a'$ equivalent if $a-a'$ is torsion.  Every equivalence
class contains a representative $a$ with $Sq^3a=P$.  
The sum over the equivalence class is 
\eqn\olpop{Z_a=\exp(-|G_a|^2)\sum_{c\in H^4_{tors}}(-1)^{f(a+c)}.}
The bilinear relation shows, given $Sq^3a=P$,
 that $f(a+c)=f(a)+f(c)+T(P,c)$.  So we can write
\eqn\nolpop{Z_a=\CN (-1)^{f(a)}\exp(-|G_a|^2),}
with 
\eqn\otopop{\CN=\sum_{c\in H^4_{tors}}(-1)^{f(c)+T(P,c)}.}
\nolpop\ expresses the contribution of an equivalence
class to the partition function in terms of an overall constant $\CN$.
We want to show that $\CN\not=0$; indeed,
  vanishing of $\CN$ would constitute a new anomaly.  
We will actually get a simple formula for $\CN$, which we hope will eventually
be useful in comparing the absolute normalization of the $M$-theory
partition function to that of Type IIA (though we will not analyze
all of the absolute normalization factors on the two sides in the present 
paper).

As we saw above, with $W_7=0$, the sign factor in \otopop\ is invariant
to $c\to c+2c'$.  So we can rewrite \otopop\ as
\eqn\npolpop{\CN=N_0\sum_{c\in L}(-1)^{f(c)+T(P,c)}}
where $L=H^4_{tors}/2H^4_{tors}$ and $N_0$ the order of the finite
group $2H^4_{tors}$.   
The definition of $P$ is such that $f(c)+T(P,c)$ is invariant under
$c\to c+c'$ for $c'\in L'={\rm ker}\,Sq^3$.
So if $N_1$ is the order of $L'$ and $L''=L/L'$, we can write
\eqn\unupop{\CN=N_0N_1\sum_{c\in L''}(-1)^{g(c)},}
where $g(c)=f(c)+T(P,c)$.

The function $g(c)$ is quadratic:
\eqn\tunupop{g(c_1+c_2)=g(c_1)+g(c_2)+\int c_1 \cup Sq^2 c_2.}
The bilinear form $<c_1,c_2>=\int c_1 \cup Sq^2 c_2$ is nondegenerate
on $L''$ (since we have divided out its kernel, which is $L'$).
Given our assumption that $W_7=0$ (and hence $f(2c)=0$), the diagonal
matrix elements of this bilinear form vanish,
since \tunupop\ implies that $0=g(2c)=g(c)+g(c)+\int c \cup Sq^2 c
=<c,c>$.
Over the field $\Z_2$, a quadratic form $<~,~>$ with vanishing diagonal
matrix elements is equivalent to an antisymmetric form, and if
nondegenerate, it can
be block-diagonalized in $2\times 2$ blocks in each of which it looks like
\eqn\omino{\left(\matrix{ 0 & 1\cr 1 & 0 \cr}\right).}
The procedure for proving this is familiar in linear algebra over $\R$.
We let $b_1$ be any element of $L''$ and (using the nondegeneracy) 
we let $b_2$ be any element of $L''$ with $<b_1,b_2>=1$.
In the subspace generated by $b_1$ and $b_2$, the form looks like \omino\
(since the diagonal elements vanish).  Repeating this procedure in the 
subspace of $L''$ orthogonal to $b_1$ and $b_2$, one gets the claimed
block diagonalization.

Suppose that $L''$ is two-dimensional (over the field $\Z_2$), and
so the quadratic form has precisely the shape \omino.
If we write $b=u_1b_1+u_2b_2$
(with $u_1,u_2\in \Z_2$), the most general quadratic
function $g(b)$ obeying \tunupop\ on $L''$ is $g(b)=u_1u_2+\epsilon_1u_1+\epsilon_2u_2$,
with constants $\epsilon_1,\epsilon_2\in \Z_2$.  A small computation
shows that in this situation
\eqn\uvu{\sum_{b\in L''}(-1)^{g(b)}=2(-1)^\alpha,}
where $\alpha=\epsilon_1\epsilon_2$.

Now suppose that $L''$ has dimension $2k$ and so the quadratic
form is the sum of $k$ blocks of the shape \omino.   The number of
elements of $L''$ is thus $N_2=2^{2k}$.
A function $g$ obeying \unupop\ is the sum of $k$ functions of the
sort considered in the last paragraph, one in each $2\times 2$ block.  Hence
\eqn\nuju{\sum_{b\in L''}(-1)^{g(b)}=2^k(-1)^\alpha.}
Here $\alpha=\sum_{i=1}^k\alpha_i$, with $\alpha_i$ being defined
as in the last paragraph for the $i^{th}$ $2\times 2$ block.  
$\alpha$ 
is called the ``Arf invariant'' of the quadratic function $g$.
The Arf invariant is a $\Z_2$-valued invariant of a quadratic function $g$, 
and can be defined invariantly as the sign of the sum \nuju. 
Up to transformations $b\to Ab+b'$ where $A$
is a linear transformation of $L''$ and $b'\in L''$, the quadratic
function $g$ is completely determined by its Arf invariant.
Indeed, it is easy to show that the number of zeroes of $g$ is 
$\half(2^{2k} \pm 2^k)$ with the sign determined by the Arf invariant. 

Our result for $\CN$ is thus
\eqn\uncon{\CN=N_0N_1\sqrt{N_2}(-1)^\alpha={N(-1)^\alpha\over \sqrt{N_2}},}
where $N=N_0N_1N_2$ is the order of $H^4(X;\Z)_{tors}$.

We should note that the factorization of $Z_a$ in \nolpop\ depended
on a specific choice of $P$.  If we transform $P\to P+Sq^3 c_0$
(with torsion $c_0$), we must take $a\to a + c_0$.  In this process
$(-1)^{f(a)}\to (-1)^{f(a)}(-1)^{f(c_0)+T(P,c_0)}$.  This sign change
of $(-1)^{f(a)}$ is compensated by a change of the Arf invariant.

\subsec{Comment on Brane Insertions} 

This paper focuses on the partition function of $M$-theory. 
Nevertheless, 
one is very interested, for a variety of reasons, in the computation of
amplitudes with  insertions of wrapped membranes.
We now briefly sketch how those are formulated on an arbitrary 
11-dimensional spin manifold $Y$. As an application we show 
that, while the partition function vanishes if $W_7\neq 0$ on 
manifolds of the form $Y=X \times \S^1$,  one could, if 
desired, always insert a torsion membrane instanton which cancels the anomaly 
described in section 6.1.   (The rest of the paper does not depend
on this construction, so the reader could omit this section.)

Let $Q$ be the worldvolume of an M2-brane instanton. The 
contribution to the path integral in the presence of the $C$ field receives
a factor 
\eqn\branecoupling{
\exp\left(i \int_{Q} C\right).
} 
This is not a topological invariant in general
when $C$ is not flat, and a careful discussion of the 
instanton amplitude involves the  calculus of ``differential 
characters'' \refs{\cheegersimons,\harris}.
We need not enter into such subtleties here because we are 
only concerned with the behavior of 
\branecoupling\ under shifts of $C$ by 
a {\it flat} field $C'$. 
 A flat $C$-field is classified
by $H^3(Y;U(1))$, and for flat $C$-fields, \branecoupling\ can
be regarded as the dual pairing $H_3(Y;\Z)\times H^3(Y;U(1))\to U(1)$.
Because this is a duality, any desired 
linear map from flat $C$-fields to $U(1)$
can be obtained as the coupling to some brane whose homology class is torsion.
In particular, by picking a suitable $Q$, the dependence of the
effective action on $C\to C+C'$ with torsion $C'$ can be canceled.
This can be done with a $Q$ whose homology class is torsion, though
the action of a brane wrapped on $Q$ is of course positive.

%

\newsec{Comparison To Type IIA}

In this section, we will analyze the RR partition function in Type IIA
and begin the process of demonstrating its relationship to the $M$-theory partition function.

\subsec{Review Of The $K(X)$ Theta Function}

First we recall \refs{\duality,\selfduality} the general construction
of a $K$-theory theta function, which serves as the RR partition function
in Type IIA.  (A precisely analogous construction based on $K^1(X)$ gives
the RR partition function of Type IIB.)  One starts on a ten-dimensional
spin manifold $X$ with the lattice
$\Gamma=K(X)/K(X)_{tors}$.  This lattice is endowed with an integer-valued unimodular antisymmetric form by the formula
\eqn\oplo{\omega(x,y)=I(x\otimes \bar y),}
where for any $z\in K(X)$, $I(z)$ is the index of the Dirac operator with
values in $z$.\foot{This antisymmetric form is $T$-duality invariant.
Indeed, if we exchange Type IIA with Type IIB and consider $\omega$
as being defined for $D$-brane states rather than for RR fields,
it becomes the $T$-duality invariant intersection form on $D$-brane
states introduced by Douglas and Fiol \dougfiol.}
  In any dimension of the form $4k+2$, one has $\omega(x,y)
=-\omega(y,x)$.
The Ramond-Ramond field $G(x)$ of a given $x$ is defined as
\eqn\noplos{{G(x)\over 2\pi}=\sqrt{\hat A(X)}\,\,\ch\,x,}
with $\hat A$ the index density of the Dirac operator and
$\ch$ the Chern character.  In \noplos, we understand the right hand side
to refer to the harmonic differential form in the specified real cohomology class.
Note that the RR fields are defined purely as differential forms.  The integral
structure is defined by deriving the RR fields from an element of $K(X)$
(which has a natural integral structure, of course), not by defining integral
cohomology classes associated with the RR fields.

\def\L{{\cal L}}
\def\T{{\bf T}}
Given a metric on $X$, one
 also endows $\Gamma$ with a metric $g(x,y)$ as follows.
One simply sets
\eqn\opop{g(x,y)=\int_X{G(x)\over 2\pi}\wedge {*G( y)\over 2\pi},}
where here $*$ is the Hodge duality operator.
Associated with the lattice $\Gamma$ is a torus $\T=A/\Gamma$ where 
$A$ is the vector space $\Gamma\otimes_\Z\R$.
The quantities $\omega$ and $g$ can be interpreted as a symplectic form
and a metric, respectively, on $\T$.  We define a complex structure $J$
on $\T$ by setting
\eqn\normop{g(x,y)=\omega(Jx,y).}
The metric, complex structure, and symplectic structure that we have defined
turn $\T$ into a K\"ahler manifold.  Suppose that $\L$ is a complex line
bundle over $\T$ with positive curvature.  Then $H^i(\T;{\cal L})=0$ for
$i>0$, and according to the index theorem, the dimension $h^0(\L)$ of
$H^0(\T;\L)$ is
\eqn\okko{h^0(\L)=\int_\T e^{c_1({\cal L})}.}
(The Todd class, which would appear in the general index theorem
for the $\bar\partial$ operator, is 1 for a complex torus.)  
Unimodularity of $\omega$ implies that
\eqn\nokko{\int_\T
e^\omega = 1,}
so if we can find an ${\cal L}$ with $c_1({\cal  L})=\omega$, then
$h^0(\L)=1$.  In this case, ${\cal L}$ has, up to a constant multiple,
a unique holomorphic section; this section, suitably normalized, is the RR partition function
(as a function of an ``external potential''). If $\T$ is endowed with a complex line bundle 
$\L$ with $c_1(\L)=\omega$, it becomes a ``principally polarized abelian
 variety.''

As was explained in detail in \imwitten, holomorphic line bundles $\L$ over 
$\T$ with {\it constant} curvature $\omega$ are in one-one 
correspondence with   $U(1)$-valued functions $\Omega$ on
$\Gamma$ such that
\eqn\unnu{\Omega(x+y)=\Omega(x)\Omega(y)(-1)^{\omega(x,y)}.}
(In brief, to define $\L$ as a unitary line bundle with connection of curvature
$\omega$, we need to specify its holonomies around noncontractible
loops in $\T$; the role of $\Omega$ is to specify these holonomies.)
While $\Omega$ cannot be taken to be identically $1$, since $\omega$ is nonvanishing, 
one can take $\Omega$ to be valued in $\Z_2$.  This is the case relevant
to constructing the RR partition function of weakly coupled Type II superstrings.

In \duality,  a natural $\Z_2$-valued
function $\Omega$, canonically associated to 
a spin manifold, and obeying \unnu,
was defined using the mod 2 index of the
Dirac operator.  The definition was as follows.
For $X$ of dimension $8k+2$, and any real vector
bundle $V$, one has a mod
2 index $q(V)$
of the Dirac operator with values in $V$.  More generally, for any
$v\in KO(X)$, one can define the mod 2 index $q(v)$ with values in $v$.
  For any $x\in K(X)$,
one has $x\otimes \bar x\in KO(X)$, so one can define $j(x)=q(x\otimes \bar x)$.
Then 
\eqn\bunnu{\Omega(x)=(-1)^{j(x)}}
can readily be shown \duality\ to obey the desired identity.

Though the identity \unnu\ does not determine $\Omega$ uniquely,
the formula \bunnu\ is distinguished because it is $T$-duality invariant,
that is, it can be described in terms of a conformal field theory with
worldsheet supersymmetry without committing oneself to a particular realization
of this theory as a sigma model with a target space $X$.  A manifestly
$T$-duality invariant 
definition of $j(x)$ is as follows.  Interpret $x$ as  a $D$-brane state
in Type IIB superstring theory in the same conformal
field theory background as the Type IIA model
under consideration (using a different GSO projection to get IIB instead
of IIA) and, 
in rough analogy with \dougfiol, define  $j(x)$ as the number,
mod 2, of zero energy states in the Ramond sector for open strings
with boundary condition $x$ at each end.\foot{The use of Type IIB
to define the $\Omega$ function for Type IIA is admittedly slightly perplexing.
Of course, to define the $\Omega$ function for Type IIB, we would similarly
look at open string boundary conditions in Type IIA.}  Though we do not
have a general proof that \bunnu\ is the unique $T$-duality invariant
solution of \unnu, this seems very likely. 

If now $\Omega(x)$ is identically 1 for torsion elements of $K(X)$,
then it can be regarded as a function on $\Gamma=K(X)/K(X)_{tors}$ and can be
used to define the line bundle ${\cal L}$ and thence the RR partition 
function.  If $\Omega$ is not identically 1 on $K(X)_{tors}$, then the
partition function of the theory vanishes upon summing over torsion.
This must be interpreted as an anomaly or inconsistency of the theory.
(In \duality, examples were given where nontriviality of the $\Omega$ function
on torsion was related by duality to more conventional anomalies.)
In our problem, we will see (in section 7.8) that the $\Omega$
function fails to be identically 1 on torsion precisely when the $M$-theory
partition function vanishes for a similar reason.

If $\Omega$ descends to a function on $\Gamma$, we can proceed to
construct a theta function that will serve as the RR partition function.
To define the theta function, we pick an arbitrary splitting of
$\Gamma$ as a sum $\Gamma_1\oplus \Gamma_2$, where $\Gamma_1 $
and $\Gamma_2$ are ``maximal Lagrangian'' sublattices, that is, $\omega(x,y)=0$
for $x,y$ both in $\Gamma_1$ or both in $\Gamma_2$, and $\Gamma_1$ and
$\Gamma_2$ are each maximal lattices with this property.  It follows
from this that $\omega$ establishes a duality between $\Gamma_1$ and 
$\Gamma_2$.  An important
 example of the use of this duality is as follows.  For $x,y\in \Gamma_2$,
we have $\Omega(x+y)=\Omega(x)\Omega(y)$.  Thus, $\Omega$
determines a homomorphism from $\Gamma_2$ to $ \Z_2$.  Duality of
$\Gamma_1$ with $\Gamma_2$ via $\omega(~,~)$ means that there exists
$\theta\in \Gamma_1/2\Gamma_1$ such that
\eqn\tuxoc{\Omega(y)=(-1)^{\omega(\theta,y)}}
for $y\in\Gamma_2$.

The theta
function is then, roughly speaking, written as a sum over $\Gamma_1$.
To be more precise, it is written as a sum over a certain coset of
$\Gamma_1$ in $\half\Gamma_1$, namely the coset containing the element
$\theta/2$, where $\theta$ was just defined.  
We introduce  a homogeneous quadratic function $\tau$ on $\Gamma_1$,
 which is the period matrix of the lattice
$\Gamma$ with respect to its decomposition as $\Gamma_1\oplus \Gamma_2$.
(We will presently explain how to compute $\tau$ explicitly.)
The theta function is then
\eqn\imocco{\Theta=\exp(-i\pi{\rm Re}\,\,\tau(\theta/2))\sum_{x\in\half\theta+\Gamma_1}
\exp\left(i\pi \tau(x)\right)
\Omega(x-\theta/2).}
The  prefactor $\exp(-i\pi{\rm Re}\tau(\theta/2))$,
which is just a constant phase multiplying the theta function,
has been chosen to cancel some of the dependence of the theta function
$\Theta$ on $\theta$.\foot{This prefactor 
is also required for the partition function 
to be well-defined in the presence of a $B$-field.}
In fact, we have defined $\theta$  as an element of
$\Gamma_1/2\Gamma_1$, but in writing the formula \imocco, $\theta$
is interpreted as an element of $\Gamma_1$.  With the prefactor
that we have chosen, under $\theta\to \theta+2b$, $\Theta$ changes by
an overall sign.  (This can be proved using formulas
we develop later.)
We do not know how to fix the overall sign of the partition
function, and in this paper, we will study only the dependence on the
RR fields, not the overall constant normalization of the partition function.  
(Note that in $M$-theory, as we saw in section 2, to get a well-defined
overall sign of the partition function requires carefully considering
the fermions as well as bosons.)

In practice, as we will see, the imaginary part of $\tau(x)$ equals
the conventional kinetic
energy of the RR fields.  The real part of $\tau$ will give an
$x$-dependent phase factor which, together with the factor 
$\Omega(x-\half\theta)$, must be compared to the phase factor coming from
$M$-theory.  With an obvious shift of the summation variable, we can
alternatively write the theta function as
\eqn\nimocco{\Theta=\exp(-i\pi{\rm Re}\,\tau(\theta/2))\sum_{x\in \Gamma_1}\exp\left(i\pi\tau(x+\theta/2)\right)
\Omega(x).}  

\subsec{ Choice of $\Gamma_1$ and $\Gamma_2$}

The $\Theta$ function can be written as in \imocco\ for {\it any}
Lagrangian decomposition $\Gamma=\Gamma_1\oplus\Gamma_2$. 
We want to make a choice that is convenient for comparing to the conventional
approach of expressing the RR partition function as a sum over $2p$-form
periods for various $p$.  First of all, the conventional approach
is not unique, as (for example)
we could treat $G_0$, $G_2$, and $G_4$ as the independent
variables, or (by duality) one could use $G_6$, $G_8$, and $G_{10}$.
It is both conventional and
much more convenient, however, to use $G_0$, $G_2$, and $G_4$ as
the independent variables.  The reason this is convenient is that, 
in comparing to $M$-theory, we want to scale up the metric $g$ of $X$
by $g\to tg$ for large positive $t$.  
As we noted in the introduction, under this scaling
the action $\int_Xd^{10}x\sqrt g |G_{2p}|^2$ scales as $t^{5-2p}$.  
Hence, for large $t$, the representation of the partition function
as a sum over fluxes of $G_0,G_2$, and $G_4$ is rapidly convergent:
the nonzero fluxes are all associated with a large 
action.  The action, in fact, is large for $G_4$, larger still for $G_2$,
and largest for $G_0$,
so we get a hierarchy of approximations: one may include $G_4$ only,
which will be the approximation of the present section;
one may include $G_2$ and $G_4$, as we do in section 9, or one
may do a complete computation, as we do in section 10.
If we would instead take $G_6$, $G_8$, and $G_{10}$ as the independent
variables, then as the action for these fields is small for large $t$,
all contributions to the path integral are important, and the 
existence of a hierarchy of successive approximations is less apparent.
Moreover, $G_2$ and $G_4$ (but regrettably not $G_0$) are the variables
that are most easily seen in $M$-theory, so in comparing to $M$-theory
it is most convenient to use a representation of the path integral
in which we sum over $G_2$ and $G_4$ (and neglect $G_0$).

With this in mind, and
with $\Gamma=K(X)/K(X)_{tors}$, there is a completely canonical
choice for $\Gamma_2$: we take $\Gamma_2$ to be the 
subgroup of $K(X)$ consisting of classes that are torsion when
restricted to the five-skeleton, modulo those that actually are torsion.
Thus, a class in $\Gamma_2$ has vanishing $G_0$, $G_2$, and $G_4$.
One might hope to take $\Gamma_1$ to be the subgroup of $K(X)$ consisting
of classes with vanishing $G_6,$ $G_8$, and $G_{10}$, but there is
no such subgroup.  Indeed, relations such as $c_3=Sq^2c_2~ \mod~ 2$ (for a
complex vector bundle with $c_1=0$) make it impossible to let $G_0$,
$G_2$, and $G_4$ vary while keeping $G_6$, $G_8, $ and $G_{10}$ zero.
There is a canonical quotient $\Gamma/\Gamma_2$, but there is no
natural way to lift this quotient to a sublattice $\Gamma_1$ of $\Gamma$. 

In practice, a choice of $\Gamma_1$ gives a recipe 
to lift a  collection of RR fields $G_0$, $G_2$, and $G_4$ (obeying
appropriate quantization conditions) to
an element of $K(X)$ mod torsion.  
There is no canonical choice of how to do this,
but, since the theta function can be computed for any decomposition
$\Gamma=\Gamma_1\oplus\Gamma_2$, we will get the same RR partition
function no matter how $\Gamma_1$ is chosen.  Actually, as we will
see shortly, given our choice of $\Gamma_2$,
the imaginary part of $\tau$ does not depend on the choice
of $\Gamma_1$, but the real part does.
The factor $\Omega(x-\theta/2)$ in the partition function also depends
on the choice of $\Gamma_1$, and when both factors are included, the 
dependence on the choice of $\Gamma_1$ cancels out.

As an aside, it is worth noting that in some situations there are other 
very natural choices of lattices $\Gamma_1,\Gamma_2$. For example
if $X= X_9 \times \S^1$ with a nine-manifold $X_9$,
 then $K^0(X) \cong K^0(X_9) \oplus K^1(X_9)$. 
One could choose $\Gamma_1 = K^0(X_9)$, $\Gamma_2=K^1(X_9)$. 
In this case, it turns out that $\tau$ is imaginary and the theta function 
is a sum of real  terms.

\subsec{Computation Of $\tau$}

Here, we will carry out the explicit computation of $\tau$.
For $x\in \Gamma=K(X)/K(X)_{tors}$, we set $ G(x)/2\pi=\sqrt{\hat A}\,\,\ch\,(x)$.
The metric and symplectic form on $\Gamma$ are defined, as we explained above,
by
\eqn\hurty{\eqalign{g(x,y) & ={1\over (2\pi)^2}
\int_X  G(x)\wedge *  G(y) \cr
                    \omega(x,y) & = 
{1\over (2\pi)^2}\int_X G(x)\wedge G(\bar y)=-{1\over (2\pi)^2}
\sum_{q=0}^5(-1)^q\int_X G_{2q}(x)\wedge G_{10-2q}(y).    \cr}}
In the last step, we use the fact that
 $y\to \bar y$ acts on the RR fields by $G_{2p}\to (-1)^pG_{2p}$.
The complex structure $J$ on $\Gamma\otimes_\Z\R$ is defined by
\eqn\purty{\omega(Jx,y) =g(x,y).}
Explicitly, this means that
\eqn\turty{(-1)^{p+1}G_{2p}(Jx)=*(G_{10-2p}(x)).}

As above, we let $\Gamma_2$ be the sublattice of $\Gamma$ with
$G_0=G_2=G_4=0$ (corresponding to $K$-theory elements whose restriction
to the five-skeleton in $X$ is torsion), and we let $\Gamma_1$ be any
complementary Lagrangian sublattice.  We pick a basis $y_i$ of $\Gamma_2$
and a dual basis $x^i$ of $\Gamma_1$:
\eqn\jumipo{\omega(x^i,y_j)=\delta^i{}_j, ~~\omega(x^i,x^j)=\omega(y_i,y_j)=0.}
Explicitly evaluating $\omega(x^i,y_j)$ from the definition of $\omega$
and the fact that $y_j\in \Gamma_2$,
we have
\eqn\umipo{\delta^i{}_j={1\over (2\pi)^2}\int_X\left(-G_4(x^i)
G_6(y_j)+G_2(x^i)G_8(y_j)-G_0(x^i)G_{10}(y_j)\right).}

The period matrix $\tau(x^i,x^j)$, also denoted $\tau^{ij}$, is
defined by requiring that
\eqn\jolopy{Z^i=x^i+\sum_j\tau^{ij}y_j} 
should obey  $J(Z^i)=\sqrt{-1} Z^i$ for all $i$, where 
$J$ is extended to act complex-linearly. Similarly, 
extending $G$ to act complex-linearly, can can use  
\turty\ to obtain: 
 \eqn\olopy{\sqrt{-1}(-1)^{p+1}G_{2p}(x^i+\sum_j\tau^{ij}y_j)=*\left(G_{10-2p}
(x^i+\sum_j\tau^{ij}y_j)\right).}
Setting $10-2p=2q$ and using $G_{2q}(y_j)=0$ for $q=0,1,2$, we get
\eqn\holopy{(-1)^{q+1}G_{10-2q}(x^i)+(-1)^{q+1}\sum_j\tau^{ij}G_{10-2q}(y_j)=
\sqrt{-1} *(G_{2q}(x^i)),~~q=0,1,2.}

If one takes the cup product of this formula with $G_{2q}(x^k)$ and
sums over $q=0,1,2$, one gets a formula for $\tau^{ij}$:
\eqn\hovopy{\tau^{ij}=\sqrt{-1} \sum_{q=0,1,2}\int_X{G_{2q}(x^i)\over 2\pi}
\wedge {*G_{2q}(x^j)\over 2\pi}+\sum_{q=0,1,2}(-1)^q\int_X
{G_{10-2q}(x^i)\over 2 \pi} \wedge {G_{2q}(x^j)\over 2\pi} .}
${\rm Im}(\tau^{ij})$ is manifestly symmetric in $i$ and $j$;
to prove symmetry of ${\rm Re}(\tau^{ij})$, one uses $\omega(x^i,x^j)=0$.

Any $x\in\Gamma_1$ has an expansion $x=\sum_i f_ix^i$ with integers $f_i$.
We define $\tau(x)=\sum_{ij}f_if_j\tau(x^i,x^j)$.
${\rm Im}\,\tau(x)$ is the conventional kinetic energy of
the RR fields $G_0$, $G_2$, and $G_4$ associated with $x$. 
For generic $x$,  $G(x)$ also has nonzero 
components $G_{2p}$ for $p\geq 3$; these depend on
the non-canonical choice of $\Gamma_1$, but  do not appear in
${\rm Im}\,\tau$.  On the other hand, ${\rm Re}\,\tau$ is
a topological invariant (independent of the metric on $X$), but
does depend on the higher components of $G(x)$.
We will show in section 7.4 that the $\Theta$ function is independent 
of the choice of $\Gamma_1$. To do this, it is 
useful to 
note that while  the function $\tau(x)$ is initially only defined for
$x\in \Gamma_1$, the explicit formula \hovopy\ makes 
sense for all $x\in \Gamma$. The resulting extension 
of $\tau(x)$ 
 has the 
the nice property that for any
$x\in\Gamma$ and for any $y\in \Gamma_2$, one has
\eqn\olopp{\eqalign{{\rm Im}\,\tau(x+y) & = {\rm Im}\,\tau(x) \cr
                     {\rm Re}\,\tau(x+y) & = {\rm Re}\,\tau(x)+\omega(x,y).\cr}}

%

\subsec{ Existence Of Description Via $G_0,G_2,$ And $G_4$}

Naively speaking, the RR partition function in Type IIA
superstring theory is defined as a sum over
$G_0$, $G_2$, and $G_4$ fields with certain quantization conditions on the
periods.  We will now show that such a description does hold in
Type IIA superstring theory, but that both the quantization conditions
on the $G_{2p}$ and the phases with which different terms contribute
to the path integral are unusual.    
Since the theta function is written as a sum over $\Gamma_1$,
the quantization condition on
the $G_{2p}$ is that there must exist  $x\in\Gamma_1$ such
that
\eqn\jurgo{{G_{2p}\over 2\pi} = \left(\sqrt {\hat A}\,\,
\ch(x+\theta/2)\right)
      _{2p}~{\rm for}~p=0,1,2.}
We will now show that the
 values of $G_{2p}$ allowed by this relation for $p=0,1,2$ are
independent of the choice of $\Gamma_1$.  This can be seen as follows.
Any change of $\Gamma_1$,
keeping $\Gamma_2$ fixed, can be implemented by selecting a map
$f:\Gamma_1\to \Gamma_2$, obeying
\eqn\oko{\omega(x_1,f(x_2))+\omega(f(x_1),x_2)=0, ~ {\rm for}~x_1,x_2\in\Gamma_1.}
Given such a map, one replaces $\Gamma_1$ by the Lagrangian lattice  
$\hat\Gamma_1$ that consists of elements $\hat x=x+f(x)$ for
$x\in \Gamma_1$.  Since $G_{2p}(f(x))=0$ for $p=0,1,2$, the condition
\jurgo\ is not affected by this transformation.
Similarly, in changing lattices, $\theta$ is mapped to $\hat\theta=
\theta+f(\theta)$ (a transformation that preserves the defining
property of $\theta$, namely that $(-1)^{\omega(\theta,y)}=\Omega(y)$ for
$y\in\Gamma_2$).  This likewise does not modify the condition
\jurgo.

Having found the quantization conditions on the $G_{2p}$,
can one forget the rest of the $K$-theory
formalism?  Not quite.  Naively, one would expect to weight
a given set of RR fields by the exponential of the classical
supergravity action.  This exponential is  positive 
 (as long as the NS $B$-field,
which would produce a phase, vanishes).  However, the $K$-theory
formalism gives us a phase.  Given $G_0$, $G_2$, and $G_4$ which
are correctly quantized -- so that a solution $x$ of \jurgo\ exists --
we pick such a solution, which is contained in $\Gamma_1$ for some choice
of $\Gamma_1$, and then we discover from \nimocco\ that the contribution
of $x$ to the partition function is
\eqn\xono{Z_x=\exp(-i\pi\Re\tau(\theta/2))\exp(i\pi\tau(x+\theta/2)) \Omega(x).}
Let us now verify that the contribution to the partition function of
a given $G_0,G_2$, and $G_4$, depends only on those fields and not
on the choice of $x$.  For this, we must show that $Z_x$ as defined
in \xono\ is invariant under 
\eqn\kilom{x\to x+f(x),~
\theta\to \theta+f(\theta),}
for any linear map $f:\Gamma_1\to\Gamma_2$ that obeys \oko.
To prove this, we must recall the multiplicative property of $\Omega$:
\eqn\obocon{\Omega(x+f(x))=\Omega(x)\Omega(f(x))(-1)^{\omega(x,f(x))}
=\Omega(x)(-1)^{\omega(x+\theta,f(x))}.}
In the second step, we have used the fact that $\Omega(f(x))=(-1)^{\omega(\theta,
f(x))}$, since $f(x)\in\Gamma_2$.  
In addition, we must use \olopp\ and \oko\ to show that
\eqn\kurry{\tau(x+\theta/2+f(x)+f(\theta/2))-\tau(x+\theta/2)=\omega(x,f(x))+
\omega(\theta,f(x))+\omega(\theta/2,f(\theta/2)).}  The last term
cancels the transformation law of the prefactor in \nimocco,
and putting the pieces together, we learn that $Z_x$ indeed has
the claimed symmetry \xono.

Thus, as one would naively expect, the RR partition function in Type IIA
can be written as a sum over a certain lattice of allowed values
of $G_{2p}$ fluxes for $p=0,1,2$, with a precise recipe for the contribution
of each lattice point.  The formula, however, is surprisingly subtle.
The phase factor coming from ${\rm Re}\,\tau$ depends on cohomology
operations such as $Sq^2$ (which constrains $G_6$ in terms of the
$G_{2p}$ with $p\leq 2$).  In addition, there is a sign factor $\Omega(x)$
coming from the mod 2 index; this factor is even more subtle, in that
there is no cohomological formula for the mod 2 index, even using
operations such as $Sq^2$.  The usual Type IIA
supergravity Lagrangian misses the phase factor in $Z_x$, and this is
quite natural since those factors really cannot be described using
conventional ingredients.

``Why'' does the RR partition function of Type IIA contain such subtle
phase factors?  One way to explain this is via $T$-duality.  Starting
with a description of the partition function as a sum over $G_0$, $G_2$,
and $G_4$ fluxes only, $T$-duality will mix in higher RR fields. To return
to a description via $G_{2p}$ for $p\leq 2$, one will have to accompany
a $T$-duality transformation with a spacetime duality.  The spacetime 
duality will generate phases.  There is consequently no $T$-duality
invariant partition function without phases.  
A manifestly $T$-duality invariant construction is the $K$-theory
theta function.
When reduced to a description via $G_0$, $G_2$, and $G_4$, it
takes the form that we have described.

Now that we have established the existence in Type IIA of a description
via $G_0$, $G_2$, and $G_4$, our main goal in the rest of this paper
will be to compare this description to what comes from $M$-theory.
In fact, in the present section, we consider only the contributions
with $G_0=G_2=0$.  After  working out the $M$-theory phase
on circle bundles in section 8  we will 
make a comparison including $G_2$ in section 9.

\subsec{Comparing $E_8$ and $K$-theory mod two indices} 

One key ingredient in comparing the $M$-theory and 
$K$-theory theta functions is the relation between the 
mod two indices used to define these two functions. 
Accordingly, 
let us consider a class $a\in H^4(X,Z)$ which has 
a $K$-theory lift $x$.  As we have seen in section 3.1, we can 
assume there is a rank 5 $SU(5)$ bundle $E$ with 
$x = E-F$   where $F$ is a trivial
rank 5 bundle. Thus, $c_1(x) = 0, c_2(x)=-a$. 
We have then $x\otimes \bar x=E\otimes \bar E\oplus
F\otimes \bar F-E\otimes \bar F-\bar E\otimes  F$.  $E\otimes\bar E$
is the same as ${\rm ad}(E)\oplus {\cal O}$, where ${\cal O}$ is a trivial
line bundle and ${\rm ad}(E)$ is the bundle derived from $E$ in the adjoint
representation of $SU(5)$.
$F\otimes \bar F$ is the same as 25 copies of ${\cal O}$.  So
for purposes of mod 2 index theory, we can replace $E\otimes \bar E
\oplus F\otimes \bar F$ by ${\rm ad}(E)$.
Likewise as $F$ is a trivial bundle of odd rank, 
we can replace $E\otimes {\bar F}\oplus\bar E\otimes F$ by $E\oplus\bar E$,
and the mod 2 index with values in this bundle is the mod 2 reduction of
$I(E)$, the ordinary index with values in $E$.  
So
\eqn\jupi{\Omega(x)=(-1)^{q({\rm ad}(E))+I(E)},}
where we recall that $q$ denotes the mod 2 index.

Now let us compare with the $M$-theory phase. 
For this, we simply construct an $E_8$ bundle $V(a)$ in the 
adjoint representation with characteristic
class $a$. We can then relate $V(a)$  to $E$ using the familiar embedding 
of $SU(5)\times SU(5)\subset
E_8$. The decomposition of the adjoint representation of $E_8$ was
given in equation \jurry.  Roughly as in that discussion, we take the first
$SU(5)$ bundle to be $E$, and the second to be the trivial bundle $F$.
Then, discarding representations that appear with even multiplicity,
the adjoint $E_8$ bundle can be expressed in terms of $E$ as
${\rm ad}(E)\oplus \wedge^2E\oplus\wedge^2\bar E$.  The
mod 2 index with values in this bundle is $q({\rm ad}(E))+I(\wedge^2E)$.
So
\eqn\popsi{{f(a)}=q({\rm ad}(E))+I(\wedge^2E).}

Now we can compare \jupi\ to \popsi\ using the index theorem. 
If we define $\ch_t(x) = \sum t^k \ch_k(x)$ for any class $x$, then 
we can use  the splitting 
principle to derive: 
\eqn\spltpr{
\ch_t(\Lambda^2 E) = \half (\ch_t(E))^2 -\half \sum_{k\geq 0} (2t)^k\ch_k(E)
}
In particular,
$\ch_5(\Lambda^2(E)) = (\ch_2\ch_3 + \ch_1\ch_4-11\ch_5)(E)$, so 
we get index densities
\eqn\indxs{
\eqalign{
i(E) & = {c_5(E) - (c_2(E) + \lambda) c_3(E) \over 24} \cr
i(\Lambda^2(E)) & = {-11 c_5(E) - (c_2(E) + \lambda) c_3(E) \over 24} \cr}
}
and it follows that  for any $SU(5)$
bundle $E$,
\eqn\nopsi{I(E)+I(\wedge^2E)=\int_X{\lambda c_3(E)+c_2(E)c_3(E)\over 2} ~\mod ~ 2.
}
(Note that it follows from \indxs\  that $c_5(E)/2$ is integral, and 
moreover that  $c_5(E) - (c_2(E) + \lambda) c_3(E)$ 
is divisible by $24$, and hence
that  
$\half (\lambda + c_2(E))c_3(E)$ is integral. )

Putting these equations together, we obtain the following key result.
If $a$ has a $K$-theory lift $x$, then 
\eqn\effomega{
(-1)^{f(a)} = \Omega(x) e^{{i \pi \over 2} \int( \lambda + c_2(x))c_3(x)} 
}
It follows, in particular,
that the right hand side of \effomega\ is independent 
of the lift $x$ of $a$.   To compare $M$-theory to Type IIA, we still
need a knowledge of the ``characteristic.''

\subsec{Evaluation Of The Characteristic}

One important ingredient in the Type IIA theta function
is the ``characteristic'' $\theta\in \Gamma_1/2\Gamma_1$, defined by the condition
$(-1)^{\omega(\theta,y)}=\Omega(y)$ for $y\in \Gamma_2$.
We will here compute $\theta$, which can be regarded as an element of
$\Gamma/(2\Gamma\oplus \Gamma_2)$, which is the same
as $\Gamma_1/2\Gamma_1$.  We will show
that
\eqn\hurryio{G_0(\theta)=G_2(\theta)=0.}
  Then we will show that
\eqn\irio{{G_4(\theta)\over 2\pi}=-\lambda+2a_0 ~~~~{\rm mod}~~2\,\,{\rm ker}\,Sq^3,}
where $a_0$ is a class encountered on the $M$-theory side in section
6.   $G_4(\theta)/2\pi$ is only
determined modulo $2\,{\rm ker}\,Sq^3$ simply because $\theta$ is only
uniquely defined mod $2\Gamma$; adding to $\theta$ an
element of $2\Gamma$ that is trivial on the three-skeleton will add
an element of $2\,{\rm ker}\,Sq^3$ to $G_4(\theta)/2\pi$.
\hurryio\ and \irio\ are approximations to the following
more precise description of $\theta$: $\theta$ is trivial on the
three-skeleton of $X$, and its image in the Atiyah-Hirzebruch
spectral sequence (see appendix C
for more detail) is the class $-\lambda+2a_0$ given in \irio.
This uniquely determines $\theta$ modulo the possibility of adding
a $K$-theory class trivial on the five-skeleton, that is, an element
of $\Gamma_2$.  So the above description completely characterizes
$\theta$ as an element of $\Gamma/(2\Gamma+\Gamma_2)$.

To verify the above properties of $\theta$, we will proceed as
far as we can with a direct, elementary computation.  This will
be done by representing $K$-theory classes in terms of branes
with even-dimensional world-volume.  Such branes enter more directly
in the physics of Type IIB superstring theory, but here we will use
them in computing the $\Omega$ function of Type IIA.  (A similar
technique was used in section 7.1 to demonstrate the $T$-duality invariance
of $\Omega$.)

The basic tool in the direct computation will be a fact explained in 
section 4 of \selfduality.  If a $K$-theory class $y$ can be
represented by a $D$-brane wrapped on a submanifold $Q_y$ of
spacetime (and endowed with some $\spinc$ structure), then the mod 2 index $j(y)$ with values in $y\otimes \bar y$
is equal to $\nu(Q_y)$, the number mod 2 of zero modes of the worldvolume
fermions of the brane wrapped on $Q_y$.  
(The worldvolume fermions are spinors of $Q_y$ with values in spinors
of the normal bundle to $Q_y$, subject to the usual chirality projection.)
Then, since $\Omega(y)$
is defined as $(-1)^{j(y)}$, we get
\eqn\tofero{\Omega(y)=(-1)^{\nu(Q_y)}.}
$\theta$,       therefore, is characterized by
\eqn\ofero{\omega(\theta,y)=\nu(Q_y)~{\rm  mod}~2}
for $y\in\Gamma_2$.

To detect $G_0(\theta)$, we take $y$ to have $G_{2p}(y)=0$ except
for $p=5$.  This means that $Q_y$ should be a $-1$-brane or a point $p$ in $X$.
The Dirac operator of a point is zero; it acts in this case on a rank
16 bundle (the spinors of the normal bundle), so the number of
zero modes is 16.  So $\omega(\theta,y)=0$ mod 2 if $y$ is dual
to a point, and we can pick $\theta$ to be trivial up to the
two-skeleton of $X$.

To evaluate $\theta$ on the two-skeleton, we must evaluate
$(\theta,y)$ where $y$ is dual to a Riemann surface $\Sigma$ in $X$
(so that $G_{2p}(y)=0$ for $p<4$).  As $X$ and $\Sigma$ are
spin, the normal bundle to $\Sigma$ in $X$ is spin.  A spin bundle
on a Riemann surface is trivial, so the normal bundle is a trivial
rank eight bundle.  The positive or negative chirality spinors of the
normal bundle are hence trivial rank eight bundles, and the number
of zero modes of the Dirac operator of the world-volume fermions is
divisible by eight.  Hence, $\omega(\theta,y)=0$ if $y$ is dual
to a Riemann surface.  So we can pick $\theta$ to be trivial up to
the four-skeleton of $X$.

To evaluate $\theta$ on the four-skeleton, we must evaluate
$(\theta,y)$ where $G_{2p}(y)=0$ for $p<3$.  Any $y$ that is the
$K$-theory class of a four-manifold $Q_y$ has this property (but as
we explain later, there are additional $y$'s, so the direct computation
we are about to make will not give a complete answer).
For such $y$'s, since $G_0(\theta)=G_2(\theta)$, we have
$(\theta,y)=\int_{Q_y}{G_4(\theta)/2\pi}$.  
If we evaluate this expression  using \irio, we find that (mod 2),
the $2a_0$ term does not contribute, and that \irio\ implies
\eqn\nini{\nu(Q_y)=(\theta,y)=\int_{Q_y}\lambda ~{\rm mod}~2.}
This formula for $\nu(Q_y)$ is correct;  it can be deduced by using
index theory to count the fermion zero modes on $Q_y$, as in
\fourflux. (The minus sign in \irio\ is a choice made for 
convenience in comparison to $M$-theory.) 

The reason that this computation does not completely determine
$\theta$ on the four-skeleton is that the condition $G_{2p}(y)=0$
for $p<3$ does not imply that $y$ is dual to a four-manifold.  It implies
that $y$ is torsion on the five-skeleton of $X$, but $y$ must actually
be trivial on the five-skeleton to be the $K$-theory class of a four-manifold
(which has codimension six in $X$).  It is perhaps helpful to recall
that the class $a_0$ entered in section 6 in considering 
the $M$-theory contribution of classes $a\in H^4(X;\Z)$ that are torsion,
and can be lifted to $K$-theory, but whose $K$-theory lift $y$
cannot be chosen to be torsion.\foot{There is no subtlety analogous
to this in the cases considered above, because there is no torsion
in $H^0(X;\Z)$, and torsion in $H^2(X;\Z)$ can always be lifted to 
torsion in $K$-theory
by finding a suitable line bundle.}
In this situation, $y$ is trivial
on the three-skeleton of $X$, $c_2(y)=-a$ and $y$ is torsion on the five-skeleton,
but $y$ is not torsion on the six-skeleton.  To completely determine
$\theta$ on the four-skeleton, we need to consider the pairing of
$\theta$ with such classes $y$.  

There is no way to make this comparison using elementary formulas,
since the definition of $a_0$ in section 6 involved a mod 2 index for which there is no explicit formula.  To proceed with branes, we would have to represent
$y$ as the $K$-theory class of a fivebrane, in which case we would
meet the mod 2 index of the worldvolume fermions in six dimensions
and would have to relate this to the $E_8$
mod 2 index considered in section 6.

Instead of proceeding precisely in this fashion,
we will take for our starting point the formula \effomega\ derived above. 
We will  apply this to our problem of deriving the characteristic 
$\theta$ by taking  
$a$ to be a  torsion class that can  be lifted to a $K$-theory class $y=E-F$
(where $F$ is trivial of the same rank as $E$, $c_1(E)=0$, and $c_2(E)=-a$).
  The $E_8$ mod
2 index $f(a)$ is
\eqn\ino{{f(a)}={\int_X Sq^2a_0\cup a}.}
This was the definition \defao\ of $a_0$.
In the present case, $\int c_2(E)c_3(E)=0$ as $c_2(E)$ is torsion.
We also have $\int_X Sq^2a_0\cup a=\int_X a_0\cup Sq^2a =\int_X a_0
\cup c_3(E)$ mod 2 (where we recall that $c_3(E)=Sq^2c_2(E)$ mod 2).
And we can identify $c_3(E)$ as $c_3(y)$.  We have then from \effomega\
\eqn\unupi{\Omega(y)=(-1)^{\half\int_X (\lambda-2a_0)\cup c_3(y)}.}
(Since the exponent is only defined mod 2, we can make choices of signs. These 
are chosen for convenience in comparing  to $M$-theory. ) 

Now, we have shown above that $\theta$ has the property that
$G_{2p}(\theta)=0$ for $p=0,1$.  Given this, it follows
that for $y$ as in the last paragraph (so that in particular
$G_{2p}(y)=0$ for $p=0,1,2$), 
we have
\eqn\kolo{\omega(\theta,y)=-\int_X{G_4(\theta)\over 2\pi}\wedge
{G_6(y)\over 2\pi}= -{1\over 2}\int_X{G_4(\theta)\over 2\pi}\wedge
c_3(y).}
Comparing the last two formulas, we see that to achieve 
$\Omega(y)=(-1)^{\omega(\theta,y)}$ for such $y$'s, we need
the result that was claimed in \irio\ for $G_4(\theta)/2\pi$.

This result has a significance that has already been explained
in the discussion of eqn. \juko. In section 6, we learned 
that the $M$-theory partition function
can be written as a sum over certain equivalence classes.  Once we pick
a solution $a_0$ of $Sq^3a_0=P$, each equivalence class contains
a representative $a=a_0+b$, where $Sq^3b=0$.  The $M$-theory four-form
is 
\eqn\igloo{{G\over 2\pi}=-{\lambda\over 2}+a={-\lambda+2a_0\over 2}+b.}
Here $b$ can be lifted to $K$-theory as an element  $x(b)\in\Gamma_1$.
In Type IIA, $G$ is interpreted as the RR form $G_4$ (with an additional
correction once we turn on $G_2$, as we  do in the next section), and
in view of our result for $\theta$, \igloo\ is equivalent to the
standard Type IIA formula
\eqn\piglo{{G_4\over 2\pi}=\left(\sqrt{\hat A} \,{\rm ch}(\theta/2 +x(b))
\right)_4.}
The $M$-theory sum over $b$ corresponds in Type IIA to the 
sum over the coset of $\Gamma_1$ in $\half\Gamma_1$ that is generated by
$\theta/2$.

\subsec{Comparison Of Phases}

As found in section 6, the phase of the contribution of a given
equivalence class to the $M$-theory partition function is
$(-1)^\alpha(-1)^{f(a_0+b)}$, where $\alpha$ is the Arf invariant
of a certain quadratic function.  By the bilinear relation this is
\eqn\uvu{(-1)^{\alpha+f(a_0)}(-1)^{f(b)+\int a_0\cup Sq^2b}.}
The factor $(-1)^{\alpha+f(a_0)}$ is independent of $a_0$, and, of course,
also independent of $b$.
In the present paper, we will not try to understand the absolute
normalization of the $M$-theory and Type IIA partition functions,
but only the dependence on RR fields.  Up to a constant factor,
the sign of the contribution to the path integral of an equivalence
class with a representative $a=a_0+b$ is
\eqn\umipi{\varphi_M(b)=(-1)^{f(b)+\int a_0\cup Sq^2 b}.}
Using \popsi\ above, if $E$ is an $SU(5)$ bundle with $c_2(E)=-b$, we can
write this as
\eqn\mumipi{\varphi_M(b)=(-1)^{q({\rm ad}(E))+I(\wedge^2E)+
\int a_0 \cup c_3(E)}.}

We want to compare this to the corresponding phase on the Type IIA side.
This is
\eqn\jumipi{\varphi_{IIA}(b)=
\exp(-i\pi{\rm Re}\tau(\theta/2))\exp(i\pi{\rm Re}\tau
(x+\theta/2))\Omega(x),}
where $x=x(b)$ is the $K$-theory class $E-F$, $F$ being a trivial rank five
bundle.  
With the help of \jupi, this becomes
\eqn\tumipi{\varphi_{IIA}(b)=\exp(-i\pi{\rm Re}\tau(\theta/2))\exp(i\pi{\rm Re}\tau
(x+\theta/2))(-1)^{q({\rm ad}(E))+I(E)}.}
Finally, we must evaluate $w={\rm Re}\,\tau(x+\theta/2)-{\rm Re}\,\tau(\theta/2)$.
This is given by 
\eqn\okaydok{
\eqalign{
w& = - \int\biggl(
{G_4(x)\over 2\pi} + \half {G_4(\theta)\over 2\pi}\biggr)
\wedge 
\biggl(
{G_6(x)\over 2\pi} + \half {G_6(\theta)\over 2\pi}\biggr)+ 
{1\over 4} \int {G_4(\theta)\over 2\pi}  \wedge 
 {G_6(\theta)\over 2\pi} \cr
& = - \int
{G_4(x)\over 2\pi}  \wedge 
{G_6(x)\over 2\pi}  -  \int {G_4(\theta)\over 2\pi}  \wedge 
 {G_6(x)\over 2\pi} \cr
& = \half \int c_2(E) c_3(E) + \half \int(\lambda-2 a_0)c_3(E)\cr}
}

With \mumipi\ and \tumipi\ as well as the last formula, we get
\eqn\zumipi{\varphi_M(b)=\varphi_{IIA}(b)(-1)^{I(E)+I(\wedge^2E)
-{1\over 2}\int_X(c_2(E)+\lambda)c_3(E)}=\varphi_{IIA}(b),}
where in the last step, \nopsi\ has been used.

This completes the proof that the $M$-theory sum over $G$-fields
reproduces the Type IIA sum over fluxes of the RR four-form,
whenever the anomalies cancel on both sides.  To complete the picture,
we will now show that the anomaly cancellation condition is also the
same on the two sides.

\subsec{Criterion For Anomaly}

In $M$-theory, we found at several points that the theory is anomalous
unless the spin manifold $X$ has $W_7=0$.
In Type IIA, we have found only one possibility of an anomaly: the theory
is anomalous if the $\Z_2$-valued function $\Omega$ on $K(X)$ is nontrivial
when restricted to torsion classes.  For then, the partition function
vanishes when summed over torsion, and the vanishing cannot be lifted
by any local observable.

So to match the two theories, we hope to show that $\Omega$ vanishes
on torsion classes if and only if $W_7=0$.  The main step is to repeat
the analysis of the class $\theta$ presented in section 7.3 without
assuming that the anomaly cancels.

First, when restricted to classes that are torsion on the
five-skeleton of $X$, $\Omega$ 
 is a homomorphism to $\Z_2$; that is, on such classes,
it obeys
$\Omega(y_1+y_2)=\Omega(y_1)\Omega(y_2)$, as $\omega(y_1,y_2)=0$.
$\Omega$ can be extended, though not canonically, to a homomorphism
$F:K(X)\to \Z_2$.

We must recall Poincar\'e duality in $K$-theory, which asserts that
there is a Pontraygin duality
\eqn\inom{K(X)\times K(X;U(1))\to U(1).}
This means that for any
 $x\in K(X)$, $y\in K(X;U(1))$, there is a $U(1)$-valued pairing
$(x,y)$, linear in each variable, such that any homomorphism
$F:K(X)\to U(1)$ is $x\to (x,f)$ for some $f\in K(X;U(1))$.  Applying
this to our homomorphism $F:K(X)\to \Z_2\subset U(1)$, we conclude
that $F(x)=(x,\bar\theta)$ for some $\bar\theta\in K(X;U(1))$.  Since $F$ maps to $\Z_2$,
$\bar\theta$ can actually be regarded as an element of $K(X;\Z_2)$.

Now from the exact coefficient sequence $0\to \Z\underarrow{2}\Z\underarrow
{r}\Z_2\to 0$ (where the first map is multiplication by 2 and the
second is mod 2 reduction), we get a $K$-theory exact sequence
\eqn\onono{\cdots \to K(X)\underarrow{r} K(X;\Z_2)\underarrow{\delta}
K^1(X)\rightarrow \cdots.}
Here $\delta$ is the ``connecting homomorphism,'' analogous to the
Bockstein map in cohomology. 

For $y$ and $z$ torsion classes in $K(X)$ and $K^1(X)$, one defines
a torsion pairing $T_K(y,z)$ analogous to the torsion pairing
in cohomology that was introduced in section 4.2.  In fact,
$T_K(y,z)=(y,w)$, where $w\in K(X;U(1))$ is such that $\delta(w)=z$.
\foot{$\delta$ is the connecting homomorphism
$\delta:K(X;U(1))\to K^1(X)$ associated with 
 a long exact sequence like \onono\  derived from the coefficient
sequence $0\to \Z\to \R\to U(1)\to 0$.  We really only need the $\Z_2$
case.}
For example, $T_K(y,\delta(\bar\theta))=(y,\bar \theta)$.  
$T_K$ is nondegenerate just like the torsion pairing in cohomology.
Thus, there is a torsion
class $y$ with $T_K(y,\delta(\bar\theta))\not= 0$ if and only if
$\delta(\bar\theta)\not= 0$.  Since $T_K(y,\delta(\bar\theta))
=(y,\bar\theta)=F(y)$, this says that $F(y)$ vanishes
on torsion classes, and thus Type IIA is anomaly-free, if and only if
$\delta(\bar\theta)=0$.

On the other hand, from exactness of \onono, vanishing of $\delta(\bar\theta)$
is precisely the condition for being able to lift $\bar\theta$ to
a class $\theta'\in K(X)$ that reduces to $\bar\theta$ mod 2.
We can calculate the condition for this in another way.

First we make a remark that holds whether $\bar\theta$ can be lifted or  not.
The restriction of $\bar\theta$ to the five-skeleton is completely
determined by the fact that $(y,\bar\theta)=\Omega(y)$ whenever
$y$ is trivial on the five-skeleton.  Proceeding exactly as in the
proof of \hurryio\ and \irio, one can show that $\bar\theta$ is trivial
on the three-skeleton and that the obstruction to trivializing it
on the four-skeleton is the class $w_4$ which is the mod 2 reduction
of $\lambda$.  Existence of the class $\bar\theta\in K(X;\Z_2)$ whose
image in the Atiyah-Hirzebruch spectral sequence for $K(X;\Z_2)$
is $w_4$ means that
the differentials in the AHSS annihilate $w_4$.  The first such
differential is $d_3'=Sq^2 Sq^1 + Sq^1 Sq^2$, regarded as a map on the $\Z_2$ cohomology.
Since $w_4$ is the reduction of an integral class $\lambda$,
we have $d_3'w_4=Sq^1Sq^2 w_4=Sq^3w_4=w_7$.
We conclude that $w_7=0$ for ten-dimensional
spin manifolds.

Now suppose that $\bar\theta$ can be lifted to a class $\theta'\in K(X)$.
Then $\theta'$ is divisible by 2 on the three-skeleton, since $\bar\theta$
vanishes there, and (by adding
to $\theta'$ two copies of a suitable sum of line bundles) 
one can assume that $\theta'$
is trivial on the three-skeleton.  On the four-skeleton, $\theta'$
is measured by a cohomology class that reduces mod 2 to $w_4$.
Any such class is $-\lambda+2a_0$ for some $a_0$.  It follows
that the class $-\lambda+2a_0$, for some $a_0$,
 is annihilated by the differentials in the
Atiyah-Hirzebruch spectral sequence for $K(X)$.  The first such differential
is $Sq^3$, regarded now as a map on the integral cohomology,
so we have $0=Sq^3(\lambda-2a_0)=Sq^3\lambda = W_7$.
Thus, $W_7=0$ precisely when $\bar\theta$ can be lifted to a class
$\theta'\in K(X)$, or in other words precisely when Type IIA is
anomaly-free.

When $\theta'$ exists,
it is in fact precisely the ``characteristic''
 that we have called $\theta$
in defining the Type IIA theta function.

We can now close a gap left open in the discussion of 
\nosolp, and show that $W_7=0$ indeed implies that 
\nojo\ always has solutions. Indeed, suppose 
$c\in \Upsilon$. Then $Sq^3(c)=0$ so $c$ has a $K$-theory 
lift $x(c)$. Moreover, since $c\in \Upsilon$, one may choose the 
class $x(c)$ to be  torsion. In this case, by \effomega, 
$f(c) = j(x(c))$. However, we have just seen that 
when $W_7=0$ we have $j(x(c))=0$. Therefore, $f(c)=0$ and so by 
\nosolp\ $Sq^3(a) = P$ has a solution.

\newsec{Including $G_2$ In $M$-Theory}

\subsec{Evaluation Of The $\eta$ Invariant}

So far we have evaluated the phase of the $M$-theory effective
action, described in section 2 in terms of $E_8$ gauge theory,
only for eleven-manifolds of the form $Y=X\times \S^1$.  Now
we are going to generalize the discussion to consider the case
that $Y$ is an $\S^1$ bundle over $X$.  We assume that the metric
on $Y$ is invariant under rotations of the $\S^1$ fibers, and that the
 $C$-field on $Y$, and hence the $E_8$ bundle,
is pulled back from $X$.  (We will later add to $C$ a topologically
trivial term that is not a pullback.)  Also, we continue 
to assume that the spin
structure on $\S^1$ is supersymmetric (unbounding).

The $\S^1$ bundle $Y\to X$ is the bundle of unit vectors in
a complex line bundle ${\cal L}$.  The basic idea will be to calculate
by Fourier transforming in the $\S^1$ direction. 
Consider functions on $Y$ that transform as $e^{-ik\theta}$ under rotations
of the $\S^1$, for some integer $k$.  
In their $X$-dependence, they can be interpreted
as sections of ${\cal L}^k$.  Thus we have a decomposition
\eqn\jutto{{\rm Fun}(Y)=\oplus_{k\in \Z}\Gamma(X,{\cal L}^k).}
Here ${\rm Fun}(Y)$ is the space of functions on $Y$, and 
$\Gamma(X,{\cal L}^k)$ the space of sections of ${\cal L}^k$.

Consider an $\S^1$-invariant Dirac operator $D_Y$ on $Y$ with real eigenvalues
$\lambda_i$.
The APS function 
\eqn\hobart{\eta(s)=\sum_i|\lambda_i|^{-s}{\rm sign}(\lambda_i),}
where the sum runs over all nonzero $\lambda_i$,
can be written 
\eqn\olpo{\eta(s)=\sum_{k\in \Z}\eta_k(s),} 
where $\eta_k(s)$
is the contribution from states that transform as $e^{-ik\theta}$ under
rotation of the circle.

We write the spin bundle $S$ of $Y$ as $S=\pi^*(S_+)\oplus \pi^*(S_-)$,
where $S_+$ and $S_-$ are the positive and negative chirality spin
bundles of $X$.  Spinors on $Y$ that transform as $e^{-ik\theta}$ under
rotations of the circle are equivalent to spinors on $X$ with values
in ${\cal L}^k$.  Let $R$ be the radius of the $\S^1$,
so the metric in the $\S^1$ direction is $R^2d\theta^2$.
We can pick a basis of eleven-dimensional gamma matrices
such that the Dirac operator reads 
\eqn\ikko{D_Y=\left(\matrix{
{i\over R}{\partial\over \partial \theta} & \bar D\cr
                                      D                  & -{i\over R}{\partial
                                            \over\partial\theta}}\right)
                                                          ,}
where we have written the Dirac equation in $16\times 16$ blocks, and
we have arranged the spinors as a column vector 
\eqn\nikko{\left(\matrix{\psi_+\cr \psi_-\cr}\right),}
with $\psi_\pm $ being sections of $\pi^*(S_\pm)$.
 $D$ and $\bar D$ are the ten-dimensional Dirac operators for
positive and negative chirality.  On spinors that transform as
$e^{-ik\theta}$ under rotations of the circle, the Dirac equation 
$D_Y\psi=\lambda\psi$ becomes
\eqn\bikko{\left(\matrix{{k\over R} & \bar D \cr
                            D        &  -{k\over R}\cr}\right)
\left(\matrix{\psi_+ \cr \psi_-\cr}\right) = 
\lambda\left(\matrix{\psi_+ \cr \psi_-\cr}\right),}
with  $\psi_\pm$ being sections of $S_{\pm}\otimes {\cal L}^k$.

We recall that the phase of the $M$-theory action comes not just
from $\eta$, but from $\eta+h$, where $h$ is the number of zero
eigenvalues.
For $k=0$, we have $\eta=0$ for the same
reason as in section 2.  (To restate the argument in the present
notation, the transformation $(\psi_+,\psi_-)\to (\psi_+,-\psi_-)$
maps $\lambda\to -\lambda$, so the nonzero eigenvalues occur in pairs.)
The phase contribution for $k=0$ therefore comes entirely
from counting the zero eigenvalues.  Since the spinors for $k=0$
are sections of $S_\pm$, regardless of what ${\cal L}$ is, the
contribution to the phase for $k=0$ is independent of ${\cal L}$ and
hence coincides with the phase of the effective action for a product
$X\times \S^1$, as investigated in section 2.

For $k\not=0$, instead, there are no zero eigenvalues, as is clear
from inspection of \ikko, so the contributions will come entirely from
the $\eta$ invariant.  The reason that it is possible to get
a simple answer is that the nonzero eigenvalues of the ten-dimensional
Dirac operator do not contribute even for $k\not= 0$.  Suppose
we have a pair of states $\psi_\pm$, which are sections of
$S_\pm \otimes {\cal L}^k$, with
$D\psi_+=w\psi_-$, $D\psi_-=\bar w\psi_+$ for some complex number $w$.
Then for these two states, the eleven-dimensional Dirac operator becomes
\eqn\plikko{\left(\matrix{ {k\over R} & \bar w \cr
                             w         & -{k\over R}}\right).}
The $\eta(s)$ function of this $2\times 2$ matrix is zero for any complex
number $w$ because the two eigenvalues have the same absolute value
and opposite sign.  So $\eta_k(s)$ for $k\not= 0$ can be computed entirely
from the zero eigenvalues of the ten-dimensional Dirac operator.

Suppose now that $\psi$ is a section of $S_+\otimes {\cal L}^k$ or
$S_-\otimes {\cal L}^k$ that is a zero mode of $D$ or $\bar D$.
We set $\chi(\psi)$ to be $1$ or $-1$ depending on whether $\psi$
has positive or negative chirality. $\psi$ is an eigenstate of the
eleven-dimensional Dirac operator with eigenvalue $k\chi(\psi)/R$.
Its contribution to $\eta_k(s)$ is hence $|k/R|^{-s}{\rm sign}(k\chi)
=|k/R|^{-s}{\rm sign}(k){\rm sign}(\chi)$.  When we sum the
quantity ${\rm sign}(\chi)$ over all zero modes, we get the index
of the ten-dimensional Dirac operator with values in ${\cal L}^k$;
we denote this as $I({\cal L}^k)$.  So we have
\eqn\olkoj{\eta_k(s)=\left|{k\over R}\right|^{-s}{\rm sign}(k)
 I({\cal L}^k).}
The function $\eta(s)$ is obtained by summing this expression over $k$.
In doing so, we can observe that $I({\cal L}^{-k})=-I({\cal L}^k)$.
So we can express $\eta(s)$ as a  sum over positive $k$ only:
\eqn\holkoj{{\eta(s)\over 2}=\sum_{k=1}^\infty\left|{k\over R}\right|^{-s}
I({\cal L}^k).}

Now, the Atiyah-Singer index theorem gives a formula that in ten dimensions
reads
\eqn\bolkoj{ I({\cal L}^k)=\alpha k+\beta k^3+\gamma k^5}
for certain rational numbers $\alpha$, $\beta$, and $\gamma$.
In particular, $I({\cal L}^k)$ is a topological invariant.  Together
with the fact that the factor $|R|^s$ in \holkoj\ will play no role
(as we will see shortly), this means that  $\eta$  will
be a topological invariant.

Using \bolkoj, we have
\eqn\nolkoj{{\eta(s)\over 2} = |R|^{s}\sum_{k=1}^\infty \left(\alpha k^{-(s-1)}
+\beta k^{-(s-3)}+\gamma k^{-(s-5)}\right).}
As expected, the series converges for sufficiently large ${\rm Re}(s)$.
In fact, in terms of the Riemann zeta function $\zeta$, we have 
\eqn\bolkoj{{\eta(s)\over 2}=|R|^{s}\left(\alpha\zeta(s-1)+\beta\zeta(s-3)
+\gamma\zeta(s-5)\right).}
This has the expected analytic continuation to $s=0$.
Since $\zeta(s)$ is regular at $s=-1,-3,-5$, the factor $|R|^{s}$ can
be dropped.  
 Using
the values of $\zeta(-1)$, $\zeta(-3)$, and $\zeta(-5)$, we get
\eqn\polko{{\eta\over 2}
= -{\alpha\over 12}+{\beta\over 120}-{\gamma\over 252}.}

The above argument was presented for the Dirac operator, but it carries
over in an obvious way to the Dirac operator coupled to any vector
bundle $V$ such that the bundle and connection are pulled back from $X$.
Instead of $I({\cal L}^k)$, we get $I(V\otimes {\cal L}^k)$ in the
above formulas.  If $V$ is an $E_8$ bundle with characteristic
class $a$, and if we set $e=c_1({\cal L})$, then we have
\eqn\molmo{I(V\otimes {\cal L}^k)=\int_X\left(248+60a+6a^2+{1\over 3}a^3\right)
\hat A(X) e^{k e}.}
(In dealing with rational or real cohomology classes, we will to keep
the formulas short sometimes omit the cup or wedge product symbol.)
Here, $\hat A(X)$ can be expanded
\eqn\otolmo{\hat A(X)=1+\hat A_4+\hat A_8=1-{\lambda\over 12}+\left(
{7\lambda^2- p_2\over 1440} \right).}
We will find it convenient to express the formulas in terms of $\lambda$
and $\hat A_8$.

The index formula \molmo\
 can be written as $\alpha k+\beta k^3+\gamma k^5$ with
\eqn\curry{\eqalign{ \alpha & = e(6a^2+60a\hat A_4+248\hat A_8) \cr
                      \beta & = {e^3\over 6}\left(60a+248\hat A_4\right)\cr
                     \gamma & = 248{e^5\over 5!}.\cr}}

We also need the corresponding values for the Rarita-Schwinger operator.
As explained in section 2, the Rarita-Schwinger operator
on an eleven-manifold $Y$ is, for our purposes, equivalent to the
Dirac operator coupled to $TY-3\CO$, and for $Y$ a circle bundle
over $X$, it is equivalent to the Dirac operator coupled to 
$TX-2\CO$.  (In string theory terms,  $-2\CO$ is the contribution
of the ghosts plus the dilatino.)
 The appropriate index formula is therefore
\eqn\koluu{I((TX-2{\cal O})\otimes {\cal L}^k)=
\int_X\left(\sum_{i=1}^5 2 \,{\rm cosh}(x_i)-2\right)\hat A(X)e^{ke},}
where $x_i$ are the Chern roots of $TX$, so $\lambda=p_1/2=\sum_i x_i^2/2$
and $p_2=\sum_{i<j}x_i^2x_j^2$.
We can evaluate the index formula as  $\alpha'k+\beta'k^3+\gamma'k^5$, with
\eqn\durry{\eqalign{\alpha' & = e(248\hat A_8-\lambda^2) \cr
                     \beta' & = {2\over 9}\lambda e^3 \cr
                    \gamma' & = 8{e^5\over 5!}.\cr}}
These formulas can be used to evaluate the phase in \olooko.

\subsec{ An Additional Phase}

The RR fields of Type IIA are expressed in terms of a $K$-theory class
$x$ by $G/2\pi =\sqrt{\hat A}\,\ch\,x$.  In comparing to $M$-theory, we will
assume that $G_0=0$ (since it has no known $M$-theory interpretation), and
hence to evaluate $G_2$ and $G_4$, we can set $\hat A$ to 1.  We then get
\eqn\intmo{\eqalign{
{G_0\over 2\pi} & = 0 \cr
{G_2\over 2\pi} & = c_1(x)  \cr
                     {G_4\over 2\pi} & = \half c_1(x)^2-c_2(x).\cr}}
In comparing $M$-theory to Type IIA, we will identify $c_1(x)$
with $e=c_1({\cal L})$, and we will identify $-c_2(x)$ with
the characteristic class $a$ of the $E_8$ bundle over $X$.
But $G_4/2\pi$ has an additional term $\half c_1(x)^2$, and if we
want to match $M$-theory with Type IIA, we need to include in
the $M$-theory description an additional term that will shift
$G/2\pi$ by $\half c_1({\cal L})^2$.

This additional term is topologically trivial, because in fact,
$c_1({\cal L})$, though nontrivial in the cohomology of $X$, pulls
back to zero in the cohomology of the circle bundle $Y\to X$.
Indeed, let $\omega$ be a one-form on $ Y$ that is $\S^1$ invariant
and restricts on each fiber of $Y\to X$ to $ d\theta/2\pi $.  The normalization
is picked so that
\eqn\polu{\int_{\S^1}\omega=1,}
where $\S^1$ is any fiber of $Y\to X$.  Such an 
$\omega$ can be written as $\omega =  (d\theta +A_idx^i)/2\pi$, where $x^i$ are coordinates on $X$ and $A$ is a connection on ${\cal L}$.  We
have $d\omega = F/2\pi$, where $F$ is the curvature of ${\cal L}$; the fact
that $F/2\pi =d\omega$  establishes (at the level of real
cohomology) that $c_1({\cal L})$ vanishes when pulled back
to $Y$.  (More generally,
 $Y$ is the bundle of  unit vectors in ${\cal L}$,
and when pulled back to $Y$, ${\cal L}$ is trivialized tautologically.)

So if we set $C'=\pi\omega\wedge d\omega$, and $G'=dC'$,
then $G'/2\pi 
=\half F\wedge F/(2\pi)^2$.  Adding $C'$ to the $C$-field on $Y$
has the effect, therefore, of shifting $G/2\pi$ by $\half c_1({\cal L})^2$.
This is the shift we want.

Since $C'$ is topologically trivial, the effect of the transformation
$C\to C+C'$ on the phase
of the $M$-theory effective action 
can be worked out from the form of the Chern-Simons coupling in
a completely naive way.  The Chern-Simons coupling is 
\eqn\killo{L_{CS}={1\over 6}\int_Y C\wedge\left(\left({G\over 2\pi}\right)^2
-{1\over 8}\left(p_2(Y)-\lambda^2\right)\right).}
If $C$ is shifted by $C\to C+C'$ with $C'$ topologically trivial,
we can calculate directly that
\eqn\ubillo{\eqalign{L_{CS}\to &L_{CS}+ 
{1\over 2}\int_Y C'\wedge \left(\left({G\over 2\pi}\right)^2
-{1\over 24}\left(p_2(Y)-\lambda^2\right)\right)\cr &
+{1\over 2}\int_Y C'\wedge {dC'\over 2\pi}\wedge {G\over 2\pi}
+{1\over 6}\int_YC'\wedge{dC'\over 2\pi }\wedge {dC'\over 2\pi}.\cr}}
Using $C'=\pi\omega\wedge d\omega$, together with \polu\ and 
the fact that $d\omega$ represents $e=c_1({\cal L})$, we can evaluate
the integral over the fibers of $Y\to X$
and find that the shift in $L_{CS}$ due to $C'$ is
\eqn\nubillo{\Delta L_{CS}=2\pi\int_X\Biggl\{{1\over 4}
e \left((a-\lambda/2)^2
-{1\over 24}(p_2-\lambda^2)\right) +{1\over 8}e^3 (a-\lambda/2)
+{1\over 48}e^5\Biggr\}.}

\bigskip\noindent{\it Aggregate $M$-Theory Phase Factor}

Combining the contributions of the $\eta$ invariants, which give
phase factors according to \olooko, with the phase we have just
found in \nubillo, the phase with which a configuration
with specified $e=c_1({\cal L})$ and characteristic class $a$ of
the $M$-theory four-form contributes to the partition function is
\eqn\oplo{\Omega_M(e,a)=(-1)^{f(a)}\exp\left[2\pi i\int_X \left(
{e^5\over 60} +{e^3 a\over 6}-{11e^3 \lambda\over 144}-{e
a \lambda\over 24}+{e \lambda^2\over 48}-{e \hat A_8\over 2}\right)
\right].}
The exponential factor in \oplo\ is unchanged if $a$ is shifted by a 
torsion class. Therefore, the sum over torsion 
projects to a sum over $a=a_0 + b$ with $Sq^3b=0$, as before. 
We will compare the formidable-looking expression \oplo\ to the Type IIA
theta function in section 9.

\subsec{Parity Symmetry}

The discussion of parity symmetry in section 3.3 can be extended to 
${\bf S}^1$ bundles $Y$ over $X$ as follows.  Parity must now be 
interpreted 
as a reversal of orientation of the ${\bf S}^1$ fiber accompanied by 
$e \rightarrow 
-e$ and $G \rightarrow -G$. 
Combined with
\intmo,
this gives in terms of integral classes 
$a\rightarrow \lambda - e^2 - a$. 
 Therefore, we have to check invariance of the phase \oplo\ 
under $(e,a) \rightarrow (-e, \lambda -e^2 -a)$. 
Using the bilinear identity \jugoxo,
we have 
\eqn\varA{
f(\lambda-e^2)= f(\lambda-e^2-a) + f(a) + \int_X (\lambda - e^2 -a) 
\cup Sq^2 a.}
The expression $\int_X e^2 \cup Sq^2 a$, 
with $e,a$ integral classes, vanishes as a consequence of
\bunvu\ 
and the Cartan formula \bormbu,\ 
taking into account 
the fact that $Sq^1$ annihilates integral classes:
\eqn\varB{
\int_X e^2 \cup Sq^2 a = \int_X  Sq^2 e^2 \cup a = 
\int_X \left((Sq^2 e)\cup e + e\cup (Sq^2 e) \right)\cup a = 0.}
Moreover, Stong's result \noko\ implies that 
\eqn\varC{\int_X (\lambda - a) \cup Sq^2 a=0.}
Therefore, the last term in \varA\ vanishes. 
Repeating these steps for $f(\lambda -e^2)$, we find 
\eqn\varD{
f(\lambda) = f(\lambda - e^2 - a) + f(e^2) + f(a).}

The variation of the additional phase factor in 
$\Omega_M(e,a)$, written in \oplo, can be evaluated 
by direct computation.  Upon doing so and using \varB, we find 
that $\Omega_M(e,a)$  transforms under parity by
\eqn\varE{
\Omega_M(-e,\lambda -e^2-a) = (-1)^{f(\lambda)+f(e^2)}\hbox{exp}\left[ 2\pi i
\int_X
\left({2e^5\over 15} -{\lambda e^3\over 18} + e{\hat A}_8\right)
\right]\Omega_M(e,a).}
The phase factor written as an exponential in \varE\ 
is in fact half the index density 
of the Dirac operator on $X$ coupled to the $K$-theory class 
${\cal L}^2 - {\one}$, 
where ${\one}$ denotes
a trivial complex line bundle and $c_1({\cal L})=e$. 
Therefore we can rewrite \varE\ as
\eqn\varF{
\Omega_M(-e,\lambda -e^2-a) = 
(-1)^{f(\lambda)+f(e^2)+I\left({\cal L}^2 - {\one}\right)}
\Omega_M(e,a).}
We will prove below, as part of a more general formula, that 
\eqn\indA{
f(e^2) = I\left({\cal L}^2 - 
{\one}\right)}
for all integral two-classes $e$. 
Therefore \varF\ reduces to 
\eqn\varH{
\Omega_M(-e,\lambda -e^2-a) =
(-1)^{f(\lambda)}\Omega_M(e,a),}
which is the same as the result found in section 3.3 for trivial
circle bundles.  Thus inclusion of $e$ does not modify the analysis
of anomaly cancellation in section 3.3.

\bigskip\noindent{\it An Elementary Formula For The Mod Two Index Of A
Product}

The identity \indA\ needed above is part of a more general 
general formula expressing the mod two index of an $E_8$ bundle 
with characteristic class $a = u\cup v$ in terms of 
elementary invariants.
Here $u,v$ are integral two-classes in $H^2(X;{\bf Z})$. 

Such a formula can be derived by constructing $E_8$ bundles using the 
embedding $SU(3)\subset E_8$, in analogy with the proof of the 
bilinear identity in section 3.1. 
Let ${\cal L}$, ${\cal M}$ be complex line bundles with 
$c_1({\cal L})=u$, $c_1({\cal M})=v$. We first construct the 
$SU(3)$ bundle 
\eqn\indB{
W = {\cal L} \oplus {\cal M} \oplus 
{\overline{\cal L}}\otimes {\overline{\cal M}}.}
A direct computation shows that 
\eqn\indC{
c_2(W) = -(u^2+v^2+u\cup v).}
Therefore, by embedding $SU(3)$ in $E_8$ (using the chain
$SU(3)\subset E_6\times SU(3)\subset E_8$) we obtain an $E_8$ 
bundle with characteristic class $a=u^2+v^2+u\cup v$. 

The decomposition of the Lie algebra of $E_8$ in terms of 
representations of $SU(3)\times E_6$ is 
\eqn\indD{
{\bf 248}=({\bf 8}, {\bf 1}) \oplus ({\bf 1}, {\bf 78}) 
\oplus ({\bf 3}, {\bf \overline{27}}) \oplus ({\bf \overline{3}}, 
{\bf 27}).}
The mod two index of the $E_8$ bundle constructed 
above is the same as the mod two index with values in the ${\bf 8}\oplus
{\bf 3}\oplus \bar {\bf 3}$ of $SU(3)$.  This can be evaluated
using the fact that the mod 2 index with values in ${\cal S}\oplus \bar
{\cal S}$ (for any ${\cal S}$) is the mod 2 reduction of the ordinary
index with values in ${\cal S}$.  We get
\eqn\indF{\eqalign{
f(u^2+v^2+u\cup v)=& I\left({\cal L}^2\otimes {\cal M}\oplus{\cal L}\otimes 
{\cal M}^2\right) + 
I\left({\cal L}\otimes{\overline{\cal M}} \oplus{\cal L}\otimes 
{\cal M}\right)\cr
& +I\left({\cal L}\oplus {\cal M}\right)\qquad \hbox{mod}\ 2.\cr}}

As an ordinary index, the right hand side of equation \indF\ can be 
expressed in terms of elementary invariants. Setting
${\cal M}={\one}$, and working mod two, we obtain 
\eqn\indG{
f(u^2) = I\left({\cal L}^2 -{\one}\right)\qquad \hbox{mod}\ 2.}
This is the formula \indA\ needed above. 

We record here a more general identity
which is easily obtained from \indF\ and \indG\  using
the bilinear identity for $f$ and the index theorem.
Applying twice the bilinear identity, and taking into account 
\varB,\ we have 
\eqn\indH{
f(u^2+v^2+u\cup v)=f(u^2) + f(v^2) + f(u\cup v).}
Combining \indF\ -- \indH,\ we arrive at 
\eqn\indJ{\eqalign{
f(u\cup v) = & I\left(({\cal L} -{\one})
\otimes({\cal M}^2 -{\one})\right)
+I\left(({\cal L}^2-{\one})\otimes
({\cal M}-{\one})\right)\cr
&+ I\left({\cal L}\otimes{\overline{\cal M}} \oplus{\cal L}\otimes 
{\cal M}\right)\qquad \hbox{mod}\ 2.\cr}}
The right hand side of \indJ\ can be evaluated using the index 
theorem, obtaining 
\eqn\fofuv{
f(uv) = \int \biggl[ u v (u+v) \left(uv - {1\over 4}\lambda\right) 
+{3\over 4}uv(u^3+v^3)
+ {1\over 12}\bigl( u v^4 + 2 u^3 v^2 - \lambda u v^2\bigr) \biggr] 
\quad\mod 2
}
The right hand side of this formula is symmetric under 
exchanging $u$ and $v$ since on a spin 10-manifold 
$I(\left({\cal L}\otimes{\overline{\cal M}}\right) = I\left(
{\overline {\cal L}} \otimes {\cal M}\right)$ mod 2.

\newsec{Generalized Comparison To Type IIA}

In this section, we show how the computation of 
section 8 is reproduced in the IIA theory. Our aim is 
to obtain the nontrivial phase \oplo\ using the 
$K$-theory formalism. The real part of the action will 
match between $M$-theory and IIA theory simply because 
the dimensional reduction of 11 dimensional supergravity 
is the IIA supergravity. 

The relation between the $M$-theory geometry and the 
RR fields has already been explained in section 8.2. 
A $K$-theory class $x$ satisfying \intmo\ corresponds to 
$M$-theory on a circle bundle $Y \rightarrow X$, where 
the Euler class of the circle bundle is $c_1(x)$.  Moreover, 
$G_4$ is pulled back to $Y$ to determine the $M$-theoretic 
$G$. We therefore must compute 
the contribution \xono\ for such $x$, and compare to 
\oplo. 

\def\L{{\cal L}}
 At $G_2=0$, we have found it necessary to compare
an $E_8$ bundle associated with $M$-theory to an 
$SU(5)$ bundle derived from $K$-theory. As in section
7, we write the characteristic class $a$ of the $E_8$ bundle
as $a=a_0+b$, where $a_0$ was defined in section 6 and $b$ has
a $K$-theory lift.  
By the relation 
\intmo, we see that we must choose our $K$-theory class to 
be represented by 
\eqn\newex{
x = E + \L - 6\one + \Delta 
}
Here $E$ is the $SU(5)$ bundle 
used  in section 7.5, with $c_2(E) = -b$. Also, $\L$ is a line 
bundle on $X$ with $c_1(\L) = e$; a rank six trivial bundle
$6\CO$ has been subtracted to ensure $G_0=0$. Finally, 
 $\Delta $ is a class 
in $\Gamma_2$ chosen so that $x\in \Gamma_1$. One can 
check, as in section 7, that the choice of $\Gamma_1$, 
i.e. the choice of $\Delta $, will 
not contribute to the phase $Z_x$, so we can ignore $\Delta$. 
With this understanding, we can 
write
\eqn\xxno{ 
x = x_0 + (\L- \one ) 
}
where $x_0=E-5\CO$ is the class used in section 7.

The evaluation of the contribution $Z_x$ to the partition function
requires evaluation 
of $\Omega(x)$ and $\Re(\tau(x))$. Let us consider first 
$\Omega(x)$. Using the bilinear identity we have 
\eqn\firsome{
\Omega(x) = \Omega(x_0) \Omega(\L-\one) e^{-i \pi I(x_0 \otimes (\bar \L - \one))}.
}
Now, $(\L - \one) \otimes (\bar \L - \one) = 2 - (\L + \bar \L) $. Therefore 
$\Omega(\L-\one)$ is elementary; it equals
the reduction modulo two of the ordinary index $I(\L)$. So
we have
\eqn\firsomeii{
\Omega(x) = \Omega(x_0)  e^{-i \pi [I(x_0 \otimes (\bar \L - \one))+I(\L)]}.
}
Now we can use the index theorem. Substituting 
\eqn\cherncl{
\ch x_0 = b + \half c_3(E) + \biggl( {b^2 - 2 c_4(E) \over 12}\biggr) + 
{ c_5(E) \over 24} 
}
and applying the result \effomega\ to evaluate $\Omega(x_0)$, we conclude 
\eqn\totaliia{
\eqalign{
\Omega(x) = 
\exp\Biggl[i \pi \biggl( f(b) & + \half (\lambda-b)c_3\cr
-\bigl[ {1\over 6} c_4 e &+ {1\over 4} c_3 e^2 
- {1\over 6}b e^3 -{1\over 12} e b^2 
+{1\over 12} e b \lambda\bigr] \cr 
-\bigl[ {e^5\over 5!} & - {\lambda e^3 \over 72} + e \hat A_8 \bigr] 
\biggr) \Biggr] \cr}
}
Here $c_4=c_4(E)$, $c_5=c_5(E)$.

Let us now turn to the contribution of $\Re(\tau(x))$. This is
a straightforward application of the general formula \hovopy:
\eqn\realpart{
\hbox{Re}\tau\left(x+{1\over 2}\theta\right)={1\over (2\pi)^2} \int
\left(G_4 G_6 - G_2 G_8\right). 
} 
In terms of Chern classes we have
\eqn\chenrexp{
\eqalign{
{G_2\over 2 \pi} & = \ch_1\left(x+{1\over 2}\theta\right)\cr
{G_4\over 2 \pi} & = \ch_2\left(x+{1\over 2}\theta\right)\cr
{G_6\over 2 \pi} & = \ch_3\left(x+{1\over 2}\theta\right)-{\lambda\over 24}
 \ch_1\left(x+{1\over 2}\theta\right) \cr
{G_8\over 2 \pi} & = \ch_4\left(x+{1\over 2}\theta\right)-{\lambda\over 24} 
 \ch_2\left(x+{1\over 2}\theta\right)\cr}
}
The contributions (proportional to $\lambda$)
from $\hat A$ cancel, and hence we get
\eqn\IIAphaseC{\eqalign{
\hbox{Re}\tau\left(x+{1\over 2}\theta\right)=
 & \int
(\ch_2(x) +\half \ch_2(\theta)) (\ch_3(x) +\half \ch_3(\theta)) \cr
&- (\ch_1(x)+\half \ch_1(\theta))(\ch_4(x) +\half\ch_4(\theta)).  \cr}
}
Now we expand the expression \IIAphaseC. 
We get three kinds of terms.
The quadratic piece in the Chern classes of $x$ is
\eqn\quadpce{\int\left(
\ch_2(x)\ch_3(x)   - \ch_1(x) \ch_4(x)\right). 
}
We can simplify the cross terms in \IIAphaseC\ using the orthogonality
relations since $\theta$ and $x$ are both in the Lagrangian lattice
$\Gamma_1$.
We use orthogonality to eliminate $\ch_3(\theta)$
and $\ch_4(\theta)$ and get the cross terms:
\eqn\crosst{\int
\left(\ch_2(\theta) \ch_3(x) - \ch_1(\theta) \ch_4(x) -
{\lambda \over 24} \bigl( \ch_1(x) \ch_2(\theta) - \ch_1(\theta) \ch_2(x)
\bigr)\right).
}
Finally there are the terms quadratic in $\theta$:
\eqn\quadthta{
{1\over 4}\left(\ch_2(\theta)\ch_3(\theta)-
\ch_1(\theta)\ch_4(\theta)\right).}

Now we write out these expressions in terms of 
Chern {\it classes} (as opposed to characters). 
The quadratic piece in the Chern classes of $x$ is 
\eqn\quadpce{
\ch_2(x)\ch_3(x)   - \ch_1(x) \ch_4(x) = 
{1\over 6} c_4 e +{1\over 4} c_3 e^2 + \half c_3 b  -{1\over 12} e b^2
+{1\over 6} b e^3 + {1\over 24} e^5
}
Here $c_i$ is an abbreviation for  $c_i(x_0)=c_i(E)$ (not the Chern
classes of $x$).

We now simplify the cross terms 
 using  $\ch_2(\theta) = - \lambda + 2 a_0 $ to get
\eqn\mxdpce{
-\lambda (\half c_3 + {1\over 6} e^3 )+ {\lambda^2 \over 24} e + a_0 \biggl(c_3 + {e^3 \over 3} - {\lambda e \over 12}
\biggr) . 
}
The piece quadratic in the Chern classes of $\theta$ cancels the first 
factor in \xono.

Combining \totaliia, \quadpce\ and \mxdpce\  and using 
$b=a-a_0$  we find 
\eqn\fnliiaphse{
Z_x = 
\exp\biggl[i \pi \biggl( 
f(b)+ \int a_0 c_3 \biggr)\biggr] \cdot
\exp\left[2\pi i\int_X \left(
{e^5\over 60} +{e^3 a\over 6}-{11e^3 \lambda\over 144}-{e
a \lambda\over 24}+{e \lambda^2\over 48}-{e \hat A_8\over 2}\right)
\right]
}
Finally, we use the bilinear identity to conclude that 
\eqn\signagain{
(-1)^{f(a)} = (-1)^{f(a_0)} e^{i \pi (f(b) + \int a_0 c_3)}
}
The sign $(-1)^{f(a_0)}$ is part of the overall manifold-dependent 
normalization which we are not trying to match (see, e.g. \uvu). 
Apart from this, comparison of \fnliiaphse\ with \oplo\ yields
 a perfect match. This completes the comparison of the 
$K$-theory and $M$-theory phase factors for nonzero values of $G_2$.


\newsec{Completing the Type IIA Theta Function} 

In this section,  we will extend the computation of sections 7 and 9 above 
to include the effects of nonzero $G_0$, thus completing the 
formula for the full Type IIA theta function. 
On the one hand, the effects of $G_0$ are the least important 
in the large volume limit, being of order 
$\exp[-G_0^2 V]$ where $V$ is the volume of $X$ in 
string units. 
On the other hand, 
while most of this paper has focused on the interesting subtleties related to 
$H^4(X;\Z)$, it is worth noting that the contribution from $G_0$ is the only
nontrivial 
contribution to the theta function for such basic manifolds as $X= {\bf S}^{10}$ and 
$X= {\bf S}^5 \times {\bf S}^5$. 
Moreover, as we leave the geometrical realm and 
make the volume smaller, these are the {\it most} important terms. 
We comment on the relation to $M$-theory at the 
end of this section. 

First, let us construct the full maximal Lagrangian 
  lattice $\Gamma_1\subset K(X)$. As we have seen, 
for all $c_1\in H^2(X;\Z)$ and $c_2\in H^4(X;\Z)$ with 
$Sq^3 c_2 = 0$, there is a $K$-theory lift in $\Gamma_1$, 
that is, there is a $K$-theory class  $x(c_1, c_2) \in \Gamma_1$ 
with 
\eqn\bdleot{
\ch(x(c_1, c_2)) = c_1 + (-c_2 + \half c_1^2) + \cdots 
}
where the higher Chern classes are such that $x$ is in a 
Lagrangian lattice.
Now, we may choose the $K$-theory lifts $x(c_1,c_2)$ such that, for all
$c_1,c_2$, the index $I(x)$ of the Dirac operator with values in
$x$ is zero.  This is possible because on any ten-manifold $X$, there
exists a $K$-theory class $y$, trivial except in a small neighborhood of a 
point in $X$, with index 1.  The Chern classes of $y$ vanish except
for ${1\over 4!} c_5(y)=1$.  By adding to $x$ a multiple of $y$, 
one can pick the
$K$-theory lifts so that $I(x)=0$ for all $x$. Similarly 
we can take $I(\theta)=0$. 
Once this is done, one can
define a complete Lagrangian
lattice $\Gamma_1$ that consists of $K$-theory classses
of the form $z = m \one + x(c_1, c_2)$ where $m\in {\bf Z}$ and 
$\one$ is a trivial complex line bundle.

We have already computed the contribution
of $x(c_1,c_2)$ to the partition function
in sections 7 and 9. 
Let us see what changes by including $m\one$. 
Now we have: 

\eqn\newgee{
\eqalign{ 
{G(z+\half \theta)\over 2\pi}& = m \sqrt{\hat A} + {G(x+\half\theta)\over 2\pi} \cr}
}

Moreover, using 
\eqn\zeroindx{
0 = I(x) = \int {G(x)\over 2\pi} \sqrt{\hat A}  
}
we get 
\eqn\newretau{
\eqalign{
\biggl(G_0 G_{10}
- G_2 G_8 + G_4 G_6\biggr)\biggr\vert_{z+\half \theta} & = 
\biggl( G_4 G_6 - G_2 G_8\biggr)\biggr\vert_{x+\half\theta}\cr
&  - 4 \pi m \bigl(\sqrt{\hat A}\bigr)_8 G_2(x) \cr}
}
Similarly, by the cocycle formula and the fact that $\omega({\one},x)=I(x)=0$,
we have
\eqn\newomeg{
\Omega_{IIA}(m \one + x) = (\Omega_{IIA}(\one))^m \Omega_{IIA}(x) . 
}

Note that, if $\Omega_{IIA}(\one) = -1$, then the dilatino has a zero mode, 
and the partition  function vanishes. 
Even when this occurs, the total number of fermion zero modes
is still even (because Type IIA has fermions coming from both left- and
right-movers on the world-sheet), and 
 by insertion of a local 
operator, we can obtain nonzero and sensible correlation functions.

The most interesting change from the previous sections  is in the kinetic energy, which now reads:
\eqn\newimtau{
{G\over 2\pi}\cdot {\rm Im} \tau \cdot {G\over 2\pi} = m^2 V 
+ \left\vert {G_2(x)\over 2\pi}\right\vert^2 + 
\left\vert {G_4(x+\half\theta)\over 2\pi} - m{\lambda\over 24}\right\vert^2 
}
where $V$ is the volume of $X_{10}$ in the string metric in 
string units. 

Thus, assuming $\Omega_{IIA}=1$ on $K^0(X)_{tors}$, the full 
IIA theta function becomes 
\eqn\newiiasum{
\eqalign{
\Theta_{IIA}= \sum_{c_1, c_2: Sq^3 c_2=0} & w(c_1,c_2) e^{-i \pi \tau \theta(c_1,c_2)^2- 2 \pi i \theta(c_1,c_2)\varphi(c_1)} 
\vartheta\biggl[\matrix{
\theta(c_1,c_2)\cr 
\varphi(c_1)\cr}\biggr](0\vert \tau) \cr}
}
where $w(c_1,c_2)$ is the weighting factor computed previously
(see equations \hovopy\  and \fnliiaphse\ 
 above). The effect of the sum over 
$G_0$ is to change the weighting factor to  
 an {\it elliptic function}, namely a theta function with 
\eqn\tauform{
\tau = i \biggl( V + \vert {\lambda\over 24} \vert^2\biggr) 
}
and characteristics
\eqn\chards{
\eqalign{
\theta(c_1,c_2) & = -{1\over 2\pi} {\int G_4(x+\half \theta) \wedge * {\lambda\over 24}  \over 
V + \vert {\lambda\over 24} \vert^2} \cr
\varphi(c_1) & = - \int (\sqrt{\hat A})_8 c_1 +\varphi_0\cr} 
}
where $\Omega(\one)= \exp[2\pi i \varphi_0] $.

\subsec{No Comparison to $M$-Theory} 

Unlike the results of sections 7 and 9, we cannot, 
unfortunately, make a comparison with $M$-theory. 
The reason is that 
the sectors with nonzero $G_0$ correspond to sectors of 
Type IIA supergravity with nonzero Romans mass
\nref\romans{L.J. Romans, ``Massive N=2a supergravity in 
ten dimensions,'' Phys. Lett. {\bf 169B} (1986) 374.}%
\nref\polwitt{J. Polchinski and E. Witten, 
``Evidence for Heterotic-Type I String Duality,'' 
Nucl. Phys. {\bf B460} (1996) 525; hep-th/9510169.}%
\refs{\romans,\polwitt}.
There is no accepted $M$-theoretic background corresponding 
to such IIA backgrounds. Nevertheless, the form of the 
above answer is somewhat suggestive, so we offer one
speculation. 

The  appearance of elliptic functions 
is suggestive of $F$-theory and  $M$-theory geometries involving 
torus bundles. Indeed, for certain manifolds, $\eta$ invariants 
are closely related to $L$-functions and modular functions \ads. 
Essentially, this arises because  $\eta(s)$ is a generalized Dirichlet series
and can be evaluated via a generalization of the Kronecker limit formula. 

A relation between $M$-theory on torus bundles and 
massive IIA string theory has in fact been suggested by 
Hull (in some special backgrounds) \hull. One way 
of interpreting Hull's result is in terms of T-duality 
which can relate IIA theory on $T^2 \times X_8$ with 
$G_0/2\pi = m$, $G_2=0$ to IIA theory on a dual torus 
with $G_0=0$ and $G_2/(2\pi) = m e_0$, where $e_0$ generates 
$H^2(T^2;\Z)$. 
\foot{It is an interesting and not entirely trivial 
exercise to demonstrate explicitly the $SO(2,2;\Z)$ 
T-duality invariance of the $K$-theory theta function 
on manifolds of the form $T^2 \times X_8$.} 
The latter geometry (for a large dual torus) has an 
M-theoretic interpretation which we have analyzed. 
Using the above results one can check that the actions 
of T-dual geometries do not agree, but appropriate 
sums of such actions do agree. In this sense we can 
confirm the suggestion of  \hull.


\newsec{Some Remarks About The $H$-Field And A Puzzle}

Though the present paper primarily focuses on the case that  the Neveu-Schwarz
three-form field $H$ (and in fact the NS potential $B$) is zero,
we will here make a few simple observations about what happens
when it is included.  

First of all, in supergravity, as explained in \ght,
 the equations of motion for the
RR fields $G_n$ can be put in the form
\eqn\impolo{dG_{n+2}= H \wedge G_n}
for all $n$.  At the level of cohomology, this implies simply that
\eqn\nimpolo{H\wedge G_n=0.}

Let us compare this to what we might expect for $D$-branes.
We simply repeat the reasoning of section 5.1, but now with $H\not= 0$.
In the presence of an $H$-field, $D$-brane charge takes values in
a twisted $K$-group $K_H$ (which can be defined straightforwardly
\uwitten\ when $H$ is torsion and less straightforwardly
\refs{\bouwknegt,\as} when it is not).  What $D$-branes represent
classes in $K_H$?
In the presence of an $H$-field, a $D$-brane can be wrapped on
a cycle $Q$ if and only if \fw\
\eqn\tilpo{H|_Q+W_3(N)=0.}
Here $H|_Q$ is the restriction of $H$ to $Q$, and $N$ is the normal
bundle to $Q$.  In terms of a cohomology class $b$ that is Poincar\'e dual
to $Q$, this equation amounts to
\eqn\uniplo{(H+Sq^3)b=0.}
This is the condition, in the presence of the $H$-field,
for $b$ to represent the lowest nonvanishing
$p$-form charge of a $D$-brane.  Of course, $b$ is of odd or even
degree for Type IIA or Type IIB and so represents an element of
$K_H^1$ or $K_H$, respectively.  We interpret the operator $H+Sq^3$ that
appears in \uniplo\ as the first differential $d_3$ of the
Atiyah-Hirzebruch spectral sequence for $K_H$; it reduces at $H=0$
to the familiar $Sq^3$.

If  RR charges have a $K$-theory interpretation, then the fields that
the charges create must also have a $K$-theory interpretation.
Using arguments along the lines of those in section
2 of \selfduality, we may expect the RR forms $G_n$ to themselves
be elements of $K_H$ -- more exactly, elements of $K_H$ or $K_H^1$ for Type IIA
or Type IIB.  Since we have identified the first AHSS differential
as $H+Sq^3$, it follows that in any physical situation in which the
$G_p$ vanish for $p<n$, we should have at the level of cohomology
\eqn\umoco{(H+Sq^3)G_n=0.}
We interpret this as the supergravity equation \nimpolo\ with a torsion
correction $Sq^3$.  (If $G_n\not= 0$, then for $m>n$, the conditions on
$G_m$ are more complicated in general 
and will involve higher order effects in the
AHSS.)

Clearly, it would be desireable to generalize the computation
in sections 7 and 9 and show that the $M$-theory
partition function on a circle bundle over $X$, with $H\not= 0$,
can be expressed in terms of RR fields obeying \umoco.
This will not be demonstrated in the present paper, although 
we have performed several computations which do in fact support
this hypothesis. In particular, for $Y=X \times {\bf S}^1$ and 
$H$ 2-torsion we have generalized the computation of section 7. 

\bigskip\noindent{\it A Puzzle}

We will now point out a puzzle that this picture raises involving
$S$-duality or $SL(2,\Z)$ symmetry for Type IIB.
First of all, to have $SL(2,\Z)$ symmetry, we must assume that
the monodromies of the Type IIB $\tau$ parameter are trivial.
This forces us to assume that the cohomology class $G_1$ (which
determines the monodromies of ${\rm Re}\,\tau$) vanishes.
This being so, the lowest RR form that may be topologically non-trivial
is $G_3$, which we will call simply $G$.  The equation \umoco\ hence
implies that at the level of cohomology we should have
\eqn\numoco{(H+Sq^3)G=0.}

Now, at least modulo torsion, the pair $\left(\matrix{G \cr H\cr}\right)$
is expected to
transform in the two-dimensional representation of $SL(2,\Z)$.
The transformation $G\to G+H$, $H\to H$ corresponds to $\tau\to\tau+1$
and is visible in string perturbation theory.  Happily, \numoco\ is
invariant under this transformation (since $H\cup H = Sq^3H$).
The problem arises because \numoco\ does {\it not} have full $SL(2,\Z)$
symmetry; it is not invariant under $G\to G$, $H\to H+G$.
\numoco\ has  an $SL(2,\Z)$-invariant extension, namely
\eqn\tmoco{H\cup G +Sq^3(G+H)=0.}
(More generally, one should allow for a $G,H$-independent 
constant on the right hand side of \tmoco\ analogous 
to $P$ introduced in section 6.2.) 
Unfortunately, it is hard to see a rationale for the $Sq^3H$ term in this equation.

The root of the problem is that the weak coupling description
in which the RR fields are classified by $K^1_H$ breaks
the symmetry between $G$ and $H$ by treating $H$ as an
``ordinary three-form field,'' while $G$ is a more subtle object
related to $K$-theory.

We do not know where the resolution of this problem may lie.
We can see at least two broad approaches to resolving the problem:

(1) Perhaps \tmoco\ is correct.  In support of this hypothesis,
we note that the sort of arguments given in \selfduality\ only
show that RR fields can be classified by $K_H(X)$ (or by $K(X)$ if
$H=0$) modulo an additive constant.  Roughly, the arguments in section
2 of 
\selfduality\
show that the RR fields created by a $D$-brane
are classified by $K_H$, but there may be a ``background field'' or
integration constant, not created by the $D$-branes and not obeying
\umoco. Thus, \umoco\ would be replaced by $(H+Sq^3)G_n=Q$,
for some class $Q$ that should be independent of $G_n$.
(In sections 7 and 9, we found a superficially similar shift by $P$,
which was interpreted in terms of the ``characteristic'' of the theta
function; it does not seem that the $H\cup H$ term has such an interpretation.)
If, for Type IIB with $n=3$, one has
$Q=H\cup H$, then we would arrive at \tmoco.
Moreover, in the special case of $M$-theory on backgrounds of 
the form $\T^2 \times X_9$ and $H$ 2-torsion we have in fact 
derived \tmoco\ from the $M$-theory phase. Unfortunately, in 
the general case
we have not been able to turn this idea into a coherent proposal,
or to find convincing support for it.

(2) Alternatively, perhaps $SL(2,\Z)$ invariance of the theory
does not come from a simple transformation law on the space of classical
fields.  On a compact ten-manifold $X$, the partition function
in the large volume limit 
is hopefully $SL(2,\Z)$-invariant.
(We take large volume on $X$ to reduce
to a situation in which supergravity, perhaps with some corrections
such as $Sq^3$, should be valid, and equations such as \umoco\ make sense.)
This invariance need not come from an $SL(2,\Z)$ action on the classical
fields.  For example, in comparing $M$-theory to Type IIA, we did not
find a simple matching between $M$-theory configurations and Type IIA
configurations; we have had to identify the contribution
of an equivalence class of $M$-theory fields (described in section 6)
with the contribution of an equivalence class of Type IIA configurations
(classified by an element of $K(X)/K(X)_{tors}$).  However,
we have not been able to find a convincing scenario for $SL(2,\Z)$
symmetry of the partition function (or of the Hilbert space in a Hamiltonian
description) without an $SL(2,\Z)$ action on the configuration space.

\newsec{Conclusions and Questions } 

In summary, let us recapitulate some of the main lessons we 
have learned from the above considerations, and raise some
questions.

One key point is that extremely subtle phases that are not generated
by any conventional supergravity Lagrangian
are essential in a careful comparison 
of Type IIA superstrings and $M$-theory.  There are descriptions
of these phases via gauge theory -- $U(N)$ gauge theory and $E_8$ gauge
theory for Type IIA and $M$-theory, respectively. By reconciling the Type IIA
and $M$-theory formulas, we have gained considerable confidence that
both are correct.

We have developed a more complete understanding of the conditions
on allowed RR fluxes coming from the $K$-theory interpretation of
RR fields.  As a byproduct, we have learned that certain apparently
stable $D$-brane configurations are actually unstable.

Is there a physical mechanism that would naturally generate the phases
required in Type IIA and in $M$-theory?
Is the use of $E_8$ gauge fields to describe $M$-theory $C$-fields
merely a technical device, or is there an underlying physical meaning
to this?  (See \horava\ for some speculations related to this question.)
Likewise, what is the deeper meaning of the $U(N)$ gauge fields
that are implicit in the $K$-theory description of RR fields?

\bigskip
\centerline{\bf Acknowledgments} 

We would like to thank M. F. Atiyah, D. Freed, M. J. Hopkins, C. Hull, 
P. Landweber, J. Morgan I. M. Singer, and G. Segal for discussions and 
explanations.
The work of GM is supported by DOE grant DE-FG02-96ER40949. 
The work of EW has been supported in part by NSF Grant
PHY-9513835 and the Caltech Discovery Fund.
DED would also like to thank D. Christensen, K. Dasgupta, J. Gomis, 
C. Rezk and especially L. Nicolaescu and S. Stolz for useful 
discussions, and Paul Taylor for the diagrams.tex package.
The work of DED has been supported by DOE grant DE-FG02-90ER40542.

\bigskip

\appendix{A}{Notation} 

 In this paper,
cohomology classes with integer coefficients or unspecified
coefficients will generally be labeled as $a,b,c$.
The symbols $\bar a,\bar b,\bar c$ will denote cohomology classes with
$\Z_2$ coefficients.    $K$-theory classes will be denoted
as $x,y,z$, and complex vector bundles as $E,F$.  The $K$-theory
class determined by a pair of bundles $E$ and $F$ will be written
as $(E,F)$ or $E-F$.

A list of selected notation used throughout the paper is:

\bigskip
$a,a',b,b',\dots$ \qquad\qquad  Generic elements of $H^4(X,Z)$. 
\medskip
$A$ \qquad\qquad\qquad   $Sq^3(H^4(X,Z))$.  Also, a gauge field. 
\medskip
$B$ \qquad\qquad  $Sq^3(H^4(X,Z)_{tors})$ 
\medskip

$c$ \qquad\qquad  A generic torsion cohomology class
\medskip
$C$ \qquad\qquad  The $M$-theory 3-field potential. 
\medskip
$e$ \qquad\qquad  The first Chern class of a circle bundle (secs. 8 and 9)
\medskip

$G$ \qquad\qquad  The $M$-theory 4-form field-strength. Also the total IIA RR field-strength. 
\medskip

$G_{2p}$ \qquad\qquad  The $2p$-form RR field-strength in IIA 
\medskip

$\Gamma$ \qquad\qquad  $K(X)/K(X)_{tors}$
\medskip

$I(\CE ;X)$ \qquad\qquad  Index of an elliptic  operator $\CE$ on $X$
\medskip

$I(x),I(x;X)$ \qquad\qquad  Index of the Dirac operator 
coupled to $x\in K(X)$.  
\medskip
$j(x)$ \qquad\qquad  Mod two index of Dirac coupled to $x\otimes \bar x$.  
\medskip

%
%

$L$ \qquad\qquad  \quad  $H^4_{tors}/2H^4_{tors} $ 
\medskip
$M$ \qquad\qquad 2-torsion subgroup of $H^7_{tors}(X,Z)$. 
\medskip
$\CO$ \qquad\qquad  A trivial  line bundle (real or complex depending
on the context).
\medskip
$P$ \qquad\qquad  A   cohomology class in $H^7(X,Z)/Sq^3(H^4_{tors})$ 
defined in  \buvu. 
\medskip
$q(x;X),q(x)$ \qquad  
The mod-two Dirac index with values in   $x\in KO(X)$. 
\medskip
$S$ \qquad\qquad    $H^4(X,Z)/H^4_{tors}(X,Z)$ 
\medskip
$S'$ \qquad\qquad    $\{ a\in H^4(X,Z)/H^4_{tors}(X,Z): Sq^3a \in Sq^3 H^4(X,Z)_{tors}\}$ 
\medskip
$T$ \qquad\qquad  $H^6/H^6_{tors}$, the torsion pairing  (or in
section 5.2 the tachyon field). 
\medskip
$\Upsilon_0$ \qquad\qquad  The kernel of $Sq^3$ on $H^4(X,Z)_{tors}$. 
\medskip

$\Upsilon$ \qquad\qquad  The subspace of $\Upsilon_0$ such that $Sq^2 b$ has a {\it torsion} integral lift. 
\medskip

$V(a)$ \qquad\qquad An $E_8$ vector bundle on $X$ determined by $a\in H^4(X,\Z)$. 
\medskip

$x,y,z$ \qquad\qquad  Generic elements of $K(X)$. 
\medskip

$X,Y,Z$ \qquad\qquad  A spin manifold, usually of dimension 10, 11, or 12
        respectively.

\appendix{B}{Computation Of Some Mod 2 Indices}

To   complete the proof 
begun in section 3.3
that the symmetry $(-1)^{F_L}$ of
Type IIA superstring theory is anomaly free, we must      prove that
the anomaly vanishes for a manifold $V_{1,1}$ defined as a hypersurface of
degree $(1,1)$ in ${\bf CP}^2\times {\bf CP}^4$.
We let $x_i$, $i=1,\dots, 3$ be homogeneous coordinates for ${\bf CP}^2$,
and we let $y_j,$ $j=1,\dots ,5$, be homogeneous coordinates for 
${\bf CP}^4$.  We can take the equation defining $V_{1,1}$ to be
\eqn\kiplo{\sum_{i=1}^3x_iy_i=0.}
For every point $(x_1,x_2,x_3)\in {\bf CP}^2$, the $x_i$ are not all zero,
and this equation defines a hyperplane in ${\bf CP}^4$, which is
isomorphic to ${\bf CP}^3$.  Hence, $V_{1,1}$ can be viewed as
a ${\bf CP}^3$ bundle over ${\bf CP}^2$.

We can pick a metric on $V_{1,1}$ in which the fibers of $V_{1,1}\to
{\bf CP}^2$ are small and
have positive scalar curvature.  In such a metric, the Dirac operator
on $V_{1,1}$ has no zero modes, 
and hence the mod 2 index $q({\cal O})$ vanishes.
We wish to show that the mod 2 indices $q(T)$ and $q(\wedge^2T)$ also vanish.

\def\LT{{\cal T}}
For this we use the fact that $V_{1,1}$ is a complex manifold, and the
complexification of $T$ splits as $T\otimes_\R\C=\LT\oplus \bar \LT$, 
where $\LT$ is the holomorphic tangent bundle to $V_{1,1}$.  Hence,
$q(T)$ is the mod 2 reduction of the ordinary index $I_\LT  $ of the Dirac
operator on $V_{1,1}$ with values in $\LT$.  Likewise $\wedge^2T\otimes_R\C
=\wedge^2\LT\oplus \wedge^2\bar\LT$, and hence $q(\wedge^2 T)$ is the mod
2 reduction of the ordinary index $I_{\wedge^2\LT}
$ of the 
Dirac operator on $V_{1,1}$ with values in $\wedge^2\LT$.
We will show that $I_\LT
 $ vanishes, and that $I_{\wedge^2\LT}$ 
is even.  The strategy will be to compute the index by first
solving the Dirac equation along the fibers of $V_{1,1}\to {\bf CP}^2$
to get an ``index bundle'' $W\to
  {\bf CP}^2$; then we compute the index of the Dirac operator on 
$V_{1,1}$ as the index of the Dirac operator on ${\bf CP}^2$ with values in $W$.
A special case of this procedure is that the Dirac
index on $V_{1,1}$ with values in any bundle $F$
that is trivial when
restricted to each fiber of $V_{1,1}\to {\bf CP}^2$
vanishes. This can be proved rather as above by picking on 
$V_{1,1}$ a metric such that the fibers are small and have positive scalar
curvature; in this metric, the Dirac operator on $V_{1,1}$ with values in $F$
has no zero modes at all, so the index bundle is trivial.

We will first show that $I_\LT=0$.
Because of the fibration of $V_{1,1} $ over ${\bf CP}^2 $ with ${\bf CP}^3$
fibers, there is an exact sequence
\eqn\omigo{0\to \LT{\bf CP}^3\to \LT V_{1,1}\to \LT{\bf CP}^2\to 0.}
This sequence does not split holomorphically, but it does split topologically
and as we are just doing index theory, we can replace $\LT V_{1,1}$ by
$\LT{\bf CP}^3\oplus \LT{\bf CP}^2$.  Now, $\LT{\bf CP}^2$ is a pullback
from the base of $V_{1,1}\to {\bf CP}^2$, so it is trivial on each fiber
of $V_{1,1}$ and hence by a remark above has zero index.  We still
have to look at the index with values in $\LT{\bf CP}^3$.  In fact,
taking a metric on ${\bf CP}^3$ that is K\"ahler,
the Dirac operator on ${\bf CP}^3$ with values in $\LT{\bf CP}^3$
has no zero modes.  This operator is indeed equivalent to the $\bar\partial$
operator acting on $\LT {{\bf CP}}^3\otimes K^{1/2}$, where $K$ is the canonical
bundle of ${\bf CP}^3$.  As $K\cong{\cal O}(-4)$, we must show that
$H^i({\bf CP}^3,\LT
 {\bf CP}^3(-2))=0$ for all $i$.  For this, we use the existence of
an exact sequence
\eqn\onol{0\to {\cal O} \to {\cal O}(1)^4\to \LT{\bf CP}^3\to 0.}
This exact sequence expresses the fact that a tangent vector field
on ${\bf CP}^3$ can be written as
\eqn\jiccx{\sum_{i=1}^4a_i{\partial\over \partial y_i},}
 where the $a_i$
are functions of the $y$'s that are homogeneous of degree 1, and 
defined up to $a_i\to a_i+w y_i$ for any complex function $w$. 
A   twisted version of this exact sequence reads
\eqn\bonol{0\to {\cal O}(-2)\to {\cal O}(-1)^4\to \LT{\bf CP}^3(-2)\to 0.}
Vanishing of the cohomology of $\LT{\bf CP}^3(-2)$ now follows from the
long exact sequence of cohomology groups derived from \bonol, together
with the standard fact that $H^i({\bf CP}^n,{\cal O}(-j))=0$ for all
$i$ and $0<j<n+1$. From absence of zero modes of the Dirac operator 
on the fibers, it follows
that the index bundle is trivial and hence that $I(\LT)=0$.

We now turn to $I(\wedge^2\LT)$.  This case is more delicate, as the 
index bundle is not trivial.  
An argument like the one surrounding \omigo\ lets us replace $\wedge^2\LT$ by $\wedge^2
\LT{\bf CP}^3$.  (In fact, for index purposes $\wedge^2\LT=\wedge^2\LT{\bf CP}^2
\oplus \LT{\bf CP}^2\otimes \LT{\bf CP}^3\oplus\wedge^2\LT{\bf CP}^3$.
Here, $\wedge^2\LT{\bf CP}^2$ can be dropped as it is a pullback from
the base, and $\LT {\bf CP}^2\otimes \LT{\bf CP}^3$ can be dropped
as its restriction to each fiber is isomorphic to $\LT{\bf CP}^3$, which
as we have just seen has zero cohomology and zero index bundle.)
To 
analyze the index with values in $\wedge^2\LT{\bf CP}^3$, 
first note that, in the last paragraph, just for computing
the Dirac index
with values in $\LT{\bf CP}^3$
(as opposed to computing
all of the cohomology groups of $\LT{\bf CP}^3(-2)$, as we actually did),
we could have assumed that the exact sequence in \onol\ splits, leading
to the $K$-theory statement $\LT{\bf CP}^3    ={\cal O}(1)^4-{\cal O}$.
Likewise, we have a $K$-theory statement $
 \wedge^2\LT{\bf CP}^3 
={\cal O}(2)^6-{\cal O}(1)^4$. 
After twisting by $K^{1/2}={\cal O}(-2)$, we have
\eqn\jupolo{
 \wedge^2\LT{\bf CP}^3(-2)
={\cal O}   ^6-{\cal O}(-1)^4.}
The index bundle of the Dirac operator with values in $\wedge^2\LT{\bf CP}^3$
is the alternating sum of the cohomology groups of $\wedge^2\LT{\bf CP}^3(-2)$.
For computing the index bundle, we can replace this by
 ${\cal O}^6-{\cal O}(-1)^4$.
The only nonvanishing cohomology group of this bundle is $H^0({\bf CP}^3,
{\cal O}^6)=\C^6$.

\def\P{{\bf P}}
The index bundle $W$ of $\wedge^2
\LT{\bf CP}^3$ is thus a rank six complex vector bundle 
over ${\bf CP}^2$.  To identify this bundle, we need to repeat the
analysis in the last paragraph more precisely, to describe
the dependence of the cohomology on the $x_i$.  First of all, 
as $(x_1,x_2,x_3)$ varies, the solution space of \kiplo\ varies
as the bundle $M=U
\oplus {\cal O}\oplus {\cal O}$ over ${\bf CP}^2$,
where $U=\LT^*{\bf CP}^2(1)$.  (A triple $y_1,y_2,y_3$ obeying \kiplo\ 
determines
a differential form $\sum_i y_i\,dx_i$ of degree 1 on  ${\bf CP}^2$;
in the description of $M$,
the summands ${\cal O}$ come from $y_4 $ and $y_5$.) 
Hence, we can regard the ${\bf CP}^3$ fiber of $V_{1,1}\to {\bf CP}^2$ as
$\P M$, the projectivization of $M$.  
In \onol, we can regard ${\cal O}(1)^4$ as $M(1)$, and in \jupolo,
${\cal O}^6$ is $\wedge^2M\otimes {\rm det}(M)^{-1/2}$ (where the last
factor will be explained in a moment). So the fiber of the index
bundle $W$ is really
\eqn\kilmopo{H^0(\P M,\wedge^2\LT \P M\otimes K^{1/2}(\P M))=\wedge^2 M\otimes
\det\,M^{-1/2}.}
(The slightly delicate factor of $\det\,M^{-1/2}$ on the right hand side
can be explained as follows.  The left hand side of \kilmopo\ is manifestly
invariant under the action of $\C^*$ on $M$; hence $\C^*$ must act trivially
on the right hand side, which is so precisely if we include
    the given power of $\det\, M$.)  Now, with $M=U\oplus {\cal O}\oplus
{\cal O}$, we find that $\wedge^2M\otimes({\rm det}\,M)^{-1/2}$ is
\eqn\turnigo{W=
(\det U)
^{1/2} \oplus(\det U)
^{-1/2} \oplus 2
U\otimes(\det U)
^{-1/2}.}
(Note that ${\bf CP}^2$ is not a spin manifold, but $(
\det U)
^{1/2}$ is
a ${\rm Spin}^c$ structure on ${\bf CP}^2$.
This is why the index bundle $W$ 
is $(\det U)^{1/2}$ tensored with a conventional
vector bundle.) Two copies of $U\otimes (\det U)^{-1/2}$ do not contribute
to the mod 2 reduction of the index.  As the index in four dimensions
with values in a bundle $E$ 
is invariant under complex conjugation of $E$, the index with values in
$(
\det U)^{1/2}$ equals that with values in $(\det U)^{-1/2}$, so these
contributions to the mod 2 reduction of the index cancel also.

\appendix{C}{The Atiyah -- Hirzebruch Spectral Sequence}

The Atiyah -- Hirzebruch spectral sequence (AHSS)
is a systematic algebraic 
algorithm relating $K$-theory to integral cohomology. In this appendix 
we will give an elementary account of this formalism, explaining 
the construction of 
\ahss\ 
in more familiar physical terms. 
In order to simplify the presentation we will focus on the 
relation between $K^0$ and even cohomology classes. The extension 
to $K^1$ and odd cohomology classes is straightforward. 

Given a manifold $X$, we can study it's topology by introducing a 
triangulation that makes $X$ look as a collection of simplexes glued 
together along their boundaries. Then we can determine the homology 
or the cohomology of $X$ in terms of the gluing data using simple 
combinatorics. In order to do this in a systematic way, it is 
often convenient to think of $X$ as a superposition of finitely many 
strata $X^p$, each stratum consisting of all simplicial cells of 
dimension $p$. $X^p$ is called the $p$-skeleton of $X$. 

When studying the $K$-theory of $X$, the stratification of $X$ 
by skeletons induces a natural filtration of $K(X)$. 
We simply 
define $K^0_p(X)$ to be the subset of $K^0(X)$ consisting of 
classes which are trivial on the $(p-1)$-skeleton. In the 
present approach, it is more convenient to think of $K$-theory 
classes as $D$-branes on $X$. For concreteness, we assume the 
dimension $N$ of $X$ to be even.
A $D(2p-1)$-brane wraps a $2p$-submanifold of $X$ which is 
Poincar\'e dual to a cohomology class in $H^{N-2p}(X, {\bf Z})$. 
Such classes are supported on the $(N-2p)$-skeleton and cannot 
be detected on lower skeletons. 
Since $K^0_N(X)$ consists of classes trivial on the $(N-1)$ 
skeleton, it follows that the worldvolume of the corresponding 
$D$-brane states must be pointlike. These are $D$-instantons in IIB. 
Similarly, $K^0_{N-1}(X)$ consists of classes which are trivial 
on the $(N-2)$ skeleton. Since there are no stable even branes in IIB, 
it follows that $K^0_{N-1}(X)=K^0_{N}(X)$.
Next, $K^0_{N-2}(X)$ parameterizes objects wrapping 
submanifolds of dimension two, i.e. $D1$-branes and so on.
The complexity of $K^0_{N-2p}(X)$ increases as we increase 
$p$ since a $D(2p-1)$-brane can have induced lower $D$-brane 
charges on it's world volume. We have accordingly the following 
sequence of inclusions 
\eqn\ahssA{
K^0_{N}(X)\subseteq K^0_{N-1}(X)\subseteq\cdots \subseteq
K^0_0(X) =K^0(X).}
Note that not all these inclusions are strict; in fact 
$K^0_{N-2p}=K^0_{N-2p-1}$ for all $p$. 
There is a similar filtration on $K^1(X)$, which can be described 
in terms of $D$-branes in IIA.

The main idea of the AHSS is that although 
$K^0_{p}(X)$ are complicated objects, it may be easier to 
determine the so-called ``successive quotients''
\eqn\ahssB{
K^0_{p}(X)/K^0_{p+1}(X).}
Physically, this means that we are trying to understand $D$-brane 
charges starting with the lowest charges -- such $D$-instantons in IIB
-- and working our way towards higher charges. For example $D$-instantons
are pointlike on $X$, so the structure of $K^0_N(X)$ 
is very simple. $D$-instantons 
can dissolve in $D1$-branes forming bound states. These are classified
by $K^0_{N-2}(X)$. Since we have already understood $D$-instantons, 
one might think that at the next step it suffices to focus only 
on $D1$-branes, regardless of their lower $D(-1)$-charge.
As explained below, this is not quite true. If we ignore 
$D(-1)$-charges, $D1$-branes are classified by 
$K^0_{N-2}(X)/K^0_{N-1}(X)$. Proceeding
similarly at each stage, we construct the ``associated graded''
\eqn\ahssC{
\hbox{Gr}K^0(X)=\oplus_{p}K^0_{p}(X)/K^0_{p+1}(X).}
Although this 
simplifies the computation of $K^0(X)$, some information is lost 
in the process. More precisely, $K^0(X)$ is not uniquely defined 
by the associated graded \ahssC.\ When we try to construct $K^0(X)$, 
given $\hbox{Gr}K^0(X)$, we have to determine a finite number of
extensions of the form 
\eqn\ahssCA{
0\rightarrow K^0_{p+1}(X)\rightarrow K^0_{p}\rightarrow 
K^0_{p}(X)/K^0_{p+1}(X)\rightarrow 0,}
which may be ambiguous. For example, if $X$ is the real projective 
plane ${\bf RP}^5$, with the stratification given by the 
linear subspaces 
$X^{5-i}={\bf RP}^i$, $0\leq i \leq 5$, the associated graded is 
\eqn\ahssCB{
\hbox{Gr}K^0({\bf RP}^5)={\bf Z} \oplus {\bf Z}_2\oplus {\bf Z}_2.}
$K^0_1({\bf RP}^5)$ is determined by the extension 
\eqn\ahssCC{
0\rightarrow {\bf Z}_2 \rightarrow K^0_1({\bf RP}^5)\rightarrow {\bf Z}_2
\rightarrow 0}
which can be either ${\bf Z}_2 \oplus {\bf Z}_2$ or ${\bf Z}_4$. 
Such problems can be solved by a more careful study of $K^0(X)$. 
In the present case, the solution is 
\nref\adams{J.F. Adams, ``Vector Fields on Spheres'', Ann. Math. 
{\bf 75} (1962) 603.}%
\refs{\adams}
\eqn\ahssCD{
K^0_1({\bf RP}^5)={\bf Z}_4.}
On the contrary, the next extension 
\eqn\ahssCE{
0\rightarrow {\bf Z}_4\rightarrow K^0_0({\bf RP}^5)\rightarrow {\bf Z}
\rightarrow 0}
admits only the trivial solution, and the final result is 
\eqn\ahssCF{
K^0({\bf RP}^5)= {\bf Z}\oplus {\bf Z}_4.}

The advantage of the associated graded \ahssC\ is that it can be
determined by successive approximations. Recall that $X$ is made of 
finitely many skeletons $X^p$, each skeleton consisting of finitely
many simplicial cells $\sigma_i^p$. Each simplicial cell $\sigma_i^p$ 
is topologically equivalent to a $p$-ball $B^p$. The boundary 
${\dot\sigma}_i^p$ of 
$\sigma^p_i$ consists of $(p+1)$ faces which are $(p-1)$-simplexes 
themselves belonging to the $(p-1)$-skeleton of $X$. 
The simplest object that can be 
formed out of this local data is 
\eqn\ahssD{
E_1^{p}=K^0(X^p, X^{p-1})=\oplus_{i}
K^0(\sigma_i^p, {\dot\sigma}_i^p).}
$E_1^p$ parameterizes $K$-theory 
classes defined on the $p$-skeleton which are trivial on the
$(p-1)$-skeleton. Note however that at this stage we do not know 
if these classes can be lifted to full $K$-theory classes on $X$. 
That is why $E_1^p$ can be thought as a zeroth order approximation 
to $K^0_{p}(X)/K^0_{p+1}(X)$.
In mathematical terms, $E_1^p$ is called the first term of the 
spectral sequence. 

Before moving on, let us rewrite \ahssD\ in a more familiar form. 
$K$-theory classes on $\sigma_i^p$ which are trivial on the boundary 
can be identified with classes on the $p$-sphere $S^p$ by collapsing 
${\dot \sigma}_i^p$. More precisely, 
$K^0(\sigma_i^p, {\dot\sigma}_i^p)$ can be identified with the 
reduced $K^0$-theory of a sphere $S^p$, which is isomorphic to 
$K^p$ of a point by Bott periodicity.
Therefore the first 
term $E_1^p$ can be identified with singular $p$-cochains on
$X$ with values in $K^p(x_0)$ ($x_0$ is an arbitrary base point of $X$)
\eqn\ahssF{
E_1^p=C^p(X; K^p(x_0))=\left\{
\matrix{&C^p(X,{\bf Z}),\qquad & p\ \hbox{even}\cr
& 0,\qquad & p\ \hbox{odd.}\cr}\right.}

Higher AHSS approximations involve a systematic refinement of \ahssD.\
We want to characterize the $K$-theory classes in $E_1^p$ which 
can be lifted to $X$. This question can be answered inductively, by 
first determining the classes which can be lifted to the 
$(p+1)$-skeleton. These will form a second term $E_2^p$, which can 
be further refined by restricting to classes which can be 
lifted to the $(p+2)$-skeleton and so on. The power of this 
approach resides in the fact that at each step, one can define 
a differential 
\eqn\ahssG{
d_r^p:E_r^p\rightarrow E_r^{p+r},\qquad d_r^p\circ d_r^{p-r}=0}
such that $E^p_{r+1}$ is the cohomology of $d_r$
\eqn\ahssH{
E^p_{r+1}=\hbox{Ker}(d_r^p)/\hbox{Im}(d_r^{p-r}).}
The spectral sequence consists of the collection $(E_r,d_r)$
obtained by summing over all $p$. 
In practice, after finitely many steps, this sequence becomes 
stationary and we obtain the successive quotient \ahssB. The spectral 
sequence is said to converge to the associated graded \ahssC.

This construction can be understood in terms of $D$-branes as well. 
As discussed in detail in sections 5.1.-- 5.2., a $D$-brane 
cannot wrap a submanifold $Q$ of $X$ unless the Poincar\'e 
dual class $b$ can be lifted to $K$-theory. In the process of 
constructing the associated $K$-theory class, one has to extend 
the tachyon condensate 
\onso\
as a unitary map between two bundles 
over the entire manifold $X$. The AHSS is simply an algorithm 
for keeping track of the possible obstructions. At each stage 
we keep only states for which $T$ can be extended a finite number 
of steps. At the same time, we have to mod out by unstable states, 
for which the extension is trivial. The whole procedure is 
reminiscent of a refined BRST quantization scheme, consisting 
of a sequence of BRST operators $d_r$, each acting on the space 
of physical states of the previous one. The true physical 
space is obtained by taking the cohomology of all operators. 

Let us work out $d_1^p$. The derivation is essentially identical to 
the extension of the tachyon field in section 5.2. We assume $p$ 
even in order to have a nontrivial $E_1^p$. Let $\sigma^{p+1}$ 
be an arbitrary $(p+1)$-simplex with faces $\sigma^p_i$. Suppose
we have $K$-theory classes $x_i$ in $K^0(\sigma_i^p,{\dot \sigma}_i^p)$.
As we saw in section 5.2, $K^1$classes are classified by homotopy 
classes of maps to the infinite unitary group $U$. Here we have 
$K^0$ classes which can be similarly classified by maps to the 
loop space\foot{The only property of $\Omega U$ needed in the following 
is that $\pi_i(\Omega U)=\pi_{i+1}(U)$.} of $U$, $\Omega U$.
So we can think of $x_i$ as maps $f:\sigma_i^p\rightarrow \Omega U$ 
mapping the boundary to a point, or equivalently as maps 
$f_i:S^p\rightarrow \Omega U$. Since the $\sigma_i^p$ are the faces 
of $\sigma^{p+1}$, the maps $f_i$ can be glued along the boundaries,
resulting in a map 
$f:{\dot\sigma^{p+1}}\rightarrow \Omega U$. With the orientations 
properly taken into account, the homotopy class of this map in 
$\pi_{p}(\Omega U) ={\bf Z}$ is 
\eqn\ahssI{
[f]=\sum_i(-1)^i[f_i].}
The map $f$ can be extended to the interior of $\sigma^{p+1}$ if 
and only if $[f]$ is trivial in $\pi_{p}(\Omega U)$. Therefore
$[f]$ is the obstruction to extending $\{x_i\}$ to the 
$(p+1)$-skeleton. 
By assigning such an $[f]$ to all $(p+1)$-simplexes, we obtain a
map from the $p$-cochains of $X$ to the $(p+1)$-cochains. This is 
the first AHSS differential
\eqn\ahssJ{
d_1^p :C^p(X; K^p(x_0))\rightarrow C^{p+1}(X; K^p(x_0)).}
According to \ahssI,\ $d_1^p$ is precisely the standard simplicial 
coboundary operator. 

Following the steps outlined above, the second term in the AHSS is 
therefore 
\eqn\ahssJ{
E_2^p = H^p(X; K^p(x_0))=\left\{\matrix{
& H^p(X;{\bf Z}),\qquad & p\ \hbox{even}\cr
& 0,\qquad & p\ \hbox{odd.}\cr}\right.}
For $K^1$, we obtain an analogous formula, with $p$ even replaced 
by $p$ odd. 
This explains the formulae 
\kipp\ and \jipp\ 
in the main text. $E_2^p$ can be regarded as a first order approximation 
to $K^0_p(X)/K^0_{p+1}(X)$. 

We can continue this process in a similar manner 
(see the discussion of the extension of $T$ in section 5.2.) 
The next differential
is trivial since $\pi_{p+1}(\Omega U)$ is trivial if $p$ is even.
This argument generalizes to all even differentials, 
showing that 
\eqn\ahssK{
d_{2r}^p = 0.}
So the next nontrivial obstruction is encountered when extending over 
the $(p+3)$-skeleton and it takes values in $\pi_{p+2}(\Omega U)$.
The corresponding AHSS differential is a ``cohomology operation''
\eqn\ahssL{
d_3^p:H^p(X;{\bf Z})\rightarrow H^{p+3}(X; {\bf Z}).}
Apparently there is no simple derivation of $d_3^p$, although 
the arguments in section 5.1. suggest that 
\eqn\ahssK{
d_3^p = Sq^3.}
This is indeed the correct answer \ahss.\

\appendix{D}{Spin Ten-Manifolds with $W_7\neq 0$}
\def\to{{{\widetilde \Omega}^{spin}}}
\def\th{{\widetilde H}}

In section 6.1 we found that $M$-theory on a spin manifold of the form 
$X \times S^1$ is inconsistent if $W_7(X)\neq 0$. This is an important 
constraint on the theory and we would like to know if spin ten-manifolds 
with $W_7(X)\neq 0$ exist. Unfortunately, there does not seem to exist 
an elementary example. However,
the existence of such manifolds can be inferred from 
an abstract cobordism argument 
explained to us by Mike Hopkins. The main idea is to use the 
Pontryagin duality 
\eqn\appdA{
H^7(X;\Z)\times H^3(X;\Q/\Z)\rightarrow \Q/\Z.}
Regarding $\Q/\Z$ as a subgroup of $U(1)$, this can be related to the 
duality discussed in section 4.2 (see formula \olop.) We have similarly 
a short exact sequence 
\eqn\appdB{
0\rightarrow \Z\rightarrow \Q \rightarrow \Q/\Z\rightarrow 0}
and a Bockstein map $\beta:H^k(X;\Q/\Z)\rightarrow H^{k+1}(X;\Z)$. 
Since the group $\Z_2$ can be embedded in $\Q/\Z$, we can regard 
the Stiefel-Whitney classes $w_k(X)$ as elements of $H^k(X;\Q/\Z)$. 
With this understanding we have 
\eqn\appdC{
W_7(X) = \beta(w_6(X)).}
The pairing \appdA\ is nondegenerate, hence $W_7(X)$ is nonzero 
if and only if there exists an element $\xi\in H^3(X;\Q/\Z)$
such that 
\eqn\appD{
I(X,\xi)=\int_X \xi\cup W_7(X)\neq 0.}
Therefore, it suffices to establish the existence of a pair $(X,\xi)$ 
such that $I(X,\xi)\neq 0$. Note that $I(X,\xi)$ is a cobordism 
invariant 
of the pair $(X,\xi)$ in the sense explained in section 3.2. The class
$\xi$ is classified by a map $f:X \rightarrow K(\Q/\Z,3)$ and 
$I(X,\xi)\in \hbox{Hom}\left(\to_{10}(K(\Q/\Z,3)),\Q/\Z\right)$. 
Now, the invariant \appD\ can be rewritten 
\eqn\appdE{
I(X,\xi)=\int_X \beta(\xi)\cup w_6(X),}
where we have to use the pairing between $H^4(X;\Z)$ and $H^6(X;\Q/\Z)$.
Note that this is in fact a familiar cobordism invariant encountered 
in section 3.2. Setting $a=\beta(\xi)\in H^4(X;\Z)$, we have 
$I(X,\xi)=v(a)$ as defined in equation (3.16). This is an invariant 
of the group $\to_{10}(K(\Z,4))$. Moreover, we can find a bordism 
class $(Y,a)$ in $\to_{10}(K(\Z,4))$ such that 
\eqn\appdEA{
\int_Y a \cup w_6(Y)=1.}
For example, pick $Y$ to be the degree $(1,1)$ hypersurface $V_{1,1}$
introduced in section 3.3 (below \eightspin), 
and $a = \lambda({V_{1,1}})$.\foot{In fact we have to be slightly 
more careful 
here. The pair $(Y,a)$ does not define an invariant of the reduced
bordism group $\to_{10}(K(\Z,4))$ since $Y$ is not a 
boundary. We have to 
replace $Y$ by a sum of two copies of $V_{1,1}$, one of them with 
reversed orientation. Then we take $a$ to be $\lambda(V_{1,1})$ 
supported on one of the two components.}
Note that $a = g^*(u)$ where $g:Y\rightarrow K(\Z,4)$ is a continuous 
map and $u\in H^4(K(\Z,4);\Z)$ is the standard generator. 

Given the existence of a such a pair $(Y,g)$, in order to find a 
pair $(X,f)$ as above it suffices to prove that the 
bordism groups 
$\to_{10}(K(\Q/\Z,3))$ and $\to_{10}(K(\Z,4))$ are isomorphic. 
Note that the exact sequence \appdB\ induces a canonical map 
$\pi:K(\Q/\Z,3)\rightarrow K(\Z,4)$. Let $C_\pi$ denote the mapping cone 
of $\pi$ so that we have a sequence of maps of the form 
\eqn\appdF{
K(\Q/\Z,3){\buildrel \pi\over\longrightarrow}
K(\Z,4)\rightarrow C_\pi.}
This induces a long exact sequence of bordism groups which reads in part 
\eqn\appdG{
\cdots \to_{11}(C_\pi)\rightarrow 
\to_{10}(K(\Q/\Z,3)){\buildrel \pi_*\over \longrightarrow}
 \to_{10}(K(\Z,4))\rightarrow \to_{10}(C_\pi)
\cdots}
We conclude that $\pi_*:\to_{10}(K(\Q/\Z,3))\rightarrow
 \to_{10}(K(\Z,4))$ is an isomorphism if one can prove that the 
bordism groups $\to_{10}(C_\pi),\to_{11}(C_\pi)$ vanish. 

This follows from the Atiyah-Hirzebruch spectral sequence 
for $C_\pi$. Note that by construction the integral homology of $C_\pi$ 
is given by $H_*(C_\pi, \Z) \simeq H_*(K(\Z,4)) \otimes_\Z \Q$. 
Using the universal coefficient theorem, one can easily establish 
that $H_*(K(\Z,4)) \otimes_\Z \Q \simeq \Q[\xi]$, where $\xi$ is a 
degree four generator. The second term in the AHSS for $C_\pi$ 
is $E^2_{p,q}=\th_p\left(C_\pi,\Omega^{spin}_q\right)$. For 
$p+q\leq 12$, we obtain the following nontrivial terms 
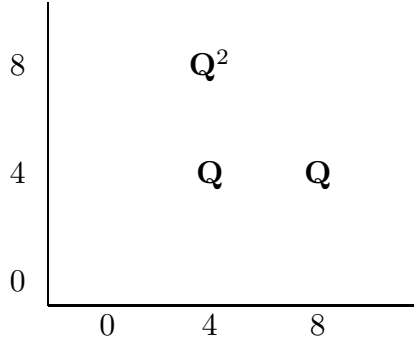
\diagram[PostScript=dvips,tight,height=1.7em,width=1.7em]
                     & \vLine &   &  &          &  & & &  \\
\lower 10pt\hbox{\hskip 15pt 8} &        &   &  &
\lower 10pt\hbox{${\bf Q}^2$} &  & & &  \\
                     &        &   &  &          &  & & &  \\
\lower 10pt\hbox{\hskip 15pt  4}&        &   &  &
\lower 10pt\hbox{\bf Q}   &  &\lower 10pt\hbox{\bf Q} & & \\
                     &        &   &  &          &  & & &  \\ 
\lower 10pt\hbox{\hskip 15pt 0} &        &   &  &          &  & & &  \\
                     &\HmeetV &  \hLine &  &      &  & & &  \\
                     &        & \raise 10pt\hbox{ 0} &  &  
\raise 10pt \hbox{4}  & 
&\raise 10pt \hbox{8} & & \\
\enddiagram

\noindent
Therefore $\to_{11}(C_\pi)=\to_{10}(C_\pi)=0$ since all terms in degrees 
$p+q=10,11$ are zero, and we obtain the desired isomorphism. 

Let us make this isomorphism more explicit. We will mainly exploit the 
surjectivity of $\pi_*$. Given the cobordism class of $(Y,g)$ defined 
above, it follows that there must exist a class $(X,f)$, $f:X\rightarrow 
K(\Q/\Z,3)$ such that
\eqn\appdI{
(Y,g) = \pi_*(X,f).}
The interpretation of this relation is quite elementary, according
to section 3.2. Namely, it means that we can find a spin manifold 
$Z$ with boundary $X-Y$ and a map $F:Z\rightarrow K(\Z,4)$
such that $F$ restricted to $X$ is $\pi\circ f :X \rightarrow K(\Z,4)$
and $F$ restricted to $Y$ is $g$. This shows that 
\eqn\appdJ{
\int_X (\pi\circ f)^*(u)\cup w_6(X) =\int_Y g^*(u)\cup w_6(Y)}
since the expressions in question are cobordism invariants. 
We have $(\pi\circ f)^*(u)=f^*\pi^*(u)$ where $\pi^*(u)$ is the pull 
back of $u$ to $H^4(K(\Q/\Z,3);\Z)$. This group consists entirely 
of elements of the form $\beta(\eta)$, with 
$\eta\in H^3(K(\Q/\Z,3);\Q/\Z)$. Therefore there exists such an element 
$\eta$ so that $\pi^*(u)=\beta(\eta)$. The pull back $f^*(\eta)$ 
defines an element $\xi\in H^3(X,\Q/\Z)$. Collecting all the facts, 
it follows that there must exist a pair $(X,\xi)$ such that 
\eqn\appK{
I(X,\xi)=\int_X\xi\cup W_7(X) =\int_Y a\cup w_6(Y) =1.}
This proves the claim. 

\listrefs
\end